\documentclass[
nofootinbib,
amsmath,amssymb,
aps,
onecolumn
]{revtex4-2}

\usepackage{multirow}
\usepackage{amsmath}
\usepackage{graphicx}
\usepackage{dcolumn}
\usepackage{bbm}
\usepackage{bm}
\usepackage{subfigure}
\usepackage{amssymb}
\usepackage{hyperref}
\usepackage{xcolor}
\usepackage{subfigure}
\usepackage{makecell}
\usepackage{mathtools}
\usepackage{booktabs}
\usepackage{xparse}
\usepackage{tikz-feynman}
\usepackage{color}
\usepackage{indentfirst}
\usetikzlibrary{decorations.pathreplacing,decorations.markings}

\stepcounter{secnumdepth}
\stepcounter{tocdepth}

\tikzset{
  wavy/.style={decorate,
  decoration={snake,amplitude=.4mm,segment length=1mm, post length=0mm,pre length=0mm}},
  on each segment/.style={
    decorate,
    decoration={
      show path construction,
      moveto code={},
      lineto code={
        \path [#1]
        (\tikzinputsegmentfirst) -- (\tikzinputsegmentlast);
      },
      curveto code={
        \path [#1] (\tikzinputsegmentfirst)
        .. controls
        (\tikzinputsegmentsupporta) and (\tikzinputsegmentsupportb)
        ..
        (\tikzinputsegmentlast);
      },
      closepath code={
        \path [#1]
        (\tikzinputsegmentfirst) -- (\tikzinputsegmentlast);
    },
   },
  },
  mid arrow/.style={postaction={decorate,decoration={
        markings,
        mark=at position .55 with {\arrow[#1]{stealth}}
      }}},
}

\begin{document}

\title{ Fixed Point Stability Switches from Attractive to Repulsive \\\vspace{1mm} at 2d Pomeranchuk/Stoner Instabilities via Field-Theoretical RG}

\author{Han Ma}
 \affiliation{Department of Physics and Astronomy, Stony Brook University, \\ Stony Brook, New York 11974, USA }
\affiliation{%
 Perimeter Institute for Theoretical Physics, Waterloo ON N2L 2Y5, Canada
}%

\date{\today}

\begin{abstract}
We study an interacting two-flavor fermionic system via field-theoretical functional renormalization group (RG). 
Each flavor, labeled by $\pm$, has a dispersion of $E^{\pm}=c k^{2\alpha}-\mu^\pm$ with tunable real exponent $\alpha>0$. The effective theory is parametrized by intra-flavor and inter-flavor interactions, preserving global U(1) $\times$ U(1) symmetry, which can be enhanced to U(2). The U(2) symmetric system has a Fermi liquid phase and two possible instabilities, leading to spontaneous spatial rotational or flavor symmetry breaking, known as the Pomeranchuk and Stoner instabilities, respectively. The key discovery of this work is the following. The Stoner instability possesses an RG fixed point that preserves the U(2) symmetry. For $\alpha<1$, this fixed point is attractive, indicating a continuous transition. Conversely, for $\alpha>1$, the fixed point becomes repulsive, and without fine-tuning, there is runaway RG flow, resulting in a discontinuous transition. The U(1) $\times$ U(1) symmetric system, with $\mu^+\neq \mu^-$, exhibits richer physics. 
This system have two Pomeranchuk instabilities. At one of them, a non-trivial RG fixed point switches its nature from attractive to repulsive as $\alpha$ increases across $1$. Notably, the runaway flow at $\alpha>1$ results in the depletion of a Fermi surface at the transition.
Collective modes in these Fermi liquids are also investigated. A universal Fermi surface deformation ratio $\delta\mu^+/\delta\mu^-$ is predicted for $\alpha<1$ at the instability as a continuous transition, which can be observed experimentally.
\end{abstract}

\maketitle

\addtocontents{toc}{\string\setcounter{tocdepth}{3}}
\addtocontents{toc}{\protect\setcounter{tocdepth}{3}}

\vspace{10pt}
\noindent\rule{\textwidth}{1pt}
\tableofcontents

\noindent\rule{\textwidth}{1pt}
\vspace{10pt}

\section{Introduction}
\label{sec:intro}

A continuous phase transition is characterized by a renormalization group (RG) fixed point, reflecting scale invariance at long wavelengths\cite{cardy1996scaling,sachdev1999quantum,fradkin2013field}. Such a fixed point encodes the universal critical behavior—properties that do not depend on microscopic details. By studying the RG flow of couplings and fields, one can pinpoint whether the system flows to a fixed point in the infrared (IR), corresponding to a continuous phase transition.

In systems with Fermi surfaces, the conventional separation between high-energy (UV) and low-energy (IR) modes based on the magnitude of momentum breaks down. Instead, due to the infinite number of zero-energy modes at the Fermi surface, low-energy modes are associated with momenta close to the Fermi surface. Consequently, the low-energy space is characterized by Fermi momentum. The scale set by this Fermi momentum plays a fundamental role in the IR physics. As a result, any effective theory describing the low-energy physics must incorporate this scale. Based on this key property, various effective theories are proposed and developed\cite{neto1994bosonization,neto1994bosonization1,dupuis1996renormalization,dupuis1998fermi,dupuis2000unified,Haldane2005luttinger,Delacretaz2022nonlinear,Khveshchenko2024pre,Khveshchenko2024strange}. From RG perspective, this appears to be at odds within the standard framework, which seeks scale-invariant fixed points. However, since this scale only reflects the size of the zero-energy space, it can effectively act as a prefactor in the total theory\cite{SHANKAR,POLCHINSKI1,borges2023field,ma2024fermi}. \textit{The low energy effective theory is thus obtained by multiplying the area of zero-energy space and the action density per unit area, which demonstrates scale invariant}. 

In addition, the presence of this extensive number of zero-energy modes results in a fundamental non-commutativity between the limits of zero momentum relative to the Fermi momentum $q\rightarrow 0$ and zero energy/frequency $E\rightarrow 0$ in physical observables\cite{landau1959theory,abrikosov1958theory,nozieres1962derivation,pines2018microscopic,abrikosov2012methods,chubukov2023nonanalytic,chubukov2023singular,DASSARMA2021168495,chubukov2018fermi,chubukov2005thermodynamics,delacretaz2025symmetry}. Consequently, for the systems with Fermi surfaces, isolating the scale invariant part of low-energy physics requires (1) focusing on the action density and (2) keeping track of the RG flow of the couplings as functions of $q$ away from the zero energy space, as the energy scale $E$ decreases. 

Using the field theoretical functional RG method developed in Ref.\cite{borges2023field,borges2023emergence,ma2024fermi,kukreja2024projective}, we study the RG flow of these coupling functions and the chemical potentials. The Fermi liquid is characterized by scale invariant IR coupling functions in the action density. They take finite values for arbitrary momentum $q$. Instability arises when an IR coupling function develops a singularity. At the mean time, chemical potentials also flow under RG. A fixed point, if reached, reveals the critical system's scale-invariant universal properties. 
\begin{figure}[h]
    \centering
    \includegraphics[width=0.42\linewidth]{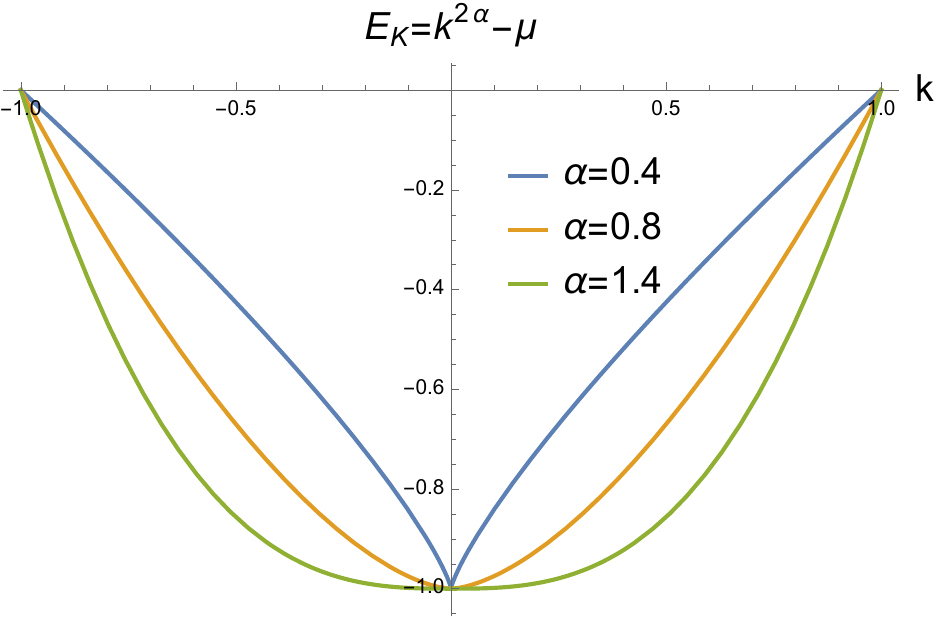}
    \caption{Dispersion relation $E=k^{2\alpha}-\mu$ for various values of $\alpha$. The chemical potential is set to $\mu=1$. 
    }
    \label{fig:dispersion}
\end{figure}

Recent work\cite{raines2024two,raines2024unconventional,raines2024unconventional1} revealed that a two-flavor fermionic system exhibits a Stoner-like instability. It spontaneously breaks flavor symmetry and was called valley polarization transition in the context where the flavors correspond to the valley index. Mean‐field calculations yield the following results. With parabola dispersions for each flavor, the susceptibility diverges on the Fermi liquid side. However, it jumps discontinuously once the system enters the symmetry‐breaking phase. Moreover, extending the dispersion to $E^{\pm}=c k^{2\alpha}-\mu^\pm$ with tunable parameter $\alpha>0$, as shown in Fig.~\ref{fig:dispersion}, the properties of the transition are modified. For $\alpha<1$, the transition is continuous, with finite $\mu^\pm$ varying smoothly across the critical point. Two Fermi surfaces exist at the instability. For $\alpha>1$, only one Fermi surface remains at the transition, suggesting a discontinuous jump of order parameter\cite{raines2024unconventional}. Thus, $\alpha$ controls whether the system undergoes a second‐order or a first‐order transition and whether both Fermi surfaces exist at the instability. 

\begin{figure}[h]
    \centering
    \includegraphics[width=0.8\linewidth]{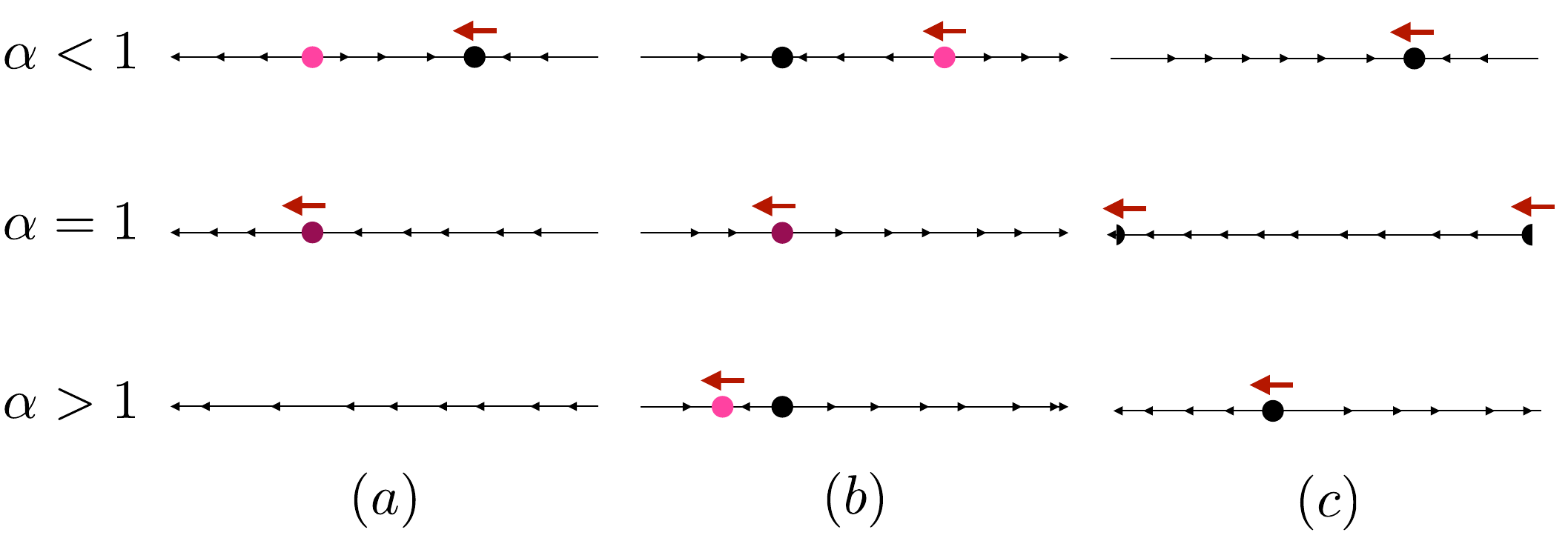}
    \caption{(a)Fixed points collision and annihilation resulting in a change of phase transition from continuous to discontinuous. (b)Fixed points merging and collision resulting in a change of fixed point stability from attractive to repulsive. (c) Fixed point stability switches without collision. Instead, the only fixed point disappears at one end of the parameter space and reappears at the other.}
    \label{fig:fp_dynamics}
\end{figure}

From the RG perspective, a potential explanation for the observed transition type change at $\alpha=1$ involves the collision of two fixed points. This phenomenon occurs in diverse systems, attracting significant attention and interest\cite{wang2022deconfined,kaplan2022conformality,Gorbenko2018a,Gorbenko2018b,Ma2019,Haldar2023,tang2024reclaiming}. In this scenario, as depicted in Fig.~\ref{fig:fp_dynamics}(a), with $\alpha$ being a parameter that does not flow under RG, we assume that two fixed points exist at the instability for $\alpha<1$: one with two Fermi surfaces (black) and the other with a single Fermi surface (red). The two-Fermi-surface fixed point is expected to be stable, representing a continuous phase transition. Upon crossing $\alpha=1$, these fixed points collide and annihilate. At the collision point, an additional exactly marginal operator emerges. When $\alpha>1$, the runaway renormalization group flow indicates a discontinuous transition. Another possible scheme involves these two fixed points passing by each other\cite{gukov2017rg,ma2023quenched}, as shown in Fig.~\ref{fig:fp_dynamics}(b). Instead of annihilation, one fixed point passes the other as $\alpha$ cross $1$. The stability shifts from the two-Fermi-surface fixed point at $\alpha<1$ to the single-Fermi-surface fixed point at $\alpha>1$. Upon crossing $\alpha=1$, these fixed points merge, exchanging their stability and potentially introducing additional marginal operators as well. At $\alpha=1$, the merged fixed point has a runaway flow towards certain direction which suggests a first‐order transition. Conversely, at the fixed point itself, we anticipate scale invariance and power-law correlation functions. If the ultraviolet coupling is within an appropriate range, this scenario also explains the first-order transition at $\alpha>1$.

Alternatively, a simpler scenario is illustrated in Fig.~\ref{fig:fp_dynamics}(c). Here, a single fixed point with two Fermi surfaces exists. It is predicted to be stable for $\alpha<1$ and unstable for $\alpha>1$. As $\alpha$ approaches $1$ from below, the fixed point moves towards parameter space boundary and vanishes there. Upon exceeding $\alpha=1$, it reappears from the opposite boundary, thereby altering its stability. This indicates a first-order transition for $\alpha\geq 1$. Given that the flow applies to the chemical potential, the discontinuous transition is accompanied by the depletion of one Fermi surface. As detailed below, this scenario precisely manifests in the two-flavor interacting fermionic system with at least U(1) × U(1) symmetry. At $\alpha=1$, although there is no fixed point at the instability, the susceptibility can still diverge near the instability on the Fermi liquid side. 

In this work, we apply the field theoretical functional RG approach to study possible instabilities of two-flavor Fermi liquid phases. Specifically, we identify two types of instabilities:
\begin{itemize}
    \item \textbf{Pomeranchuk instability}\cite{pomeranchuk1958stability,oganesyan2001quantum,dell2006fermi,wu2007fermi,chubukov2009spin,maslov2010fermi,mross2010controlled,hartnoll2014transport,metlitski2010quantum,wu2018conditions}: arising from density fluctuations, which lead to rotational symmetry breaking of the circular Fermi surface;
    \item \textbf{Stoner instability}\cite{stoner1938collective,stoner1939collective,raines2024two}: caused by fluctuations associated with the flavor index, resulting in spontaneous flavor symmetry breaking.
\end{itemize}
It is noteworthy that both of them occur at $q=0$, and thus do not induce translational symmetry breaking.
In a U(2) symmetric system, we are able to find both types of instabilities\cite{wu2007fermi}. Furthermore, RG flow calculations of the chemical potentials reveal that the Stoner instability in the U(2) system undergoes a continuous transition for $\alpha<1$ and a discontinuous transition for $\alpha>1$. 
In a U(1) $\times$ U(1) system, we find two Pomeranchuk instabilities. One of them corresponds to a fixed point that is attractive for $\alpha<1$ and repulsive for $\alpha>1$.
Similar to the U(2) case, this Pomeranchuk instability in the U(1) $\times$ U(1) system exhibits a transition type change. But distinctly, one Fermi surface vanishes when $\alpha>1$. In total, our study is not only consistent with mean-field results, but also delineates the parameter regime where the phenomena of interest manifest. Furthermore, a universal Fermi surface deformation is obtained from the study of collective modes at the instabilities, offering a testable prediction for future experiments.

Our analysis of the Stoner instability in the U(2) symmetric system elucidates multiple phenomena. It not only accurately describes the conventional onset of ferromagnetism in interacting spinful fermionic systems with $\alpha=1$\cite{stoner1938collective,stoner1939collective}, but also extends to the isospin polarization transitions proposed in Moir\'{e} systems\cite{ghazaryan2021unconventional,zhou2021half,zhou2022isospin,chatterjee2022inter,liu2022isospin,dong2023isospin,raines2024unconventional,raines2024unconventional1}. As we only focus on $q=0$ instabilities, inter-valley momentum is irrelevant to their low energy physics. In addition, it provides a mathematical framework to study the mechanism of interaction-induced altermagnetism\cite{wu2007fermi,jungwirth2024supefluid}, attributed to density fluctuations that vary across flavors. More generally, our results on the Pomeranchuk instabilities in a U(1) $\times$ U(1) Fermi liquid can address the physics of interacting fermions with two valleys, where chemical potentials $\mu^\pm$ may differ. Multilayer Moir\'{e} systems offer an ideal platform\cite{balents2020superconductivity,andrei2021marvels}. Firstly, rhombohedral graphene exhibit very flat valence and conduction bands with dispersions beyond the conventional quadratic form, such as $|k|^n-\mu$ in $n$-layer graphene. Consequently, as the dispersion varies with $\alpha \neq 1$, quantum phase transition changed from continuous to first order can be observed. Secondly, as a highly tunable system, the parameter regime of interest would be accessible in Moir\'{e} materials. This allows future experiments to test our interesting findings that, for $\mu^+\neq \mu^-$, one Fermi surface may vanish as the number of layers increases in these Moir\'{e} systems at the Pomeranchuk instability.

In Sec.~\ref{sec:summary}, we provide a summary of the results. We then elaborate the details of the effective theory as our starting point in Sec.~\ref{sec:theory}. Tunable parameters are identified. In Sec.~\ref{sec:kinetic_eqn}, the action is discussed at the mean field level. Next, using the field theoretical functional renormalization group outlined in Sec.~\ref{sec:RG_flow}, we study the Fermi liquid phases and their instabilities in Sec.~\ref{sec:FPs}, to check if these transitions out of Fermi liquid phases correspond to RG fixed points. Finally, in Sec.~\ref{sec:collective_modes}, we investigate the collective modes as the pole of IR coupling functions and explore their behavior at different parameter regimes. 

\section{Summary of Results \label{sec:summary}}

We begin with a summary of results for interacting two-flavor fermionic systems. The flavor index , denoted $\pm$, can represent spin, valley, or other distinct quantum numbers. We generalize the dispersion to $E^{\pm}=c k^{2\alpha}-\mu^\pm$ where $\alpha>0$ is a tunable exponent. Galilean invariance holds at $\alpha=1$. Systems with real-space locality are of physical interest, which can be achieved when $2\alpha$ is an integer. For theoretical exploration, we allow arbitrary real $\alpha$ to study continuous interpolation between integer values of $2\alpha$. In the low energy limit, we consider nearly forward scatterings processes, preserving at least U(1) $\times$ U(1) symmetry. Quartic interactions (Eq.~\eqref{eq:action}) include intra-flavor couplings $\lambda_\pm$ and inter-flavor couplings:  $\lambda_{1}$ for density-density interactions defined in Eq.~\eqref{eq:density-density} and $\lambda_2$ for flavor-exchange scatterings defined in Eq.~\eqref{eq:flavor_exchange}. Correspondingly, we define generalized Landau parameters as $F_\pm = \frac{(\mu^\pm)^{1/\alpha-1}}{\pi \alpha c^{1/\alpha}  } \lambda_\pm $ and $F_{1,2}= \lambda_{1,2} \frac{(\mu^+ \mu^-)^{\frac{1-\alpha}{2\alpha}}}{\pi \alpha c^{1/\alpha}  }$. Employing the field-theoretical functional RG method, we extend these couplings to coupling functions $F_{\pm,1,2}(\vec{q})$ depending on the momentum transfer of the nearly forward scattering processes. We analyze the RG flow of these functions and the chemical potentials. Finite IR coupling functions at arbitrary $q$ define the universal physics of the Fermi liquid phase in the low energy limit. Instability arises when an IR coupling function diverges. By tracking the RG flow of chemical potentials, we identify the RG fixed point associated with this instability.

When $\mu \equiv \mu^+=\mu^-$ and $F \equiv F_+=F_-=F_1+F_2$, the global symmetry is enhanced to U(2). The system is in Fermi liquid phase provided $F\pm F_1 >-1$. There can be two types of instabilities: the Pomeranchuk instability leading to rotational symmetry breaking at $F+ F_1 \leq -1$ due to density fluctuations, and the Stoner instability spontaneously breaking flavor symmetry at $F- F_1 \leq -1$. The Stoner instability is associated with the diverging quantum fluctuation tied to flavor index, which serves as an order parameter distinguishing the symmetry breaking phase. We identify a one-loop RG fixed point of the chemical potentials at the Stoner instability. The fixed point's stability changes from attractive to repulsive as $\alpha$ increases across $1$, a scenario depicted in Fig.~\ref{fig:fp_dynamics}(b). At the transition, U(2) symmetry is still preserved implying that $\mu^+=\mu^-$ holds at the transition point for any value of $\alpha$.

\begin{figure}[h]
    \centering
    \includegraphics[width=0.9\linewidth]{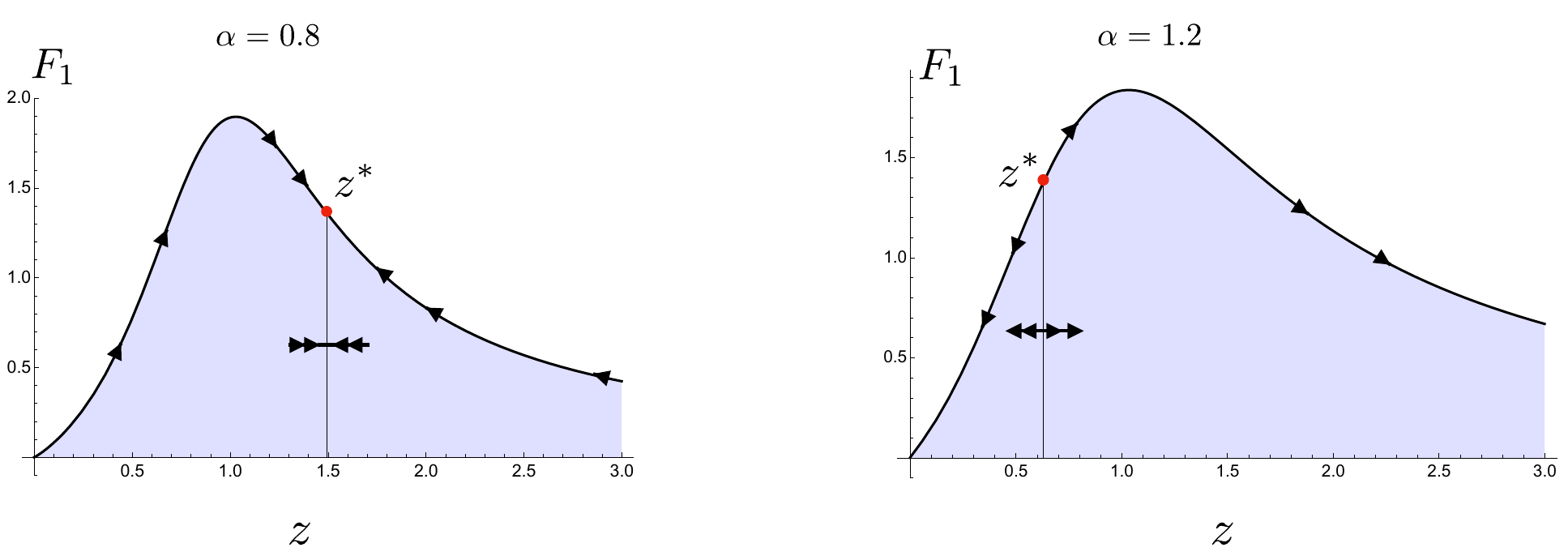}
    \caption{Phase diagrams in the $F_1$-$z$ plane with $z=\mu^+/\mu^-$ at a given $\alpha$. Blue regions denote the Fermi liquid phase, bounded by the Black curve representing the Pomeranchuk instability at $F^{+}_{1,c}$. The RG fixed point is located at $z^\ast$.}
    \label{fig:phase_diagram}
\end{figure}

The U(1) $\times$ U(1) symmetric Fermi liquid in general have $\mu_+\neq \mu_-$, which has richer physics. We firstly identify a special interaction ratio being $\lambda_\pm/\lambda_1=- \mathcal{N}_\mp/\mathcal{N}_\pm$ with the particle number of each flavor $\mathcal{N}_\pm=\int \frac{d^2p}{(2\pi)^2} n^\pm_{\vec{p}} = \frac{1}{4\pi c^{1/\alpha}} (\mu^\pm)^{1/\alpha}$. At this ratio, the net interaction on a single fermion cancels at the mean-field level. However, the intra-flavor and inter-flavor interaction remain locked via the chemical potentials. As a result, the Fermi surfaces couple to each other beyond the mean-field approximation. This holds even if we adjust the ratio to $\lambda_\pm/\lambda_1=- \mathcal{N}_\mp/\mathcal{N}_\pm +\Delta_\pm$ with constant $\Delta_\pm$. Alternatively, we can express it in terms of Landau parameters as $F_\pm/F_1 = -z^{\mp \frac{1+\alpha}{2\alpha}}+z^{\pm \frac{1-\alpha}{2\alpha}}\Delta_\pm$ with $z=\mu^+/\mu^-$. At $\Delta_\pm=0$, there is only one Pomeranchuk instability due to density fluctuation. At this instability, the chemical potentials exhibit a flow identical to the free system at one loop order, as there is no net interaction effect. More generally, for non-zero $\Delta_\pm$, two instabilities can emerge, driven by distinct patterns of density fluctuations, each located at interaction $F_{1,c}^\pm$. The one at $F_{1,c}^+$ has a fixed point at particular chemical potentials. For $\alpha<1$, this fixed point is attractive, indicating a continuous transition with two Fermi surfaces. On the contrary, for $\alpha>1$, this fixed point becomes repulsive. Remarkably, without fine-tuning, the runaway RG flow leads to the depletion of one Fermi surface. This means that this instability is a discontinuous transition where one of the Fermi surface already vanishes at the transition point. The phase diagrams at different $\alpha$ are shown in Fig.~\ref{fig:phase_diagram}. As a conclusion to this section, Table~\ref{tab:summary} offers an overview of these key results.

\begin{table}
\centering
    {\setcellgapes{1.2ex}\makegapedcells
    \begin{tabular}{p{2.3cm}|p{3.8cm}|p{8cm}}\hline 
    \makecell{ Symmetry \\in UV} & \makecell{Conditions on \\chemical potentials, 
    \\$z=\frac{\mu^+}{\mu^-}$ \\
    and quartic couplings 
    \\$F_\pm$ and $F_{1,2}$
    } 
    & \makecell{Fixed points at Pomeranchuk/ \\Stoner instability }\\\hline\hline
      \hspace{0.6cm} U(2)    & \makecell{$\mu\equiv \mu^+=\mu^-$ \\ \\$F\equiv F_+=F_-$\\$=F_1+F_2$ } & \makecell[l]{
      $\circ$~~ Tunable parameters: $F\pm F_1$\\
    $\circ$~~ Two possible instabilities at \\
           \hspace{0.5cm}\textbullet~~ $F + F_1 = -1$: density fluctuation \\ \hspace{1.0cm}breaks rotational symmetry \\
           \hspace{0.5cm}\textbullet~~ $F - F_1 = -1$:  fluctuation leads to \\ \hspace{1.0cm}flavor symmetry breaking; \\ \hspace{1.0cm}Fixed point at  $\tilde{\mu}^\ast = (\frac{\lambda(0)+\lambda_1(0)}{ 2\pi \alpha c^{1/\alpha}})^{\frac{\alpha}{\alpha-1}}$: \\ \hspace{1.5cm}$\square$ is stable (attractive) for $\alpha<1$; \\ \hspace{1.5cm}$\square$ is unstable (repulsive) for $\alpha>1$. }  \\\hline
       \multirow{3}{*}[-2cm]{ U(1) $\times$ U(1)} &  \makecell{$F_{1,2}=0$} &  \makecell[l]{$\circ$~~Tunable parameters: $F_\pm$\\ $\circ$~~Phase separation for each fermion\\\hspace{0.5cm}flavor independently at $F_\pm \leq -1$  
       } 
       \\\cline{2-3}
       &\makecell{$F_\pm/F_1 = -z^{\mp \frac{1+\alpha}{2\alpha}}$} &  \makecell[l]{ $\circ$~~Tunable parameters: $F_{1}$ (fix UV $F_2$)\\$\circ$~~Fermi liquid phase within $F_1< F_c=\frac{1}{z^{ \frac{1+\alpha}{2\alpha}}+z^{- \frac{1+\alpha}{2\alpha}}}$\\$\circ$~~Pomeranchuk Instability  at $F_c$ \\$\circ$~~ $\mu^\pm$ are not affected by interactions} \\ \cline{2-3}
       & \vspace{-1.2cm}\makecell{$R_\pm=F_\pm/F_1 $\\\\$= -z^{\mp \frac{1+\alpha}{2\alpha}} +z^{\pm \frac{1-\alpha}{2\alpha}}\Delta_\pm$} & \makecell[l]{$\circ$~~ Tunable parameters: $\Delta_\pm$, $F_{1}$ (fix UV $F_2$) \\$\circ$~~ Fermi liquid within $\textrm{Min}[F_{1,c}^\pm]<F_1<\textrm{Max}[F_{1,c}^\pm]$\\\hspace{0.5cm} with $ F_{1,c}^\pm =\frac{1}{2} \frac{(R_++R_-) \pm \sqrt{4+(R_+-R_-)^2}}{1-R_+R_-}$\\$\circ$~~ Pomeranchuk Instabilities at $ F_{1,c}^\pm $; \\\hspace{0.5cm}a fixed point locates at $\lbrack F_{1,c}^+ \rbrack^\ast$ and \\\hspace{0.5cm} $z^\ast = (\Delta_+^\ast/\Delta_-^\ast)^{-\frac{\alpha}{1-\alpha}}$ with $\Delta_\pm^\ast$ \\\hspace{0.5cm} satisfying $\sqrt{\Delta_+^\ast\Delta_-^\ast}\lbrack F_{1,c}^+ \rbrack^\ast=2$. \\ \hspace{1cm}\textbullet~~For $\alpha<1$, it is attractive; \\ \hspace{1cm}\textbullet~~For $\alpha>1$, it is repulsive. 
       }  \\\cline{2-3}\hline\hline
    \end{tabular}}
    \caption{Fermi liquids with different global symmetries have various instabilities in the forward scattering channel. }
    \label{tab:summary}
\end{table}

\section{System and Effective Theory \label{sec:theory}}

The action in Euclidean space-time can be written as a summation of free part and interacting part $S=S_0+S_\lambda$, where
\begin{eqnarray}
S_0 &=&\int d\omega d^2k ~\sum_\sigma \psi^\dag_{\omega,\vec{k},\sigma} (i \omega -\varepsilon^\sigma_{\vec{k}})\psi_{\omega,\vec{k},\sigma}  ,\nonumber\\
S_\lambda &=& \int d\omega d\Omega d\Omega' d^{2} \vec{k}_1 ~d^{2} \vec{k}_2 ~d^2 \vec{q} ~\sum_{\sigma_{1,2},\sigma'_{1,2}}\lambda_{\vec{k}_1+\frac{\vec{q}}{2},\sigma_1;\vec{k}_2-\frac{\vec{q}}{2},\sigma_2}^{\vec{k}_1-\frac{\vec{q}}{2},\sigma'_1;\vec{k}_2+\frac{\vec{q}}{2},\sigma'_2} ~\psi^{\dag}_{K_1,\sigma_1}\psi^{\dag}_{K_2,\sigma_2}\psi_{K'_2,\sigma'_2 }\psi_{K'_1,\sigma'_1}.
\label{eq:action}
\end{eqnarray}
For simplicity, we have used the short hand notation $K_1=(\Omega+\frac{\omega}{2},\vec{k}_1+\frac{\vec{q}}{2} )$, $K_2 =( \Omega-\frac{\omega'}{2},\vec{k}_2-\frac{\vec{q}}{2} )$, $K'_2 =( \Omega+\frac{\omega'}{2},\vec{k}_2+\frac{\vec{q}}{2})$ and $K'_1=(\Omega-\frac{\omega}{2},\vec{k}_1-\frac{\vec{q}}{2} )$. $\sigma=\pm$ is the flavor index. In $S_0$, the dispersion is  $\varepsilon^\pm_{\vec{k}}=c k^{2\alpha}-\mu^\pm$ with chemical potential $\mu^\pm$. In general, $\mu^+\neq \mu^-$. Here we consider the generic dispersion with $\alpha$ tunable. Accordingly, the system has two circular Fermi surfaces with Fermi momentum $k_F^\pm=(\mu^\pm/c)^{1/(2\alpha)}$, respectively. In the vicinity of the Fermi surface, the dispersion can be linearized as
\begin{eqnarray}
\varepsilon^\pm_{\vec{q}+\vec{k}_F^\pm}= v_F^\pm q \cos \theta
\label{eq:linear_dispersion}
\end{eqnarray}
with $\theta$ being the angle between small momentum $\vec{q}$ and the Fermi momentum $\vec{k}_F^\pm$ along particular angular direction. 
Fermi velocities are $v_F^\pm=2\alpha c^{1/(2\alpha)} (\mu^\pm)^{1-1/(2\alpha)} $. The interacting part of the action can be expanded as $S_\lambda=S_++S_-+S_1+S_2$ with 
\begin{eqnarray}
S_{+} &=& \int d\omega d\Omega d\Omega' d^{2} \vec{k}_1 ~d^{2} \vec{k}_2 ~d^2 \vec{q} ~ (\lambda_+)_{\vec{k}_1,\vec{k}_2}(\vec{q})~\psi^{\dag}_{K_1,+}\psi^{\dag}_{K_2,+}\psi_{K'_2 ,+}\psi_{K'_1,+} , \label{eq:interaction_1} \\
S_{-} &=& \int d\omega d\Omega d\Omega' d^{2} \vec{k}_1 ~d^{2} \vec{k}_2 ~d^2 \vec{q} ~(\lambda_{-})_{\vec{k}_1,\vec{k}_2}(\vec{q}) ~
\psi^{\dag}_{K_1,-} \psi^{\dag}_{K_2,-} \psi_{K'_2,-} \psi_{K'_1,-},\label{eq:interaction_2}\\
S_{1} &=& \int d\omega d\Omega d\Omega' d^{2} \vec{k}_1 ~d^{2} \vec{k}_2 ~d^2 \vec{q} ~(\lambda_1)_{\vec{k}_1,\vec{k}_2}(\vec{q}) ~\sum_{\sigma_{1,2}} \Big[\psi^{\dag}_{K_1,+}  \psi^{\dag}_{K_2,-} \psi_{K'_2,-}\psi_{K'_1,+} + 1 \leftrightarrow 2
\Big] ,\label{eq:density-density}
\\
S_{2} &=& \int d\omega d\Omega d\Omega' d^{2} \vec{k}_1 ~d^{2} \vec{k}_2 ~d^2 \vec{q} ~(\lambda_2)_{\vec{k}_1,\vec{k}_2}(\vec{q}) ~\sum_{\sigma_{1,2}} \Big[\psi^{\dag}_{K_1,+}  \psi^{\dag}_{K_2,-} \psi_{K'_2,+}\psi_{K'_1,-} + 1 \leftrightarrow 2
\Big] .\label{eq:flavor_exchange}
\end{eqnarray}
\footnote{ 
The U(2) invariant interactions are constructed as 
\begin{eqnarray}
V_\rho&=&\delta_{\sigma_1,\sigma'_1}\delta_{\sigma_2\sigma'_2}\psi^{\dag}_{K_1,\sigma_1}\psi^{\dag}_{K_2 ,\sigma_2} \psi_{K'_2,\sigma'_2}\psi_{K'_1 ,\sigma'_1} =V_0,
\nonumber\\
V_m &=&\delta_{\sigma_1\sigma'_2}\delta_{\sigma_2\sigma'_1}\psi^{\dag}_{K_1,\sigma_1}\psi^{\dag}_{K_2 ,\sigma_2} \psi_{K'_2,\sigma'_2}\psi_{K'_1 ,\sigma'_1} =\frac{1}{2}( V_x+V_y+V_z +V_\rho),\nonumber
\end{eqnarray} 
where $V_\mu= \left[\psi^{\dag}_{K_1,\sigma_1}~ \sigma^\mu_{\sigma_1\sigma_2}~ \psi_{K'_1,\sigma_2} \right]\left[\psi^{\dag}_{K_2,\sigma'_1} ~\sigma^\mu_{\sigma'_1\sigma'_2}~\psi_{K'_2,\sigma'_2}\right]$ with $\sigma^0=I$ and $\sigma^\mu$ being Pauli matrices. If $V_{\rho}$ and $V_m$ have couplings $\lambda_{\rho}$ and $\lambda_m$, then we have $\lambda_\rho+\lambda_m = \lambda_+=\lambda_-$ with $\lambda_\pm$ defined in the main context. Notice that $\lambda_\rho=\lambda_1$ and $\lambda_m=\lambda_2$. When $\lambda_+=\lambda_-=\lambda_1+\lambda_2$, the system has U(2) symmetry. In the limit of low energy and small momentum transfer, $\lambda_2$ is decoupled from the RG flow of couplings $\lambda_1$ and $\lambda_\pm$. For simplicity, our discussion at the level of mean field and kinetic equation below considers the case where $\lambda_2=0$. But we will relax the $\lambda_2$ to be finite later in the discussion from the perspective of RG.}
The total action $S$ has at least U(1)$\times$U(1) symmetry. In the low energy limit, the quartic interaction has two channels. The one contains the nearly forward scatterings with momentum transfer $|\vec{q}|\ll k_F $ and the other contains the BCS scatterings with $|\vec{k}_1+\vec{k}_2|=|\vec{Q}| \ll k_F$. In this work, we focus solely on the former, as the system is restricted to be at least U(1)$\times$U(1) symmetric and any BCS instability is suppressed. The corresponding vertex diagrams are
\begin{eqnarray}
&&\begin{tikzpicture}[baseline={([yshift=-4pt]current bounding box.center)}]
\coordinate (v1) at (-15pt,-15pt);
\coordinate (v2) at (-15pt,15pt);
\coordinate (v3) at (15pt,-15pt);
\coordinate (v4) at (15pt,15pt);
\coordinate (v5) at (0pt,0pt);
\draw[thick,postaction={mid arrow=red} ](v1)--  (v5);
\draw[thick,postaction={mid arrow=red} ](v5)--  (v2);
\draw[thick,postaction={mid arrow=red} ](v3)--  (v5);
\draw[thick,postaction={mid arrow=red} ](v5)--  (v4);
\node at (-22pt,-22pt) {\scriptsize $K_1,+$};
\node at (-22pt,20pt) {\scriptsize $K'_1,+$};
\node at (22pt,-22pt) {\scriptsize $K_2,+$};
\node at (22pt,20pt) {\scriptsize $K'_2,+$};
\node at (0pt, -40pt) {\scriptsize $\lambda_+$};
\node at (v1)[circle,fill,inner sep=1pt]{};
\node at (v2)[circle,fill,inner sep=1pt]{};
\node at (v3)[circle,fill,inner sep=1pt]{};
\node at (v4)[circle,fill,inner sep=1pt]{};
\end{tikzpicture},\quad 
\begin{tikzpicture}[baseline={([yshift=-4pt]current bounding box.center)}]
\coordinate (v1) at (-15pt,-15pt);
\coordinate (v2) at (-15pt,15pt);
\coordinate (v3) at (15pt,-15pt);
\coordinate (v4) at (15pt,15pt);
\coordinate (v5) at (0pt,0pt);
\draw[thick,postaction={mid arrow=red} ](v1)--  (v5);
\draw[thick,postaction={mid arrow=red} ](v5)--  (v2);
\draw[thick,postaction={mid arrow=red} ](v3)--  (v5);
\draw[thick,postaction={mid arrow=red} ](v5)--  (v4);
\node at (-22pt,-22pt) {\scriptsize $K_1,-$};
\node at (-22pt,20pt) {\scriptsize $K'_1,-$};
\node at (22pt,-22pt) {\scriptsize $K_2,-$};
\node at (22pt,20pt) {\scriptsize $K'_2,-$};
\node at (0pt, -40pt) {\scriptsize $\lambda_- $};
\node at (v1)[circle,fill,inner sep=1pt]{};
\node at (v2)[circle,fill,inner sep=1pt]{};
\node at (v3)[circle,fill,inner sep=1pt]{};
\node at (v4)[circle,fill,inner sep=1pt]{};
\end{tikzpicture},\quad\begin{tikzpicture}[baseline={([yshift=-4pt]current bounding box.center)}]
\coordinate (v1) at (-15pt,-15pt);
\coordinate (v2) at (-15pt,15pt);
\coordinate (v3) at (15pt,-15pt);
\coordinate (v4) at (15pt,15pt);
\coordinate (v5) at (0pt,0pt);
\draw[thick,postaction={mid arrow=red} ](v1)--  (v5);
\draw[thick,postaction={mid arrow=red} ](v5)--  (v2);
\draw[thick,postaction={mid arrow=red} ](v3)--  (v5);
\draw[thick,postaction={mid arrow=red} ](v5)--  (v4);
\node at (-22pt,-22pt) {\scriptsize $K_1,+$};
\node at (-22pt,20pt) {\scriptsize $K'_1,+$};
\node at (22pt,-22pt) {\scriptsize $K_2,-$};
\node at (22pt,20pt) {\scriptsize $K'_2,-$};
\node at (0pt, -40pt) {\scriptsize $\lambda_{1} $};
\node at (v1)[circle,fill,inner sep=1pt]{};
\node at (v2)[circle,fill,inner sep=1pt]{};
\node at (v3)[circle,fill,inner sep=1pt]{};
\node at (v4)[circle,fill,inner sep=1pt]{};
\end{tikzpicture}
,\quad \nonumber\\
&&\begin{tikzpicture}[baseline={([yshift=-4pt]current bounding box.center)}]
\coordinate (v1) at (-15pt,-15pt);
\coordinate (v2) at (-15pt,15pt);
\coordinate (v3) at (15pt,-15pt);
\coordinate (v4) at (15pt,15pt);
\coordinate (v5) at (0pt,0pt);
\draw[thick,postaction={mid arrow=red} ](v1)--  (v5);
\draw[thick,postaction={mid arrow=red} ](v5)--  (v2);
\draw[thick,postaction={mid arrow=red} ](v3)--  (v5);
\draw[thick,postaction={mid arrow=red} ](v5)--  (v4);
\node at (-22pt,-22pt) {\scriptsize $K_1,-$};
\node at (-22pt,20pt) {\scriptsize $K'_1,-$};
\node at (22pt,-22pt) {\scriptsize $K_2,+$};
\node at (22pt,20pt) {\scriptsize $K'_2,+$};
\node at (0pt, -40pt) {\scriptsize $\lambda_{1} $};
\node at (v1)[circle,fill,inner sep=1pt]{};
\node at (v2)[circle,fill,inner sep=1pt]{};
\node at (v3)[circle,fill,inner sep=1pt]{};
\node at (v4)[circle,fill,inner sep=1pt]{};
\end{tikzpicture} 
,\quad \begin{tikzpicture}[baseline={([yshift=-4pt]current bounding box.center)}]
\coordinate (v1) at (-15pt,-15pt);
\coordinate (v2) at (-15pt,15pt);
\coordinate (v3) at (15pt,-15pt);
\coordinate (v4) at (15pt,15pt);
\coordinate (v5) at (0pt,0pt);
\draw[thick,postaction={mid arrow=red} ](v1)--  (v5);
\draw[thick,postaction={mid arrow=red} ](v5)--  (v2);
\draw[thick,postaction={mid arrow=red} ](v3)--  (v5);
\draw[thick,postaction={mid arrow=red} ](v5)--  (v4);
\node at (-22pt,-22pt) {\scriptsize $K_1,+$};
\node at (-22pt,20pt) {\scriptsize $K'_1,-$};
\node at (22pt,-22pt) {\scriptsize $K_2,-$};
\node at (22pt,20pt) {\scriptsize $K'_2,+$};
\node at (0pt, -40pt) {\scriptsize $\lambda_2$};
\node at (v1)[circle,fill,inner sep=1pt]{};
\node at (v2)[circle,fill,inner sep=1pt]{};
\node at (v3)[circle,fill,inner sep=1pt]{};
\node at (v4)[circle,fill,inner sep=1pt]{};
\end{tikzpicture}, \quad 
\begin{tikzpicture}[baseline={([yshift=-4pt]current bounding box.center)}]
\coordinate (v1) at (-15pt,-15pt);
\coordinate (v2) at (-15pt,15pt);
\coordinate (v3) at (15pt,-15pt);
\coordinate (v4) at (15pt,15pt);
\coordinate (v5) at (0pt,0pt);
\draw[thick,postaction={mid arrow=red} ](v1)--  (v5);
\draw[thick,postaction={mid arrow=red} ](v5)--  (v2);
\draw[thick,postaction={mid arrow=red} ](v3)--  (v5);
\draw[thick,postaction={mid arrow=red} ](v5)--  (v4);
\node at (-22pt,-22pt) {\scriptsize $K_1,-$};
\node at (-22pt,20pt) {\scriptsize $K'_1,+$};
\node at (22pt,-22pt) {\scriptsize $K_2,+$};
\node at (22pt,20pt) {\scriptsize $K'_2,-$};
\node at (0pt, -40pt) {\scriptsize $\lambda_2$};
\node at (v1)[circle,fill,inner sep=1pt]{};
\node at (v2)[circle,fill,inner sep=1pt]{};
\node at (v3)[circle,fill,inner sep=1pt]{};
\node at (v4)[circle,fill,inner sep=1pt]{};
\end{tikzpicture},
\end{eqnarray}
where the momentum transfer $\vec{q}$ is considered to be small.

\section{Landau Fermi Liquid Theory and Kinetic Equation \label{sec:kinetic_eqn}}

Within the framework of Landau's theory\cite{landau1959theory,abrikosov1958theory,nozieres1962derivation,fradkin2013field,pines2018microscopic,abrikosov2012methods}, we can understand the Fermi liquid phase of this system. 
In terms of Wigner density operator for each species at particular real frequency $n^\pm_{\vec{p}}(\vec{q},\omega_R)= \int \frac{d\Omega_R}{2\pi}\psi^{\dag}_{\vec{p}+\vec{q}/2, \pm}(\Omega_R+\omega_R)\psi_{\vec{p}-\vec{q}/2,\pm}(\Omega_R)$ defined in Appendix \ref{app:FL_2_flavor}, the energy function $\mathcal{H}_{\vec{k}}^\pm(\omega_R)$ defined as 
\begin{eqnarray}
-S_{\omega \rightarrow -i\omega_R}=\int d\omega_R d^2k \sum_{\sigma=\pm}\left[\psi^\dag_{\omega_R,\vec{k},\sigma} ( \omega_R )\psi_{\omega_R,\vec{k},\sigma}  +\mathcal{H}^\sigma_{\vec{k}}(\omega_R)\right]     
\end{eqnarray}
can be expressed to be $\mathcal{H}_{\vec{k}}^\pm(\omega_R) = \int \frac{d^2q}{(2\pi)^2}n^\pm_{\vec{k}}(\vec{q},\omega_R) \mathcal{E}_{\vec{k}}^\pm(\vec{q},-\omega_R)$ with
\begin{eqnarray}
  \mathcal{E}_{\vec{k}}^\pm(\vec{q},\omega_R)=  \varepsilon_{\vec{k}}^\pm \delta^2(\vec{q})\delta(\omega_R)  + \int ~\frac{ d^{2} \vec{p}  }{(2\pi)^2}~ \left[(\lambda_\pm)_{\vec{k},\vec{p}}(\vec{q})  n^\pm_{\vec{p}}(-\vec{q},\omega_R) 
  + (\lambda_1)_{\vec{k},\vec{p}}(\vec{q})  n^\mp_{\vec{p}}(-\vec{q},\omega_R) \right],
  \label{eq:energy_per_particle}
\end{eqnarray}
where small $\vec{q}$ is considered.
Note that the $\lambda_2$ term is absent. This is because the flavor-changing scattering processes at small $\vec{q}$ do not contribute to the low energy physics when $\mu^+ \neq \mu^-$.
In general, the system has U(1) $\times$ U(1) symmetry and all the couplings can be different and nonzero. Bare particles are dressed by both $\lambda_\pm$ and $\lambda_1$ interactions, forming quasiparticles. Specifically, the energy density function at zero frequency and zero momentum transfer $\mathcal{E}_{\vec{k}}^\pm(0,0)$ represents the energy per quasiparticle at mean-field level. Each quasiparticle's energy combines the bare energy $\varepsilon^\pm_{\vec{k}}$ with its renormalization due to interaction. 
Within this framework, one can obtain quasiparticle density of states at Fermi level, polarization associated with flavor (see Appendix \ref{app:thermodynamics}) and so on. 

In the very simple case where $\lambda_\pm$ and $\lambda_1$ are independent of momenta $\vec{k}$, $\vec{p}$ and $\vec{q}$, if $\lambda_\pm/\lambda_1$ is equal to ratio 
\begin{eqnarray}
    \frac{\lambda_\pm}{\lambda_1} =\left(\frac{\lambda_\pm}{\lambda_1}\right)_c\equiv - \frac{ \int ~\frac{ d^{2} \vec{p}  }{(2\pi)^2} ~n^\mp_{\vec{p}}(0,0)}{\int ~\frac{ d^{2} \vec{p}  }{(2\pi)^2} ~n^\pm_{\vec{p}}(0,0)} =- \left(\frac{\mu^\mp}{\mu^\pm}\right)^{1/\alpha},
    \label{eq:special_ratio}
\end{eqnarray}
the energy per particle is not renormalized by interaction since the net interaction is zero. If $\frac{\lambda_\pm}{\lambda_1} = \left(\frac{\lambda_\pm}{\lambda_1}\right)_c +\Delta_\pm$, the quasiparticle energy gains a correction due to inter-flavor coupling,
\begin{eqnarray}
\mathcal{E}_{\vec{k}}^\pm(0,0) &=&  \varepsilon_{\vec{k}}^\pm   + \lambda_1 ~  \Delta_\pm   \int ~\frac{ d^{2} \vec{p}  }{(2\pi)^2} n^\pm_{\vec{p}}(0,0) .
\end{eqnarray}
This reproduces the standard formalism of Landau Fermi liquid theory for each species. Bare particles are dressed by the interaction $\lambda_1 \Delta_\pm$ to form quasiparticles. The renormalization of each flavor depends solely on its own density. This particular interaction ratio $\left(\frac{\lambda_\pm}{\lambda_1}\right)$ suggests that at mean field level fermions are not aware of the other flavor if $\Delta_\pm$ is independent of $\mu^\pm$. Beyond mean-field, the two flavors remain coupled. This is evident in the study of the RG flow in Sec.~\ref{sec:RG_flow} and collective modes derived from the kinetic equation in Sec.~\ref{sec:collective_modes}.

Next, we present the kinetic equation (derived in Appendix \ref{app:Kinetic_eqn}), which is equivalent to the linearized saddle point equation of the action in Eq.~\eqref{eq:action}, revealing the collective modes. These modes consist of a group of particle-hole excitations in the system.
In the domain of real frequency, it is
\begin{eqnarray}
   \left[-\omega_R  +      \vec{q}  \cdot \nabla_{\vec{k}}\varepsilon^\pm_{\vec{k}} \right] \delta n^\pm_{\vec{k}}(\vec{q},\omega_R)
    - 4 \vec{q} \cdot \nabla_{\vec{k}}  (n^0_{\vec{k}})^\pm   \int  \frac{d^2\vec{p} }{(2\pi)^2}   \left[\lambda_\pm    \delta n^\pm_{\vec{p}}(\vec{q},\omega_R)+\lambda_1    \delta n^\mp_{\vec{p}}(\vec{q},\omega_R) \right] =0,
    \label{eq:EOM_q_omega}
\end{eqnarray}
up to the first order of distribution away from equilibrium $\delta n^\pm_{\vec{k}}$.
The distribution at equilibrium is $(n^0_{\vec{k}})^\pm (\vec{q},\omega_R)= N(-\varepsilon_{\vec{k}}^\pm)\delta(\omega_R)\delta^2(\vec{q}) $, where $N(x)$ is the fermion distribution, with $N(-E)=1$ if $E\geq 0$, and $N(-E)=0$ otherwise. In terms of 
$u_{\vec{k}}^\pm (\vec{q},\omega_R)$ defined as $\delta n^\pm_{\vec{k}}(\vec{q},\omega_R) =  (\mu^\pm/c)^{\frac{1}{2\alpha}} \delta[k-(\mu^\pm/c)^{\frac{1}{2\alpha}} ] u^\pm_{\vec{k}}(\vec{q},\omega_R)$, the equation\footnote{We have used 
\begin{eqnarray}
    \nabla_{\vec{k}}\varepsilon^\pm_{\vec{k}} &=& 2\alpha c(\mu^\pm/c)^{1-1/(2\alpha)}\hat{k}, \nonumber\\
    \nabla_{\vec{k}}  (n^0_{\vec{k}})^\pm(\vec{q},\omega_R) &=& - \nabla_{\vec{k}}\varepsilon^\pm_{\vec{k}} \delta (-\varepsilon^\pm_{\vec{k}})\delta(\omega_R)\delta^2(\vec{q})= \hat{k}  (\mu^\pm/c)^{1/(2\alpha)} \delta[k-(\mu^\pm/c)^{1/(2\alpha)}] \delta(\omega_R)\delta^2(\vec{q}), \nonumber\\
    \delta n^\pm_{\vec{k}}(\vec{q},\omega_R) &=&  (\mu^\pm/c)^{1/(2\alpha)} \delta[k-(\mu^\pm/c)^{1/(2\alpha)}] u^\pm_{\vec{k}}(\vec{q},\omega_R). \nonumber
\end{eqnarray}
} is
\begin{eqnarray}
\left[-\omega_R  +  v_F^\pm q \cos \theta_k\right] u^\pm_{\theta_{k}}(\vec{q},\omega_R)
+v_F^\pm q \cos \theta_k  \int \frac{d\theta_p}{2\pi} \left[F_\pm  u^\pm_{\theta_{p}}(\vec{q},\omega_R)+ \left(\frac{\mu^\pm}{\mu^\mp}\right)^{\frac{1-\alpha}{2\alpha}} F_1    u^\mp_{\theta_{p}}(\vec{q},\omega_R) \right] =0,
\end{eqnarray}
in terms of the Landau parameters as $F_\pm = \frac{(\mu^\pm)^{1/\alpha-1}}{\pi \alpha c^{1/\alpha}  } \lambda_\pm $ and $F_1= \lambda_1 \frac{(\mu^+ \mu^-)^{\frac{1-\alpha}{2\alpha}}}{\pi \alpha c^{1/\alpha}  }$. The solution is given by Fermi surface deformations as
\begin{eqnarray}
u^\pm_{\theta_{k}}(\vec{q},\omega_R) 
=\frac{(\mu^\pm)^{1/\alpha-1}}{2\alpha c^{1/\alpha}  }\frac{v_F^\pm q \cos \theta_k  }{-\omega_R  +     v_F^\pm q \cos \theta_k} \mathcal{A}^\pm
\label{eq:u_pm}
\end{eqnarray}
with their amplitudes $\mathcal{A}^\pm$. The collective mode velocity $v_{zs}\equiv \omega_R/q$ satisfies
\begin{eqnarray}
&&\left[1+ F_+ \mathcal{I}_R\left(\frac{v_{zs}}{v_F^+ }\right) \right] \left[1+F_- \mathcal{I}_R\left(\frac{v_{zs}}{v_F^- }\right)\right]-F_1^2 \mathcal{I}_R \left(\frac{v_{zs}}{v_F^- } \right) \mathcal{I}_R \left(\frac{v_{zs}}{v_F^+} \right)=0
\label{eq:LKE_imbalanced}
\end{eqnarray}
and the ratio $\mathcal{A}^-/\mathcal{A}^+$ satisfies 
\begin{eqnarray}
\left[1+F_\pm \mathcal{I}_R \left(\frac{v_{zs}}{v_F^\pm }\right)\right] + \left(\frac{\mu^\mp}{ \mu^\pm} \right)^{\frac{1-\alpha}{2\alpha}} F_1\mathcal{I}_R\left(\frac{v_{zs}}{v_F^\mp }\right) \frac{\mathcal{A}^\mp }{\mathcal{A}^\pm}=0 ,
\label{eq:amplitude}
\end{eqnarray}
where the function $\mathcal{I}_R(x)$ has the form of
\begin{eqnarray}
\mathcal{I}_R(x) = \int_0^{2\pi} \frac{\cos \theta}{-x +\cos \theta} = 
  1-\frac{1}{\sqrt{1-x^{-2}}}  , \textrm{ when } x>1 
.
\label{eq:mathcal_I_integral}
\end{eqnarray}
We will find out in Sec.~\ref{sec:RG_flow} that the poles of the IR coupling functions satisfying the same equation with $F_\pm$ and $F_1$ being the UV interaction. In other words, the poles correspond to solutions of the collection modes. We then study these solutions in Sec.~\ref{sec:collective_modes}.
At the Pomeranchuk instabiltiy, the collective mode velocity vanishes, i.e. $v_{zs}=0$, on the surface of couplings defined by $F_1^2=(1+F_+)(1+F_-)$. At this transition point, the deformation of Fermi surface does not cost energy, and the bosonic collective mode is gapless.

Next section, in order to study these instabilities and to explore the physics of their associated RG fixed points, we apply the field theoretical functional renormalization method to this system. We analyze the RG flow of coupling functions and chemical potentials.

\section{Field Theoretical Functional Renormalization Group Flow}
\label{sec:RG_flow}

The RG study begins with the establishment of an effective action and performs a scaling analysis of fields and couplings. From the generic action in Eq.~\eqref{eq:action}, we can derive the effective action in the low energy limit. Near the Fermi surface, fermionic fields $\psi_{\omega,\vec{k}_F+\vec{q},\sigma}$ with small $\vec{q}$ are the low energy fields, expressed as $\psi_{\omega,\theta,\vec{q},\sigma}$, where $\vec{k}_F$ aligns with the angular direction $\theta$. For free fermions, the low energy dispersion is linear near the Fermi surface as displayed in Eq.~\eqref{eq:linear_dispersion}. Thus, the quadratic part of the action is
\begin{eqnarray}
S_0 &=& \int d\omega dq d\theta ~\sum_\sigma ~k_F^{\sigma}~\psi^\dag_{\omega,\theta,\vec{q},\sigma} (i \omega -v^\sigma_F {q} \cos \theta)\psi_{\omega,\theta,\vec{q},\sigma}  .
\end{eqnarray} 
Moreover, since the low energy limit only permits nearly forward scattering, the interacting part of the action is
\begin{eqnarray}
S_{\lambda} = \sqrt{k_F^{\sigma_1}k_F^{\sigma_2}} \int d\omega d\Omega d\Omega' d {\kappa}_1 d\theta_1 ~d {\kappa}_2 d \theta_2 ~d^2 \vec{q} ~ \lbrack F_{\sigma_1,\sigma_2}^{\sigma'_1,\sigma'_2} \rbrack_{\theta_1,\theta_2}(\vec{q})~\psi^{\dag}_{\mathcal{K}_1,\sigma_1}\psi^{+,\dag}_{\mathcal{K}_2,\sigma_2}\psi_{\mathcal{K}'_2 ,\sigma'_2}\psi_{\mathcal{K}'_1,\sigma'_1}  ,
\end{eqnarray}
where $\mathcal{K}_1=(\Omega+\frac{\omega}{2},\theta_1,\vec{\kappa}_1+\frac{\vec{q}}{2} )$, $\mathcal{K}_2 =( \Omega-\frac{\omega'}{2},\theta_2,\vec{\kappa}_2-\frac{\vec{q}}{2} )$, $\mathcal{K}'_2 =( \Omega+\frac{\omega'}{2},\theta_2,\vec{\kappa}_2+\frac{\vec{q}}{2})$ and $\mathcal{K}'_1=(\Omega-\frac{\omega}{2},\theta_1,\vec{\kappa}_1-\frac{\vec{q}}{2} )$. 
Thus, the action density $(S_0+S_\lambda)/(\textrm{Max}[k_F^+,k_F^-])$, instead of the total action, is scale invariant. 
From $S_0$, the frequency has engineer dimension $[\omega]=[\mu]=1$ and the momentum relative to the Fermi surface scales as $[q]=1$. Besides, $[c]=1-2\alpha$. This yields a dimensionless Fermi velocity $v_F^\pm = 2\alpha c^{1/(2\alpha)} (\mu^\pm)^{1-1/(2\alpha)} $. The field scales with dimension $[\psi]=-2$. As a result, the quartic interaction has $[\lambda]=-1$. It is easy to check that the extended Landau parameters $F_{\pm} = \frac{2}{\pi}\frac{(\mu^\pm)^{1/\alpha-1}}{2\alpha c^{1/\alpha}} {\lambda}_\pm $ and $F_{1,2}= \frac{2}{\pi}\frac{({\mu}^+ {\mu}^-)^{1/(2\alpha)-1/2}}{2\alpha  {c}^{1/\alpha}}  {\lambda}_{1,2}$ are dimensionless. The scale transformation thus acts as
\begin{eqnarray}
&&\omega_s = \omega /s, \quad k_s= k/s, \quad \mu^\sigma_s= \mu^\sigma/s ,\nonumber\\
&& c_s = c s^{2\alpha-1} ,\quad \psi_s=s^2\psi,\quad \lambda_{s;i}= s \lambda_i ,\quad F_{s;i}=F_i,
\end{eqnarray}
where $i=\pm,1,2$. 
We can define dimensionless couplings as
\begin{eqnarray}
  \tilde{\lambda}_i = \lambda_i \omega, \quad   \tilde{c}= c \omega^{2\alpha-1} ,\quad \tilde{\mu}^\sigma=\mu^\sigma \omega^{-1}.\label{eq:dim_less}
\end{eqnarray}
The detailed formalism of the field-theoretic functional RG is presented in Appendix \ref{app:RG}. Here we directly provide the RG equations for the Landau parameters and chemical potentials. Our discussion centers on the couplings $F_\pm$ and $F_{1,2}$ as functions of momentum transfer $\vec{q}$, and we will not consider their dependent on the angles $\theta_{1,2}$ in this work, which will be addressed in future work. For small $\vec{q}$, one can do Taylor expansion as
\begin{eqnarray}
    F_{\pm,1,2} (\vec{q}) = F_{\pm,1,2} + \vec{q}\cdot \nabla_{\vec{q}}F_{\pm,1,2} (\vec{q}) \Big|_{q=0} + \frac{1}{2} q^2 \nabla^2_{\vec{q}}F_{\pm,1,2} (\vec{q}) \Big|_{q=0}+\dots,
\end{eqnarray}
if the coupling functions are analytic within the Fermi liquid phases. 
Basically, including $\vec{q}$ dependence brings irrelevant operators into our effective theory. This is crucial for identifying the Pomeranchuk instability, which happens at finite attractive interaction $F_{\pm,1,2}$.

\subsection{Chemical Potential}

At one loop order, the self-energy adjusts the Fermi level. As a result, the dimensionless chemical potentials $\tilde{\mu}^\pm$ follows the RG flow given by
\begin{eqnarray}
\frac{d \tilde{\mu}^+}{d\ln E } &=& 
-\tilde{\mu}^+ +\frac{\tilde{\mu}^+}{2} \left[ F_+(0) + z^{-\frac{1+\alpha}{2\alpha}} F_1(0)\right] , \nonumber\\
\frac{d \tilde{\mu}^-}{d\ln E} &=& 
-\tilde{\mu}^- + \frac{\tilde{\mu}^- }{2} \left[F_-(0)  + z^{\frac{1+\alpha}{2\alpha}} F_1(0)\right] .
\label{eq:beta_function_chemical}
\end{eqnarray}
For convenience, we use the notation $z= \tilde{\mu}^+/\tilde{\mu}^-$.
Notice at this order, these $\beta$ functions only depend on coupling functions at zero momentum transfer. 
In general, the fixed point structure depends on $F_{\pm}(0)$ and $F_{1}(0)$, which may depend on $z$. Before diving into the detailed analysis, we present the RG equations for quartic interactions.

\subsection{Quartic Interaction in Particle-Hole Channel}

At one loop order, we can have $\beta$ functions for quartic couplings in the particle-hole channel as following.
\begin{eqnarray}
\frac{d F_\pm (\vec{q})}{d \ln E} &=& \kappa_\pm (\vec{q},E) [F_\pm(\vec{q})]^2 + \kappa_\mp (\vec{q},E) [F_1 (\vec{q})]^2,\nonumber\\
\frac{d F_1 (\vec{q})}{d\ln E} &=& \left[\kappa_{+}(\vec{q},E)F_+(\vec{q}) + \kappa_{-}(\vec{q},E)F_-(\vec{q})\right] F_1(\vec{q}), \nonumber\\
\frac{d F_2 (\vec{q})}{d\ln E} &=& \kappa_{+-}(\vec{q},E) [F_2(\vec{q})]^2 ,
\label{eq:beta_function_quartic}
\end{eqnarray}
where 
\begin{eqnarray}
\kappa_\pm(\vec{q},E) &=& \frac{d}{d\ln E} \int \frac{d\Omega dk d\theta}{(2\pi)^2}
\frac{1}{-i\Omega+\varepsilon^\pm_{\vec{k}+\vec{q}/2}}\frac{1}{-i(E+\Omega)+\varepsilon^\pm_{\vec{k}-\vec{q}/2}} \nonumber\\
&=& 
\frac{d}{d\ln E} \left[\frac{1}{\sqrt{\frac{(v_F^\pm)^2q^2}{E^2}+1}}\right] =  \frac{(v_F^\pm)^2 q^2 /E^2}{\left[1+ \frac{(v_F^\pm)^2 q^2}{E^2}\right]^{3/2}}, \nonumber\\
\kappa_{+-}(\vec{q},E) &=& \frac{d}{d\ln E} \int \frac{d\Omega dk d\theta}{(2\pi)^2}
\frac{1}{-i\Omega+\varepsilon^+_{\vec{k}+\vec{q}/2}}\frac{1}{-i(E+\Omega)+\varepsilon^-_{\vec{k}-\vec{q}/2}} \nonumber\\
&=& -\frac{1}{2}\frac{d}{d\ln E} \int \frac{ dk d\theta}{2\pi}
\frac{\Theta(\varepsilon^+_{\vec{k}+\vec{q}/2})-\Theta(\varepsilon^-_{\vec{k}-\vec{q}/2})}{iE+\varepsilon^+_{\vec{k}+\vec{q}/2}-\varepsilon^-_{\vec{k}-\vec{q}/2}},
\end{eqnarray}
where $\Theta(x)$ yields the sign of $x$. 
Notice that $F_2(\vec{q})$ flows independently as processes with small momentum transfer involving flavor exchange do not contribute to the low-energy physics once $\mu^+ \neq \mu^-$. 
Below we study the U(1) $\times$ U(1) cases and the U(2) symmetric case where the differential equations are largely simplified.

\section{Fermi Liquid Phase and Instabilities \label{sec:FPs}}

\subsection{Decoupled Fermi Liquid}

When $F_1(\vec{q})=F_2(\vec{q})=0$, the two flavors are completely decoupled at any energy scale. Thus, the physics of a single Fermi surface fully characterizes the behavior of the system with two decoupled Fermi surfaces.
Each $\beta$ function of chemical potential is reduced to the one for single-flavor fermion 
\begin{eqnarray}
\frac{d \tilde{\mu}^\pm}{d\ln E } = 
-\tilde{\mu}^\pm + \frac{1}{2}  \tilde{\mu}^\pm  F_\pm(0) .
\label{eq:mu_single}
\end{eqnarray}
As studied in Ref.~\cite{SHANKAR}, the RG flow of the dimensionless chemical potential reveals the effect of interactions. For free fermions, where $F_\pm(0)=0$, the physical Fermi surface remains unrenormalized. In the low energy limit, the Fermi surface measured in the unit of the energy scale increases divergently.  
In the presence of interactions, renormalization induces a shift in the zero-energy level due to non-trivial self-energy, and also modifies the scaling behavior, both of which are captured in the flow of $\tilde{\mu}^\pm$. In the scenarios where particle number is conserved, the flow of $\mu^\pm $ maintains $k_F$. Here we do not impose this constraint; instead, the chemical potential selects the particle number that minimizes the energy. Assuming constant Landau parameters, the $\beta$ equation can be solved, yielding $\tilde{\mu}^\pm \sim E^{-(1-\frac{1}{2}F_\pm(0))}$. Specifically, if $F_\pm(0)=0$, then $\tilde{\mu}^\pm \sim E^{-1}$, which implies $\mu^\pm \sim \mathcal{O}(1)$ to be a constant. For non-zero $F_\pm(0)$, we obtain $\mu^\pm \sim E^{\frac{1}{2}F_\pm(0)}$. This means that the Fermi surface expands for $F_{\pm}(0)<0$ (attractive) and shrinks for $F_{\pm}(0)>0$ (repulsive). Pictorially, repulsive interactions $F_{\pm}(0)>0$ force fermions apart in real space, causing them to cluster more in momentum space. Conversely, attractive interactions $F_{\pm}(0)<0$ draw fermions together in real space, leading to a broader momentum-space distribution.
Fixed points exist only at $F_\pm(0)=2$. At this point, the interaction-induced flow precisely counterbalance the intrinsic energy-scale flow of the free-fermion chemical potential, resulting in $\tilde{\mu}^\pm \sim \mathcal{O}(1)$, indicating a scaling distinct from that of free fermions. Given fixed $\lambda_\pm(0)>0$, there are fixed points located at $(\tilde{\mu}^\pm)^\ast = (\frac{\lambda_\pm(0)}{ 2\pi \alpha c^{1/\alpha}})^{\frac{\alpha}{\alpha-1}}$. While the stability changes as $\alpha$ surpasses 1, they only exist within the Fermi liquid phase and do not affect the instability at $\lambda_\pm(0)<0$. The search for fixed points at the instability typically requires the consideration of higher-loop diagrams. In the U(2) and U(1) $\times$ U(1) symmetric system where the Fermi surfaces are coupled together, a fixed point exhibiting a similar $\alpha$-dependence does affect the nature of the instability. Thus, we will defer a detailed discussion on this $\alpha$ dependence to those sections later.

As $F_{1,2}=0$, we have simple RG equations for the quartic interactions
\begin{eqnarray}
\frac{d F_\pm (\vec{q})}{d \ln E} &=& \kappa (\vec{q},E) [F_\pm(\vec{q}) ]^2 .
\end{eqnarray}
These equations can be solved directly\cite{ma2024fermi}, yielding the coupling functions as 
\begin{eqnarray}
\lbrack {F}_\pm({q};E) \rbrack^{-1}=[ F_\pm(q;\Lambda)]^{-1} + \mathcal{I}^\Lambda_E \left(\frac{E}{v_F^\pm {q}} \right)  .
\end{eqnarray}
The function 
\begin{eqnarray}
    \mathcal{I}^\Lambda_E (x) = \frac{1}{\sqrt{\frac{ E^2}{\Lambda^2}x^{-2}+1}}-\frac{1}{\sqrt{x^{-2}+1}}
\end{eqnarray}
is related to $\mathcal{I}_R(x)$ defined in Eq.~\eqref{eq:mathcal_I_integral} through analytic continuation $\mathcal{I}^{\Lambda \rightarrow \infty}_E (i x)=\mathcal{I}^{\Lambda }_{E\rightarrow 0} (i x)=\mathcal{I}_R(x)$.
The comprehensive analysis of this result is in Ref.~\cite{ma2024fermi}. In terms of scale invariant variable $\tilde{q}$, the IR coupling functions are $[\tilde{F}_\pm (\tilde{q})]^{-1} 
= [ {F}_\pm (0)]^{-1} +  \mathcal{I}^\infty_{0} [1/(v_F^\pm \tilde{q})]  $. At the UV scale $\Lambda$, given constant couplings $F_\pm(0) \geq -1$, the IR coupling functions $\tilde{F}_\pm(\tilde{q})$ develops a non-trivial $\tilde{q}$-dependence. They characterize the low energy universal property of the Fermi liquid phase. If $F_\pm(0) \leq -1$, ${F}_\pm({q};E)$ would develop a singularity at 
\begin{eqnarray}
q^\pm_c 
&=&\Lambda \frac{\sqrt{2\sqrt{4[F_\pm(0)]^2-1}
\left(\sqrt{4[F_\pm(0)]^2-1}-\sqrt{3} \right)}}{\sqrt{4[F_\pm(0)]^2-1}+\sqrt{3}}, \nonumber\\
E_c^\pm&=&  \frac{v_F^\pm q_c^\pm}{\sqrt{\frac{1}{\left[\frac{1}{F_\pm(0)} +\frac{1}{\sqrt{1+\frac{(v_F^\pm q_c^\pm)^2}{\Lambda^2}}}\right]^2}-1}}. 
\end{eqnarray}
When $F_\pm(0)=-1$, $q^\pm_c=0$ and $E_c^\pm=0$ indicate that such singularity corresponds to an instability at zero momentum. It is a Pomeranchuk instability arising from the deformation of the Fermi surface. It spontaneously breaks continuous rotational symmetry in space.

\subsection{U(2) Symmetric Fermi Liquid }

\begin{figure}[h]
    \centering
    \includegraphics[width=0.6\linewidth]{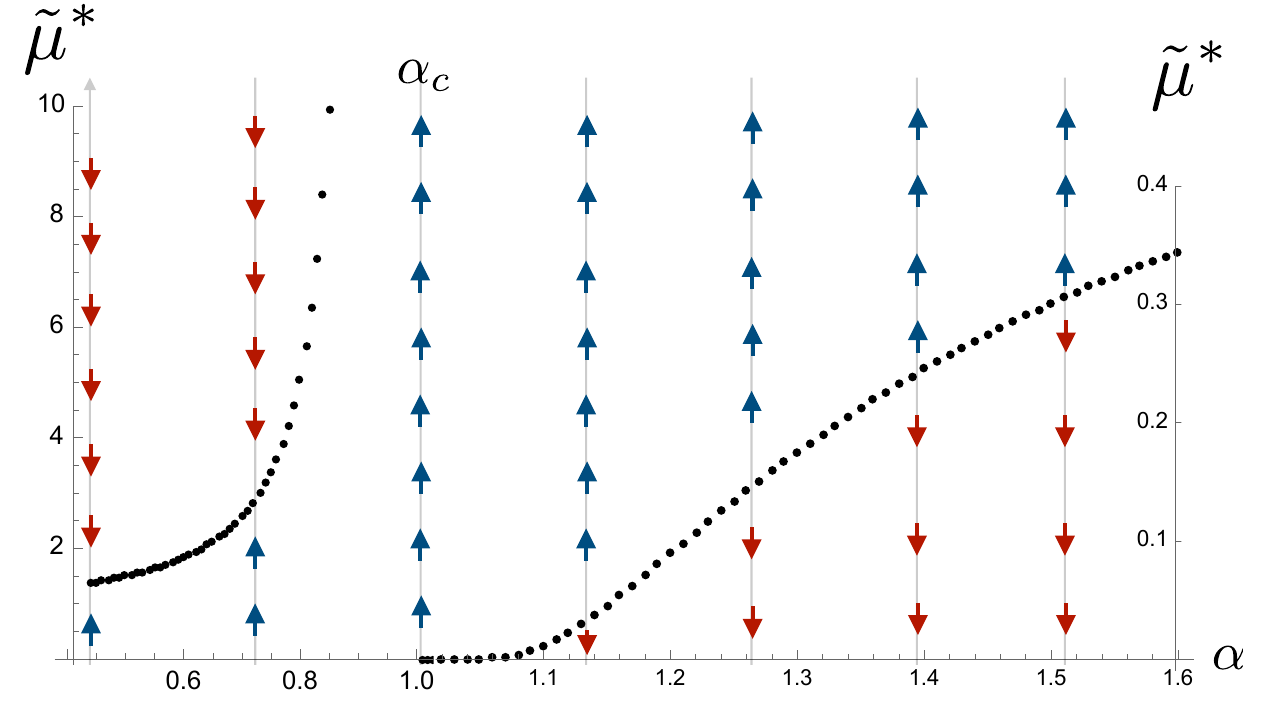}
    \caption{RG flow and fixed point structure are shown for $\frac{\lambda(0)+\lambda_1(0) }{2\pi \alpha c^{1/\alpha}}=0.67$ at different $\alpha$ values. Black dots are fixed points. To effectively visualize the significant variation in $\tilde{\mu}^\ast$ across the range of $\alpha$, the plot employs dual y-axes: the left y-axis for $\alpha<1$ while the right y-axis for $\alpha\geq 1$.}
    \label{fig:fp_collision}
\end{figure}

When $\tilde{\mu}^+=\tilde{\mu}^- \equiv \tilde{\mu}$, the systems has U(2) symmetry if the quartic interactions satisfy relation $F_+(\vec{q})=F_-(\vec{q}) \equiv F(\vec{q})=F_1(\vec{q})+F_2(\vec{q})$ in the UV.
This drives the chemical potentials to flow as following
\begin{eqnarray}
\frac{d \tilde{\mu}}{d\ln E } &=& -\tilde{\mu}+
\frac{\tilde{\mu}}{2} \left[ F(0) +  F_1(0)\right]  .
\end{eqnarray}
This resembles the same form in Eq.~\eqref{eq:mu_single} with Landau parameter $F_{\pm}(0)$ replaced by $F(0) +  F_1(0)$. The Fermi surface expands for $F(0)+F_1(0) <0$ while contraction occurs for $F(0)+F_1(0) >0$. Along the RG flow, the two chemical potentials stay equal. As a result, $v_F^+=v_F^-\equiv v_F$ and $\kappa_+ (q,E)=\kappa_- (q,E)=\kappa_{+-}(q,E) \equiv \kappa (q,E)$ holds under RG. The $\beta$ functions for quartic interactions become
\begin{eqnarray}
\frac{d \lbrack F (\vec{q})\pm F_1(\vec{q}) \rbrack}{d \ln E} &=& \kappa (\vec{q},E) \left([F(\vec{q}) \pm F_1 (\vec{q})]^2 \right),\nonumber\\
\frac{d F_2 (\vec{q})}{d \ln E} &=& \kappa(\vec{q},E)  [F_2(\vec{q})]^2.
\label{eq:interactions_SU(2)}
\end{eqnarray}
One can check that $\frac{d [F (\vec{q}) - F_1(\vec{q}) -F_2(\vec{q})]}{d \ln E} = \kappa (\vec{q},E)  [F(\vec{q})- F_1 (\vec{q}) -F_2(\vec{q})][F(\vec{q})- F_1 (\vec{q}) +F_2(\vec{q})] =0$ if $F(\vec{q})- F_1 (\vec{q}) -F_2(\vec{q})=0$ at any particular energy scale. This means under RG flow, although $F_2(\vec{q})$ flows separately, the U(2) symmetry is preserved. Then, we can solve these equations to get 
$[\mathfrak{F}_\pm (\vec{q},E)]^{-1}
= [\mathfrak{F}_\pm (\vec{q},\Lambda)]^{-1} +\mathcal{I}^\Lambda_E \left(\frac{E}{v_F q } \right)    $ where $\mathfrak{F}_+=F+ F_1$ and $\mathfrak{F}_-=F-F_1=F_2$. In terms of scale invariant variable $\tilde{q}$, they can be written as $[\tilde{\mathfrak{F}}_\pm (\tilde{q})]^{-1} 
= [ {\mathfrak{F}}_\pm (0)]^{-1} +  \mathcal{I}^\infty_{0} [1/(v_F \tilde{q})]  $. The coupling functions $\tilde{\mathfrak{F}}_\pm (\tilde{q})$ are analytic in $\tilde{q}$ as long as 
$ {\mathfrak{F}}_\pm (0) > -1 $. They define the IR physics of two-flavor Fermi liquid phase. As long as one of $\mathfrak{F}_\pm(0)$ violates this condition, the system would develop an instability and go through a phase transition out of Fermi liquid phase. 
Similarly, both instabilities occur at $q=0$. At $F(0)+F_1(0)=-1$, the chemical potential exhibits unbounded expansion. Conversely, the instability at $F(0)-F_1(0)=-1$ features a fixed point at $\tilde{\mu}^\ast=\lbrack \frac{\lambda(0)+\lambda_1(0) }{2\pi \alpha c^{1/\alpha}}\rbrack^{\frac{\alpha}{\alpha-1}} $ in this one-loop order if $\lambda(0)+\lambda_1(0)>0$. 
Around this fixed point, the $\beta$ function for the chemical potential is approximately 
\begin{eqnarray}
\frac{d\tilde{\mu}}{d\ln E} = \left(\frac{1}{\alpha}-1\right)(\tilde{\mu}-\tilde{\mu}^\ast).
\end{eqnarray}
Thus, this fixed point is attractive if $\alpha<1$ and is repulsive if $\alpha>1$. Furthermore, for $\alpha>1$, without fine tuning, the RG flow leads to either infinity or zero depending on its UV value. The fixed point location and nearby flow are illustrated in Fig.~\ref{fig:fp_collision}.

Next, let us discuss the physics of the two instabilities. Notice that the interactions are $\tilde{\mathfrak{F}}_+(\tilde{q})= 2\tilde{F}_1(\tilde{q})+\tilde{F}_2(\tilde{q})$ and $\tilde{\mathfrak{F}}_-(\tilde{q})=\tilde{F}_2(\tilde{q})$. The renormalized couplings are proportional to the 4-point vertex functions of fermions in the following way,
\begin{eqnarray}
\langle S^{\mu}_{K_1}(q,E) S^{\mu}_{K_2}(-q,-E)\rangle &=& \langle \left[\psi^{\dag,\sigma_1}_{K_1}~ (\sigma^{\mu})_{\sigma_1}^{\sigma_2}~ \psi_{K'_1,\sigma_2} \right]\left[\psi^{\dag,\sigma'_1}_{K_2} ~(\sigma^{\mu})_{\sigma'_1}^{\sigma'_2}~\psi_{K'_2,\sigma'_2}\right] \rangle \sim \tilde{F}_2(q/E), \nonumber\\
\langle \rho_{K_1}(q,E) \rho_{K_2}(-q,-E)\rangle &=& \langle \left[\psi^{\dag,\sigma_1}_{K_1}~ I_{\sigma_1}^{\sigma_2}~ \psi_{K'_1,\sigma_2} \right]\left[\psi^{\dag,\sigma'_1}_{K_2} ~I_{\sigma'_1}^{\sigma'_2}~\psi_{K'_2,\sigma'_2}\right] \rangle \sim \tilde{F}_1({q}/E),
\label{eq:4pt_F12}
\end{eqnarray}
where $\mu=x,y$, repeated indices are summed over. Therefore, for the transition driven by $\mathfrak{F}_+(0)\leq -1$ while $\mathfrak{F}_-(0)>-1$, the IR $\tilde{\mathfrak{F}}_+(\tilde{q})$ develops a singularity and IR $\tilde{\mathfrak{F}}_-(\tilde{q})$ keeps finite, indicating that $\tilde{F}_1(\tilde{q})$ diverges while $\tilde{F}_2(\tilde{q})$ is finite. 
According to Eq.~\eqref{eq:4pt_F12}, the divergence of $\tilde{F}_1(\tilde{q})$ describes dramatic density fluctuation of the Fermi surfaces, giving rise to the Pomeranchuk instability. Concurrently, the distortions of the two Fermi surfaces are nearly indistinguishable, a point we will address in detail via the study of collective modes in Sec.~\ref{sec:collective_modes}. Thus, this instability induces only the rotational symmetry breaking, and not symmetry breaking among flavors. In contrast, the transition driven by $\mathfrak{F}_-(0)\leq -1$ with $\mathfrak{F}_+(0)>-1$ has only singular $\tilde{F}_2(\tilde{q})$ in IR. It results in a symmetry breaking in flavor. If the flavor stands for spin of fermions, then a spontaneous ferromagnetism is developed. If the flavor index labels different valley, the system develops a valley polarization order through this instability.
The study of the RG flow can be straightforwardly generalized to the case where interactions are angular dependent. We leave this generalization for future study. 

\subsection{U(1) $\times$ U(1) Fermi Liquid}

We now consider the case where $F_+\neq F_-$, which breaks the U(2) symmetry down to U(1) $\times$ U(1).
In this case, both chemical potentials and quartic couplings follow distinct RG flow. To proceed, we solve the RG equations for the quartic couplings. Subsequently, we study whether $\tilde{\mu}^\pm$ flow to a fixed point given various UV quartic interactions, including coupling strengths that render the systems unstable. If an attractive fixed point exists, the instability corresponds to a RG fixed point, indicating a continuous phase transition. In addition, as $F_2$ flows independently, we assume that $F_2(\vec{q})$ remains analytic in the IR. We also fix the UV coupling $F_2(0)$ to a constant, ensuring the instability is not induced by $F_2(0)$. 

The RG equations can be rewritten in terms of  $F_\pm(\vec{q})/F_1(\vec{q}) = R_\pm (\vec{q})$ as 
\begin{eqnarray}
\frac{dR_+(\vec{q})}{d\ln E} &=& \kappa_-(\vec{q},E) F_1(\vec{q}) \left[1-R_+(\vec{q})R_-(\vec{q})\right] ,\nonumber\\
\frac{dR_-(\vec{q})}{d\ln E} &=& \kappa_+(\vec{q},E) F_1(\vec{q}) \left[1-R_+(\vec{q})R_-(\vec{q})\right] ,\nonumber\\
\frac{dF_1(\vec{q})}{d\ln E} &=& \left[\kappa_+ (\vec{q},E)R_+(\vec{q}) + \kappa_-(\vec{q},E)R_-(\vec{q})\right][F_1(\vec{q})]^2.
\label{eq:quartic_beta_R}
\end{eqnarray}
The solution of these differential equations yields scale-dependent functions $R_\pm(\vec{q},E)$ and $F_1(\vec{q},E)$. 
In the low energy limit, the fixed point coupling functions, denoted as $\tilde{R}_\pm(\tilde{q}) $ and $\tilde{F}_1(\tilde{q})$, depend only on dimensionless variable $\tilde{q}\equiv q/E$. 
They satisfy the fixed point equations $\frac{d \tilde{R}_\pm(\tilde{q})}{d\ln E}=0$ and $\frac{d \tilde{F}_1(\tilde{q})}{d\ln E}=0$, leading to
\begin{eqnarray}
-\frac{d \tilde{R}_+(\tilde{q})}{d\ln \tilde{q}} &=& \tilde{\kappa}_-(\tilde{q}) \tilde{F}_1(\tilde{q}) \left[1-\tilde{R}_+(\tilde{q})\tilde{R}_-(\tilde{q})\right] ,\nonumber\\
-\frac{d\tilde{R}_-(\tilde{q})}{d\ln\tilde{q}} &=& \tilde{\kappa}_+(\tilde{q}) \tilde{F}_1(\tilde{q}) \left[1-\tilde{R}_+(\tilde{q})\tilde{R}_-(\tilde{q})\right] ,\nonumber\\
-\frac{d\tilde{F}_1(\tilde{q})}{d\ln \tilde{q}} &=& \left[\tilde{\kappa}_+ (\tilde{q})\tilde{R}_+(\tilde{q}) + \tilde{\kappa}_-(\tilde{q})\tilde{R}_-(\tilde{q})\right][\tilde{F}_1(\tilde{q})]^2,
\label{eq:fp_quartic}
\end{eqnarray}
where $\tilde{\kappa}_\pm (\tilde{q}) = \frac{(v_F^\pm)^2 \tilde{q}^2 }{\left[1+ (v_F^\pm)^2 \tilde{q}^2\right]^{3/2}}$. 
The details of solving these equations can be found in Appendix \ref{app:solving_quartic_RG}.
The solutions are
\begin{eqnarray}
\tilde{R}_+(\tilde{q}) &=&  {R}_+(0) - {F}_1(0) \left[1- R_+(0) R_-(0)\right]\mathcal{I}^\Lambda_0 [1/(v_F^- \tilde{q}) ] ,
\nonumber\\
\tilde{R}_-(\tilde{q}) &=& {R}_-(0) -{F}_1(0) \left[1- R_+(0) R_-(0)\right]\mathcal{I}^\Lambda_0 [1/(v_F^+ \tilde{q})]  ,
\end{eqnarray}
and
\begin{eqnarray}
[\tilde{F}_1(\tilde{q}) ]^{-1}  &=&  [{F}_1(0) ]^{-1} + R_+(0) \mathcal{I}^\Lambda_0 \left[\frac{1}{v_F^+\tilde{q}}\right] + R_-(0) \mathcal{I}^\Lambda_{0}\left[\frac{1}{v_F^-\tilde{q}}\right] \nonumber\\
&+& F_1(0) \left(R_+(0)R_-(0)-1\right) \mathcal{I}^\Lambda_0 \left[\frac{1}{v_F^+\tilde{q}}\right]\mathcal{I}^\Lambda_0 \left[\frac{1}{v_F^-\tilde{q}}\right] .
\label{eq:F1_U1}
\end{eqnarray}
For finite $R_\pm(0)$ and $F_1(0)$, $\tilde{R}_\pm(\tilde{q})$ stays finite for all $\tilde{q}$.
However, $\tilde{F}_1(\tilde{q})$ may diverge, which indicates an instability. 
The pole of the renormalized coupling function after a Wick rotation physically corresponds to a collective mode. Its velocity satisfies
\begin{eqnarray}
&& [{F}_1(0) ]^{-1} + R_+(0) \mathcal{I}^\Lambda_0 \left[\frac{-iE}{v_F^+ {q}}\right] + R_-(0) \mathcal{I}^\Lambda_{0}\left[\frac{-iE}{v_F^- {q}}\right] \nonumber\\
&+& F_1(0) \left(R_+(0)R_-(0)-1\right) \mathcal{I}^\Lambda_0 \left[\frac{-iE}{v_F^+ {q}}\right]\mathcal{I}^\Lambda_0 \left[\frac{-iE}{v_F^- {q}}\right]  =0.
\label{eq:general_poles}
\end{eqnarray}
Notice that this equation is identical to Eq.~\eqref{eq:LKE_imbalanced}. Its solution will be presented in Sec.~\ref{sec:collective_modes}. At the transition point, the collective mode has zero velocity and characterizes the critical deformation of the Fermi surfaces.
Next we study the IR coupling functions based on whether $\tilde{R}_+ (\vec{q})\tilde{R}_- (\vec{q})$ equals one in the UV.

\subsubsection{Mean-Field Interactions Cancellation
}

\begin{figure}[h]
    \centering
    \includegraphics[width=0.5\linewidth]{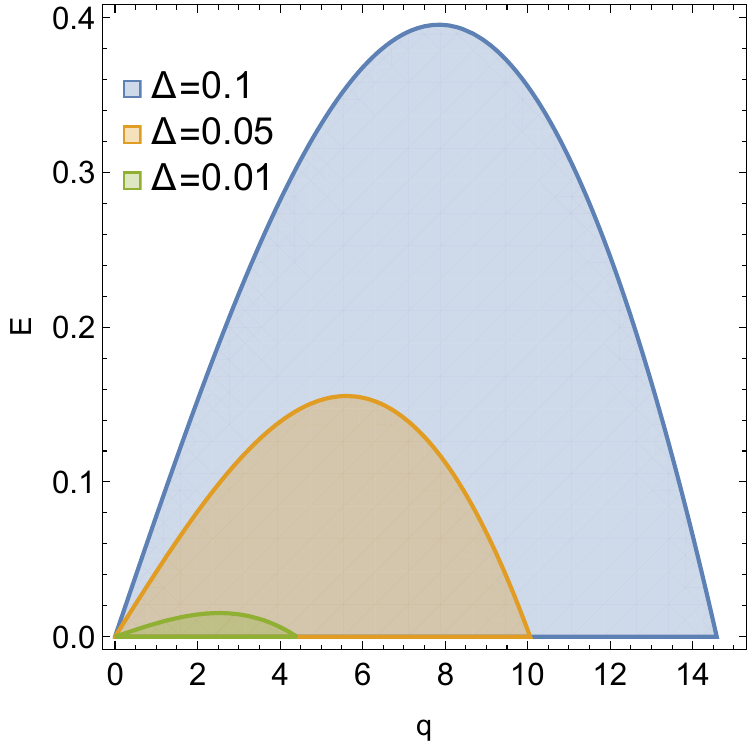}
    \caption{Region where the left-hand side of Eq.~\eqref{eq:KEQ_U1_special} is negative for $\alpha=1.5$, $v_F^-=0.5$, $z=0.8$, $\Lambda=10$ and $F_1(0)= F_c+\Delta$ with various values of $\Delta$. The plot is qualitatively similar for $\alpha<1$.}
    \label{fig:U1_special_instability}
\end{figure}

We consider the special interactions satisfying Eq.~\eqref{eq:special_ratio}. Accordingly, the Landau parameters have ratio $F_\pm(0)/F_1(0)= -z^{\mp \frac{1+\alpha}{2\alpha}}$, which satisfies $R_+(0)R_-(0) =1$. The expression of $\tilde{F}_1(\tilde{q})$ can be largely simplified to be
\begin{eqnarray}
[\tilde{F}_1(\tilde{q}) ]^{-1}  &=& [F_1(0) ]^{-1}- z^{-\frac{1+\alpha}{2\alpha}}\mathcal{I}^\Lambda_0 [1/(v_F^+ \tilde{q})]   - z^{\frac{1+\alpha}{2\alpha}} \mathcal{I}^\Lambda_0 [1/(v_F^- \tilde{q}) ]   .
\label{eq:KEQ_U1_special}
\end{eqnarray}
When $F_1(0)<F_c\equiv \frac{1}{z^{-\frac{1+\alpha}{2\alpha}}   + z^{\frac{1+\alpha}{2\alpha}}  }$, $\tilde{F}_1(\tilde{q})$ is analytic and has the same sign of $F_1(0)$, meaning that the system is in the Fermi liquid phase. 
Conversely, for $F_1(0)\geq F_c$, the system experiences an instability as the IR coupling function $\tilde{F}_1(\tilde{q})$ becomes singular. As shown in Fig.~\ref{fig:U1_special_instability}, the system develops a singularity in $\tilde{F}_1(\tilde{q})$ within the shaded region for positive $\Delta$ defined as $F_1(0)= F_c+\Delta$. As $\Delta$ decreases to zero, the boundary of the shaded region shrinks and ultimately vanishing at $q=0$ and $E=0$ when $\Delta$ is tuned to zero. This implies that the instability occurs at $q=0$ and zero energy when $F_1(0)$ reaches $F_c$.

Furthermore, the $\beta$ function for $\tilde{\mu}$ is given by $\frac{d \tilde{\mu}^\pm}{d\ln E}=- \tilde{\mu}^\pm$, identical to that of free fermions with a Fermi surface. 
Any Fermi surface flows according to $\tilde{\mu}^\pm \sim E^{-1}$, resulting in a physical Fermi surface with $\mu^\pm \sim \mathcal{O}(1)$, which remains invariant under RG. This is an interesting property, as the interactions have no net impact on the Fermi surface size. Nonetheless, the Fermi surfaces remain coupled. Later in Sec.~\ref{sec:collective_modes}, we investigate the collective mode as the pole of IR coupling function $\tilde{F}_1(\tilde{q})$, which reveals the synchronized oscillation of the two Fermi surfaces due to this coupling.

\subsubsection{Generic Interactions }

In general, for $R_+(0)R_-(0) \neq 1$, two Pomeranchuk instabilities can happen when $F_1(0)$ is tuned to 
\begin{eqnarray}
    F_{1,c}^\pm(0)=\frac{1}{2} \frac{\lbrack R_+(0)+R_-(0))\rbrack \pm \sqrt{4+\lbrack R_+(0)-R_-(0) \rbrack^2}}{1-R_+(0)R_-(0)},
    \label{eq:instability_U1_LKE}
\end{eqnarray}
where the onset of singularity occurs in the coupling function $\tilde{F}_1(\tilde{q})$.
This indicates a Fermi liquid phase within the range $\textrm{Min}\left[F_{1,c}^+(0),F_{1,c}^-(0)\right]<F_1(0)<\textrm{Max}\left[F_{1,c}^+(0),F_{1,c}^-(0)\right]$. Beyond this range, a transition out of the Fermi liquid phase happens at $q=0$, and the coupling function features a singularity, qualitatively similar to the behavior shown in Fig.~\ref{fig:U1_special_instability}.

\begin{figure}[h]
    \centering
    \includegraphics[width=0.48\linewidth]{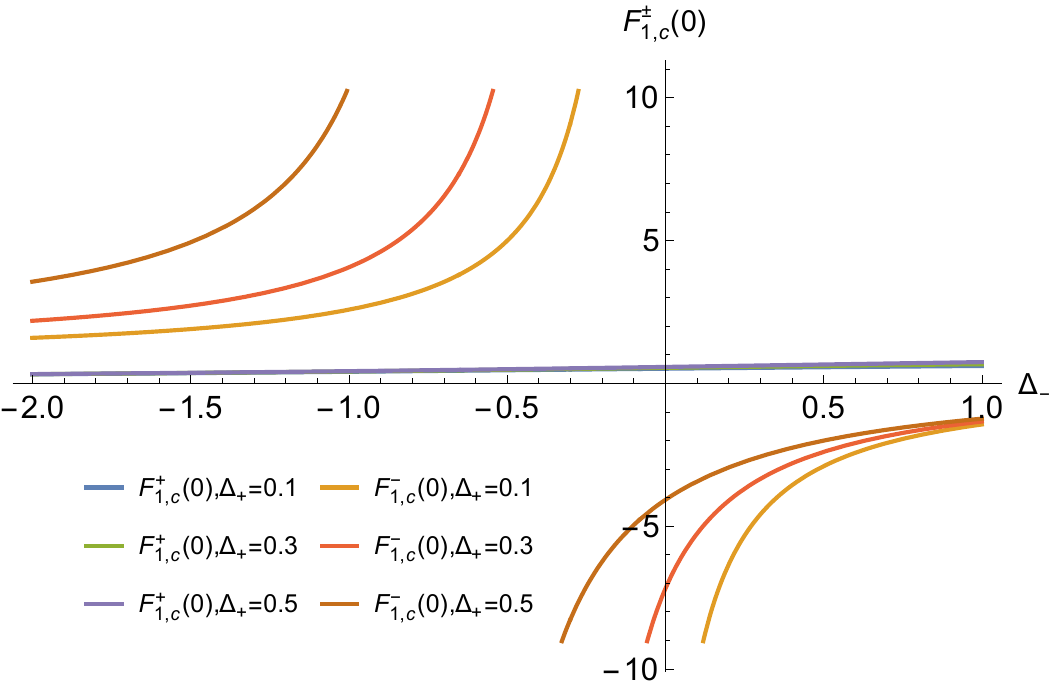}
    \includegraphics[width=0.48\linewidth]{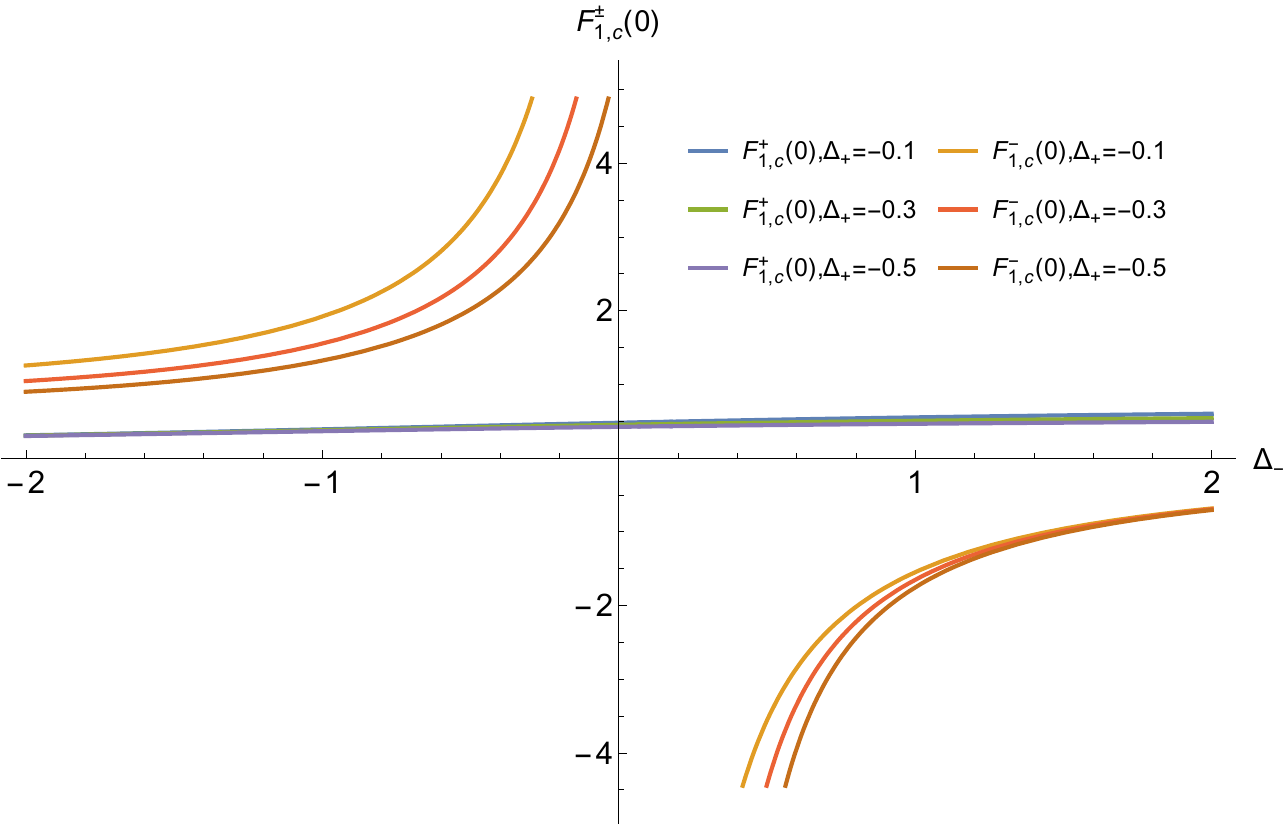}
    \caption{$F_{1,c}^\pm(0)$ versus $\Delta_-$ for varying $\Delta_+$. We set $\alpha=1.5$ and $z=0.8$.}
    \label{fig:F1c_U1}
\end{figure}

We now specifically assign the couplings to be $R_\pm(0)= -z^{\mp \frac{1+\alpha}{2\alpha}} + z^{\pm \frac{1-\alpha}{2\alpha}} \Delta_\pm$, where $\Delta_\pm \neq 0$ are constant.
As displayed in Fig.~\ref{fig:F1c_U1}, $F_{1,c}^+(0)$ remains finite for all $\Delta_\pm$. Similarly, $F_{1,c}^-(0)$ is finite if $\Delta_\pm$ are finite. For finite $z$, varying $\alpha$ only qualitatively changes the dependence of $F_{1,c}^\pm(0)$ on $\Delta_\pm$. As $\Delta_\pm \rightarrow 0^\pm$, $F_{1,c}^- (0)$ diverges to $\mp \infty$. 

One can further study the RG equation for the chemical potentials, including where the instabilities occur. Individual chemical potential flows as
\begin{eqnarray}
\frac{d \tilde{\mu}^+}{d\ln E } &=& 
-\tilde{\mu}^+ + \frac{1}{2}\tilde{\mu}^+  z^{\frac{1-\alpha}{2\alpha}}\Delta_+  F_{1}(0), \nonumber\\
\frac{d \tilde{\mu}^-}{d\ln E} &=& 
-\tilde{\mu}^-  +\frac{1}{2} \tilde{\mu}^-  z^{-\frac{1-\alpha}{2\alpha}}\Delta_- F_{1}(0).
\end{eqnarray} 
For finite $F_1(0)$,
a fixed point exists at $z^\ast=(\frac{\Delta_-}{\Delta_+})^{ \frac{\alpha}{1-\alpha}}$ with specific $\Delta_{\pm}$ satisfying the condition
\begin{eqnarray}
(\Delta_+\Delta_-)^{1/2} F_{1}(0)=2
\label{eq:fp_condition}
\end{eqnarray}
for any $\alpha \neq 1$. At $\alpha=1$, if $\Delta_+=\Delta_-=2/F_{1}(0)$, a fixed point can be found at $(\tilde{\mu}^+)^\ast=(\tilde{\mu}^-)^\ast$ with an emergent U(2) symmetry. Away from this fixed point, we have the flow equation as 
\begin{eqnarray}
\frac{d \tilde{\mu}^\pm}{d\ln E } &=& 
\pm \frac{1-\alpha}{4\alpha}  \tilde{\mu}^\pm \left[  (\Delta_+/\Delta_-)^{  \mp\frac{1}{2}+\frac{\alpha}{1-\alpha}}\right] (z-z^\ast)\Delta_\pm F_{1}(0).
\end{eqnarray}
\begin{figure}[h]
    \centering
    \includegraphics[width=0.95\linewidth]{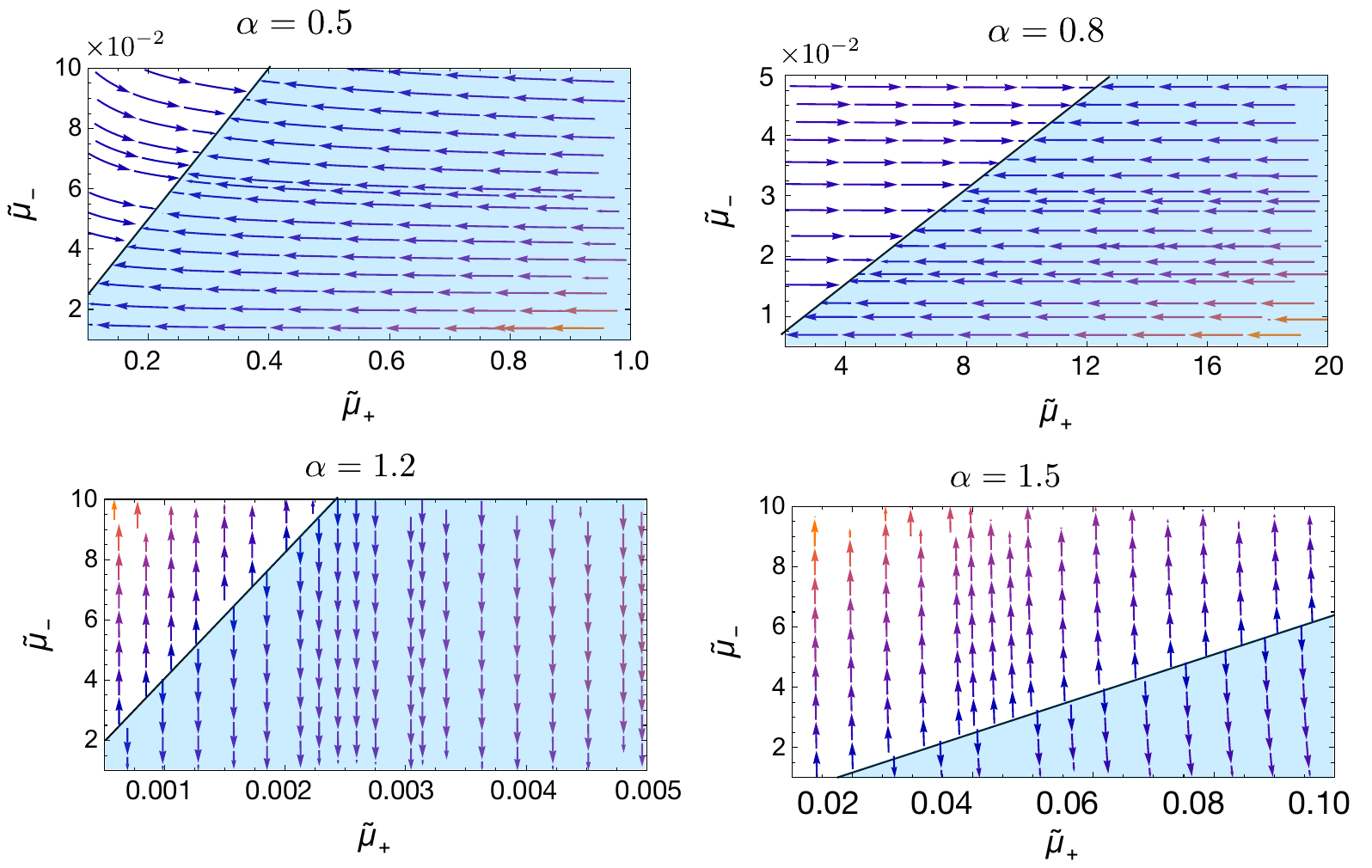}
    \caption{RG flow of $\tilde{\mu}^\pm$ at various $\alpha$ near the fixed point at $z^\ast=(\Delta_-/\Delta_+)^{\alpha/(1-\alpha)}$ with $\Delta_-  F_{1}(0)=4$ and $\Delta_+ F_{1}(0)  =1$ satisfying Eq.~\eqref{eq:fp_condition}.}
    \label{fig:zast_flow_both_mu}
\end{figure}
It is obvious that the stability of the fixed point reverses as $\alpha$ cross $1$.
The RG flow of $\tilde{\mu}^\pm$ is presented in Fig.~\ref{fig:zast_flow_both_mu}, assuming $\Delta_-  F_{1}(0)=4$ and $\Delta_+ F_{1}(0)  =1$.
We observe that when $\alpha<1$, the fixed point at $z^\ast$ is attractive whereas it becomes repulsive when $\alpha>1$. As $\alpha$ approaches one from below, the fixed point $z^\ast$ converges to zero if $\Delta_-<\Delta_+$. Conversely, when $\alpha$ exceeds one, the fixed point reappears from $z^\ast \rightarrow \infty$. Thus, the fixed point's stability change is due to crossing a boundary of the parameter space. This analysis holds even when $F_1(0)$ falls into the regime $\left(\textrm{Min}\left[F_{1,c}^+(0),F_{1,c}^-(0)\right],\textrm{Max}\left[F_{1,c}^+(0),F_{1,c}^-(0)\right]\right)$. It further suggests that within the Fermi liquid phase, one Fermi surface is already depleted.

\begin{figure}
    \centering
    \includegraphics[width=0.5\linewidth]{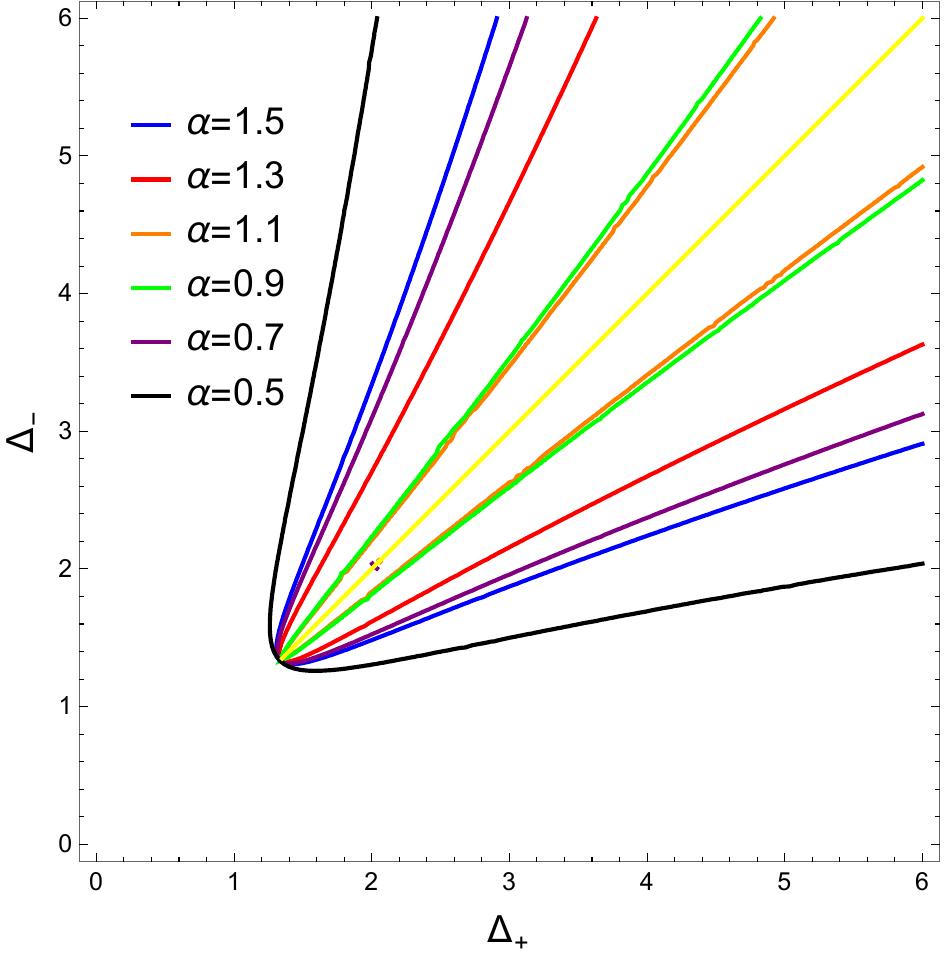}
    \caption{The points $(\Delta_+^\ast,\Delta_-^\ast)$ satisfy the condition in Eq.~\eqref{eq:fp_condition} for various $\alpha$. As $\alpha \rightarrow 1$, the curve converges to the yellow line. }
    \label{fig:fp_delta}
\end{figure}

At the phase boundary, $F_1(0)$ takes its critical values $F_{1,c}^\pm(0)$, which are functions of $\Delta^\pm$ and $z$.
In order to satisfy the fixed point condition in Eq.~\eqref{eq:fp_condition}, setting $z=z^\ast$ enforces the dependence of $z$ on $\Delta^\pm$ and $\Delta^\pm$ must adopt specific values.
We observe that $\lbrack F_{1,c}^{-}(0) \rbrack^\ast = -1/\sqrt{\Delta_+\Delta_-}$, which fails to meet the fixed point condition for any $\Delta_\pm$, since $(\Delta_+\Delta_-)^{1/2} \lbrack F_{1,c}^{-}(0) \rbrack^\ast =-1 \neq 2$. Consequently, at this instability, $\tilde{\mu}^\pm$ continue to flow and no fixed point is reached, at least at this one-loop order. On the contrary, for $\lbrack F_{1,c}^{+}(0) \rbrack^\ast $, a curve of points $(\Delta_+^\ast,\Delta_-^\ast)$ exists that satisfies the condition in Eq.~\eqref{eq:fp_condition}, as depicted in Fig.~\ref{fig:fp_delta}.
Near the true fixed point associated with the Pomeranchuk instability, the RG flow diagrams are qualitatively identical to Fig.~\ref{fig:zast_flow_both_mu} when $F_1(0)$ is replaced by $\lbrack F_{1,c}^{+}(0) \rbrack^\ast $ determined by $(\Delta_+^\ast,\Delta_-^\ast)$. This indicates that the transition is continuous for $\alpha<1$, corresponding to a stable fixed point. While the transition becomes discontinuous for $\alpha>1$, as the only fixed point becomes unstable and the chemical potentials flow to either infinity or zero under RG.

\section{Collective modes \label{sec:collective_modes}}

With the knowledge of phases and transitions of this interacting fermionic system, we proceed to investigate the collective modes. This is accomplished by solving the pole equation of the IR coupling functions, as given in Eq.~\eqref{eq:general_poles}, or equivalently the linearized kinetic equation in Eq.~\eqref{eq:LKE_imbalanced}. The constant interactions $F_{\pm,1,2}$ parametrizing the kinetic equation are identified with the UV interactions at $\vec{q}=0$, namely $F_{\pm,1,2}(0)$. We thus use these terms interchangeably.

\subsection{Two Flavors are Decoupled}

We begin by considering the simplified case where $F_1=F_2=0$, in which each of the individual Fermi surfaces $\pm$ has a zero sound mode with a velocity $v_{zs}\equiv E/q$ satisfying
\begin{eqnarray}
1+F_\pm \mathcal{I}_R\left(\frac{E}{v_F^\pm q}\right) = 0.
\end{eqnarray}
The amplitude $\mathcal{A}_\pm$ is not fixed. The two modes have velocities depending on the interactions $F_\pm$, as
\begin{eqnarray}
v_{zs}^\pm  \equiv \left( \frac{E}{ q} \right)_{\pm}  =v_F^\pm \frac{  F_\pm +1  }{\sqrt{1+2 F_\pm  }}.
\end{eqnarray}
For each Fermi surface, a well-defined propagating zero sound mode exists with $v_{zs}^\pm >v_F^\pm$ when $F_\pm >0$. It manifests as a hidden mode when $-1/2<F_\pm<0$ and a decaying mode when $-1<F_\pm<-1/2$. If $F_\pm<-1$, these modes become unstable, growing exponentially and resulting in Pomeranchunk instability. The condition $F_\pm=-1$ corresponds to the transition point out of the U(1) $\times$ U(1) Fermi liquid phase. This result aligns with findings from studies on single-flavor interacting spinless fermions with a circular Fermi surface\cite{lucas2018electronic,DASSARMA2021168495,klein2020hidden}.
With finite $F_1$, the Fermi surfaces become coupled and additional collective modes emerge.

\subsection{U(2) Symmetric Fermi Liquid \label{sec:U2_collective_mode}}

\begin{figure}[h]
    \centering
    \includegraphics[width=0.45\linewidth]{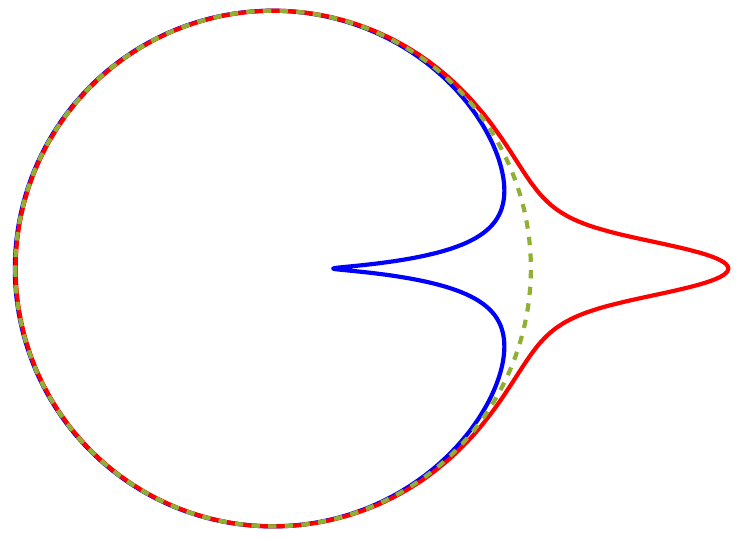}
    \includegraphics[width=0.35\linewidth]{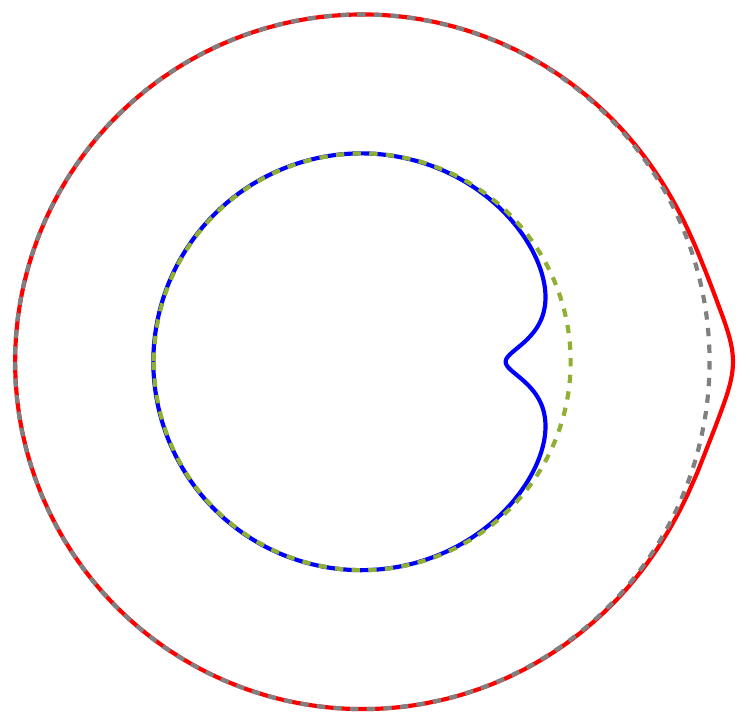}
    \caption{Collective mode configurations $\pm \frac{1}{50} \mathcal{A}_\pm u_{\theta_k}^\pm(\vec{q},\omega)$ at $\omega/q=v_{zs}$ are shown for a U(2) symmetric system (left) and a U(1)$\times$U(1) Fermi liquid with a special interaction ratio $F_\pm/F_1=- z^{\mp\frac{1+\alpha}{2\alpha}} $ (right). In both cases, the dashed circle is the original Fermi surfaces and a normalization factor of $1/50$ is chosen for visual clarity. The U(2) symmetric system (left) features out-of-phase deformation of the two Fermi surfaces, with an amplitude ratio independent of $\alpha$. Parameters are $F=0.2$, $F_1=0.1$, $\mu^\pm=1$ and $v_F^\pm=1$. For the U(1) $\times$ U(1) system (right), parameters are $\alpha=0.5$ and $z=0.6$. The resulting amplitude ratio $|\mathcal{A}^-/\mathcal{A}_+|=z^{1/\alpha}<1$ causes a smaller deformation for larger Fermi surface with chemical potential $\mu^-$.}
    \label{fig:collective_mode_conf}
\end{figure}

When the two Fermi surfaces are of equal size with $\mu^+=\mu^-$ and $v_F^+ =v_F^- \equiv v_F$, two modes exist with velocities
\begin{eqnarray}
(v_{zs}^\pm)^2=\left(\frac{E}{ q}\right)^2_\pm = \frac{v_F^2}{1-\frac{1}{\left[ 1-\frac{F_+ + F_-\pm \sqrt{4F_1^2+(F_+-F_-)^2}}{2(F_1^2-F_+F_-)} \right]^2}}. 
\end{eqnarray}
Each of them represents a specific manner in which the $\pm$ Fermi surfaces can be deformed, according to $u^\pm_{\theta_{k}}(\vec{q},\omega) $ in Eq.~\eqref{eq:u_pm}
with the amplitude ratio
\begin{eqnarray}
\left(\frac{\mathcal{A}_-}{\mathcal{A}_+} \right)_\pm =- \frac{F_+-F_-\pm \sqrt{4F_1^2+(F_+-F_-)^2}}{2 F_1}.
\end{eqnarray}
In the U(2) symmetric system, where $F_+=F_- \equiv F$, the resulting velocities are
\begin{eqnarray}    
v_{zs}^+ \equiv \left(\frac{E}{q}\right)_+ &=& v_F \frac{  F-|F_1|+1  }{ \sqrt{2(F-|F_1|)+1}} , \quad \textrm{ with } \left(\frac{\mathcal{A}_-}{\mathcal{A}_+} \right)_+= -\frac{|F_1|}{F_1} ,\nonumber\\
  v_{zs}^- \equiv \left(\frac{E}{q}\right)_- &=& v_F \frac{  F+|F_1|+1  }{ \sqrt{2(F+|F_1|)+1}}, \quad \textrm{ with } \left(\frac{\mathcal{A}_-}{\mathcal{A}_+} \right)_-= \frac{|F_1|}{F_1}.
\end{eqnarray} 
Therefore, these two mode velocities share the same expression as those of the decoupled Fermi liquid, with $F_\pm$ replaced by $F\pm F_1$. This is what we expect based on the RG equations for the U(2) symmetric system. As we analyzed before,  
the renormalized interaction $F- F_1=F_2$ is proportional to the 4-point function $\langle \psi^\dag_{K_1} \sigma^{x,y} \psi_{K'_1}\times \psi^\dag_{K_2} \sigma^{x,y} \psi_{K'_2} \rangle$ as a flavor-flavor correlation function, and the renormalized interaction $F_1$ is proportional to the density-density correlation function $\langle \psi^\dag_{K_1} I \psi_{K'_1}\times \psi^\dag_{K_2} I \psi_{K'_2}\rangle$. Thus, if $F_1>0$($F_1<0$), $v_{zs}^+$($v_{zs}^-$) is the velocity of the collective mode inducing flavor symmetry breaking as the two Fermi surfaces have out-of-phase deformation, i.e. $\frac{\mathcal{A}_-}{\mathcal{A}_+} = -1$. One example is shown in Fig.~\ref{fig:collective_mode_conf}(left).
The other velocity is associated with the mode responsible for the density fluctuation with in-phase Fermi surfaces deformation $\frac{\mathcal{A}_-}{\mathcal{A}_+} = 1$.

Similarly, as the decoupled Fermi liquid discussed above, a system with $-1/2<F\pm F_1 <0$ has hidden modes, while $-1<F\pm F_1 <-1/2$ indicates decaying modes. 
When $F\pm |F_1|=-1$, the collective mode has zero velocity $v_{zs}^\pm=0$. This indicates that, beyond this point, the proliferation of the mode is energetically favored leading to an instability of the U(2) Fermi liquid with spontaneous rotation or flavor symmetry breaking. This is the Pomeranchuk or Stoner instability for the two-flavor interacting fermion system with global U(2) symmetry.

\subsection{U(1) $\times$ U(1) Fermi liquid}

For unequal chemical potentials, $\mu^+\neq \mu^-$, Eq.~\eqref{eq:LKE_imbalanced}
and
Eq.~\eqref{eq:amplitude} can only be solved numerically. 
With fixed $\mu^\pm$, varying $F_\pm$ and $F_1$ enables the exploration of the system's behavior, specifically the collective mode velocity and Fermi surface deformation amplitude. 

\begin{figure}[h]
    \centering
    \includegraphics[width=0.7\linewidth]{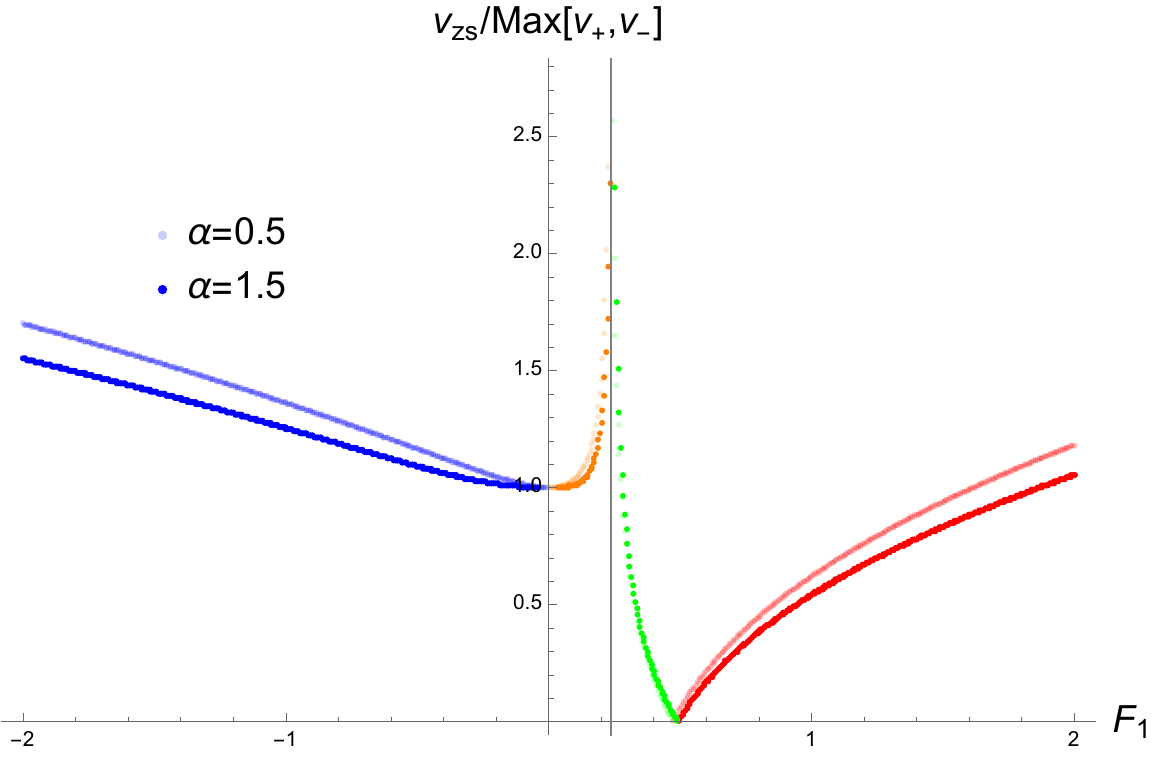}
    \caption{Collective mode velocities in the U(1)$\times$U(1) Fermi liquid with interaction ratio $F_\pm/F_1=- z^{\mp\frac{1+\alpha}{2\alpha}} $. The plot contains a propagating mode (blue), a hidden mode (orange), a decaying mode (green), and an unstable growing mode (red) depending on $F_1$. The vertical gray line represents $F_1=1/2F_c$, where $v_{zs}$ diverges. At $F_1=F_c$, $v_{zs}=0$. Other parameters are $\alpha=0.8$, $z=0.8$, and $v_F^-=0.5$.}
    \label{fig:all_collective_modes_U1_special}
\end{figure}

We start with the specific case where the quartic interactions satisfy Eq.~\eqref{eq:special_ratio}. Accordingly, the generalized Landau parameters have the relation $F_\pm/F_1=- z^{\mp\frac{1+\alpha}{2\alpha}} $.
Therefore, the mode velocity satisfies 
\begin{eqnarray}
\frac{1}{F_1} - \left(\frac{\mu^+}{\mu^-}\right)^{-\frac{1+\alpha}{2\alpha}}\mathcal{I}_R\left(\frac{E}{ v_F^+ q }\right)- \left(\frac{\mu^-}{\mu^+}\right)^{-\frac{1+\alpha}{2\alpha}} \mathcal{I}_R\left(\frac{E}{ v_F^- q }\right) =0.
\label{eq:mode_U1_special_ratio}
\end{eqnarray}
A well-defined propagating
collective mode exists when $F_1<0$, with a velocity $v_{zs}\equiv E/q > \textrm{Max}(v_F^+,v_F^-)$.
As $|F_1|$ increases, the velocity increases. For positive interactions within the region $0<F_1<\frac{1}{2}F_c$, where $F_c=\frac{1}{z^{-\frac{1+\alpha}{2\alpha}}+z^{\frac{1+\alpha}{2\alpha}}}$, a hidden mode is found. When $F_1>\frac{1}{2}F_c$, only imaginary solutions for $v_{zs}$ exist. The solutions within the range $1/2F_c<F_1<F_c$ correspond to a decaying mode while that for $F_1>F_c$ is an unstable growing mode. 
The velocities of these distinct modes at different value of $F_1$ are presented in Fig.~\ref{fig:all_collective_modes_U1_special}.
The critical value $F_c$ signifies the Pomeranchuk instability of the two-flavor Fermi liquid phase. At $F_c$, the collective mode is gapless with $v_{zs}=0$. For any value of $F_1$, the collective mode exhibits Fermi surface deformation with an amplitude ratio $\mathcal{A}_-/\mathcal{A}_+  = - z^{1/\alpha}$, given fixed $z$ and $\alpha$ taking any value. The sign of this ratio indicates an out-of-phase density oscillation between the two Fermi surfaces, as illustrated in Fig.~\ref{fig:collective_mode_conf}(right).
Consequently, despite the interaction having no impact on single fermion properties or Fermi surface size, a single collective mode dictates the relative density oscillation between the two Fermi surfaces, revealing coupling beyond the mean-field approximation. 

In general, $F_+F_--F_1^2 \neq 0$. 
Particularly, with $R_\pm =F_\pm/F_1
= -z^{\mp \frac{1+\alpha}{2\alpha}} + z^{\pm\frac{1-\alpha}{2\alpha}} \Delta_\pm
$, the system still has a single mode when $\Delta_\pm$ are small. 
\begin{figure}[h]
    \centering
    \includegraphics[width=0.4\linewidth]{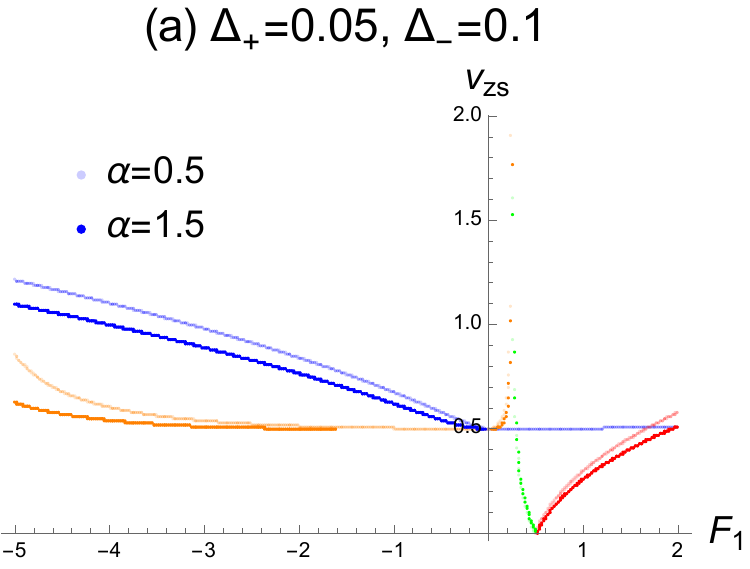}
    \includegraphics[width=0.4\linewidth]{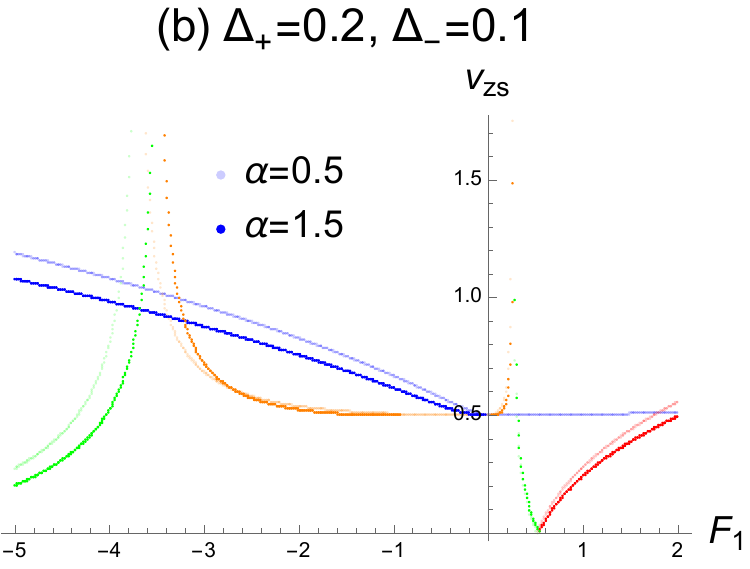}
    \includegraphics[width=0.4\linewidth]{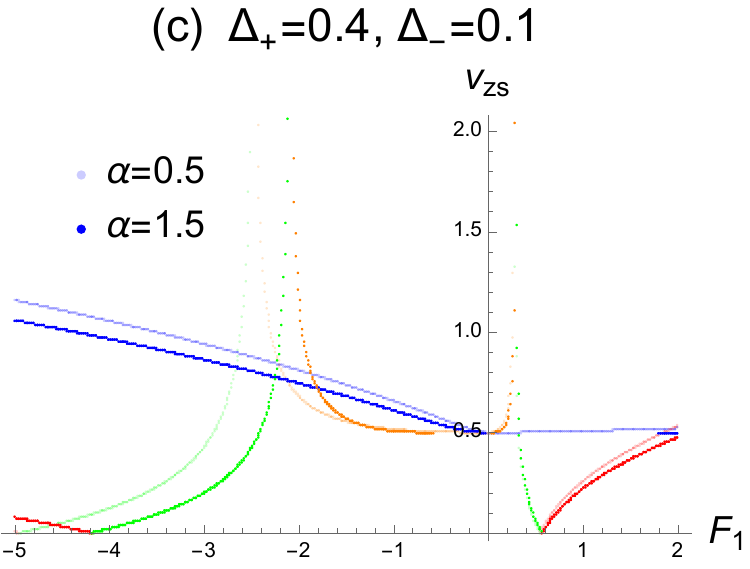}
    \includegraphics[width=0.4\linewidth]{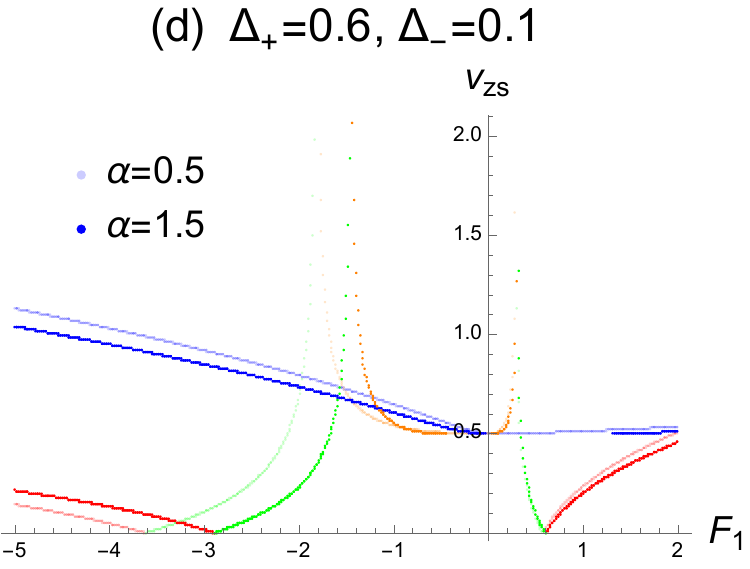}
    \caption{Collective mode velocities in the U(1) $\times$U(1) Fermi liquid with interaction ratio $F_\pm/F_1=- z^{\mp\frac{1+\alpha}{2\alpha}} +z^{\pm\frac{1-\alpha}{2\alpha}}  \Delta_\pm$. For a fixed $\Delta_-$, increasing $\Delta_+$ tends to favor the simultaneous existence of two collective modes and reduces the Fermi liquid phase region. Instability occurs when an unstable growing mode (red) develops. Parameters are $z=0.8$ and $v_F^-=0.5$.}
    \label{fig:modes_Delta}
\end{figure}
For $\Delta_+$ and $\Delta_-$ sharing the same sign, the system evolves continuously as a function of them.
As shown in Fig.~\ref{fig:modes_Delta}, with $z$, $\alpha$ and $\Delta_-$ fixed, an increase in $\Delta_+$ leads to a wider range of $F_1$ where the system can simultaneously support two collective modes. In addition, this increase in $\Delta_+$ shrinks the Fermi liquid region within the parameter range defined by $F^\pm_{1,c}$, presented in Eq.~\eqref{eq:instability_U1_LKE}, where there are two Pomeranchuk instabilities.
Furthermore, we can study the amplitude ratio of Fermi surface deformations, $\mathcal{A}_-/\mathcal{A}_+$, using Eq.~\eqref{eq:amplitude}. Given that $\tilde{\mu}^\pm$ flows and $z^\ast=(\Delta_-^\ast/\Delta_+^\ast)^{\alpha/(1-\alpha)}$, the amplitude ratio at the instability fixed point occurring at $F_{1,c}^+(0)$ is determined to be 
\begin{eqnarray}
(\mathcal{A}_-/\mathcal{A}_+)_+=\frac{ 1+ R_- \lbrack F^+_{1,c} \rbrack^\ast}{-\lbrack F^+_{1,c} \rbrack^\ast} (z^\ast)^{\frac{1-\alpha}{2\alpha}} =  -(z^\ast)^{1/\alpha}  .
\label{eq:amplitude_U1_solution}
\end{eqnarray}
Thus, $\Delta_\pm^\ast$ turns out to be the only variables parameterizing the fixed point.

\section{Discussion and Conclusion}

\paragraph{Fixed point for Free fermions} -- While we can define fixed points for interacting fermionic systems with Fermi surfaces by identifying vanishing $\beta$ functions for the chemical potentials, it fails to identify the fixed point for free fermions with Fermi surface, where the $\beta$ function of the chemical potential remains non-zero as
\begin{eqnarray}
    \frac{d \tilde{\mu}^\pm}{d\ln E } &=& 
-\tilde{\mu}^\pm .
\label{eq:mu_free}
\end{eqnarray}
This implies that $\tilde{\mu}^\pm$ diverge under RG. However, this does not suggest that a system of free fermions with a Fermi surface fails to correspond to a fixed point. Rather, it indicates that this fixed point resides infinitely far from the origin, coinciding with the fixed point for free fermions lacking a Fermi surface. 
The fixed points discovered in this work, which occur at finite $\tilde{\mu}^\pm$, are located closer to the origin.
Below we explain this in details.

An analogy can be drawn using free bosonic theory
\begin{eqnarray}
    S_0= \int d^dx \left[\kappa (\partial \phi_x)^2 +m^2 \phi^2_x \right].
\end{eqnarray}
Starting with the Gaussian fixed point at $m^2=0$, and assigning the kinetic term to be marginal, i.e. $[\kappa]=0$, the scaling dimensions are $[x]=-1$ and $[\phi]=\frac{d-2}{2}$. Upon introducing a mass term, which is relevant due to $[m^2]=2$, the mass coupling diverges under RG flow. This does not imply that the massive theory lacks a corresponding fixed point. On the contrary, it corresponds to a well-defined fixed point, where the system's ground state is a trivial product state. At this fixed point, we observe a distinct set of scaling dimensions: $[m^2=0]$ and $[\phi]=d/2$. The mass term now has zero scaling dimension and does not flow, while the kinetic term, with $[\kappa]=-2$, becomes irrelevant. 

\begin{figure}[h]
    \centering
    \includegraphics[width=0.5\linewidth]{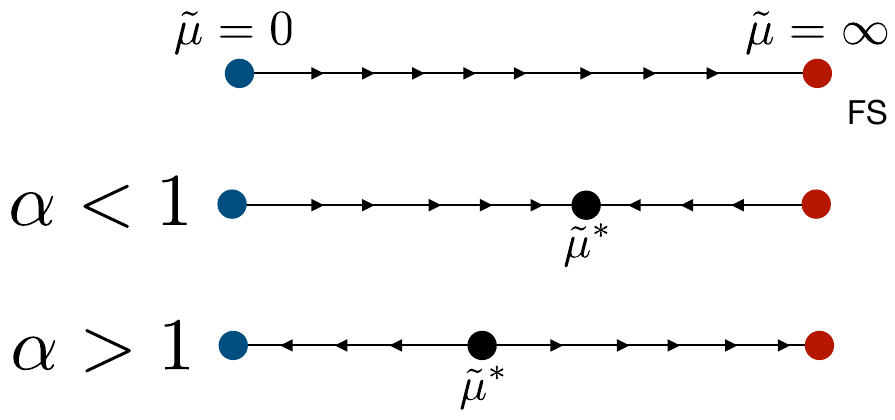}
    \caption{Fixed point structures along the $\tilde{\mu}$ axis. The fixed point for free fermion with a Fermi surface is located at infinity away from origin.}
    \label{fig:fp_free}
\end{figure}

The fixed point for free fermions with a Fermi surface can be interpreted analogously. If we express the action as
\begin{eqnarray}
S_0 &=& k_F\int_{\rm shell} d\omega dk d\theta ~\sum_\sigma \psi^\dag_{\omega,\theta,\vec{q},\sigma} \left[ \kappa(i \omega -ck^{2\alpha} )+  \mu \right]\psi_{\omega,\theta,\vec{q},\sigma} ,
\end{eqnarray}
we define the origin as the fixed point for free fermions without a Fermi surface at $\mu=0$. The integration over $k$ is restricted to the momentum shell around the Fermi surface. The RG flow of $\tilde{\mu}$ due to $[ \mu]=1$ leads to infinity where the fixed point with a Fermi surface is located, as shown in Fig.~\ref{fig:fp_free}. At origin, the scaling dimensions are $=[\omega]=[k]=1$, $[c]=1-2\alpha$, and $[\kappa]=0$. For this fixed point at $\infty$, the system acquires a distinct set of scaling dimensions: the chemical potential becomes marginal, $[ \mu]=0$ and $[\psi]=-1$, rendering the kinetic term irrelevant, i.e. $[\kappa]=-1$. This is consistent with the notion that IR physics is solely determined by fermions residing on the Fermi surface.

Alternatively, we can define the origin as the fixed point for free fermions with a Fermi surface. The action is thus
\begin{eqnarray}
S_0 &=& k_F\int d\omega dq d\theta ~\sum_\sigma \psi^\dag_{\omega,\theta,\vec{q},\sigma} (i \omega -v^\sigma_F {q} \cos \theta + \delta \mu)\psi_{\omega,\theta,\vec{q},\sigma} .
\end{eqnarray}
The changes in the chemical potentials, $\delta\tilde{\mu}^\pm$, flow according to $\frac{d \delta\tilde{\mu}^\pm}{d\ln E } = 
-\delta\tilde{\mu}^\pm $.
Now there is a fixed point of Fermi sea at $\delta\tilde{\mu}^\pm=0$, i.e. the origin. With interactions present, the flow of $\tilde{\mu}$ and $\delta\tilde{\mu}^\pm$ cannot be treated independently, resulting in the flow equations previously mentioned in Sec.~\ref{sec:RG_flow}.
In comparison to the fixed point of free fermions with a Fermi surface, our newly identified fixed point in an interacting system occurs at finite $\tilde{\mu}^\ast$, suggesting a modified scaling of the chemical potential. Whether it leads to non-Fermi liquid behavior is not clear yet. A detailed correspondence to non-Fermi liquid characteristics will be explored in future work.

\paragraph{No other relevant terms} --
In a system lacking a Fermi surface, the dispersion $k^{2\alpha}$ with $\alpha>1$ is less relevant than $k^2$. Consequently, if we initiate the action with $k^{2\alpha}$, the $k^2$ term emerges at lower energy scales. The low-energy properties are then dictated by the emergent $k^2$ dispersion, rather than the initial $k^{2\alpha}$ dispersion.

However, in systems possessing a Fermi surface, the low-energy dispersion is governed by the linearized form near the Fermi surface, irrespective of the dispersion at $k=0$. Therefore, no further relevant operators emerge as the energy scale is lowered as long as the Fermi surface exists.
This also explains that, in the presence of a Fermi surface, the transport property is independent of $\alpha$; however, $\alpha$ becomes relevant upon Fermi surface depletion.

\paragraph{Conclusion and outlook} --
In summary, we investigated a two-flavor interacting fermionic system with Fermi surfaces. The U(2) symmetric system exhibits both Pomeranchuk and Stoner instabilities, leading to spontaneous spatial rotational and flavor symmetry breaking, respectively. The Stoner instability features a RG fixed point whose stability switches from attractive to repulsive as $\alpha$ increases across 1, indicating a change from a continuous to a discontinuous transition. At the transition, regardless of the value of $\alpha$, $\mu^+=\mu^-$ maintains the U(2) symmetry. The U(1) $\times$
U(1) symmetric system, with $\mu^+\neq \mu^-$, displays more complex physics. This system possesses two Pomeranchuk instabilities. At one of them, a non-trivial RG fixed point demonstrates analogous behavior as $\alpha$ is varied. The key distinction here is that $\mu^+/\mu^-$ is finite at the fixed point for $\alpha<1$ but flows to infinity or zero at the transition when $\alpha>1$ depending on the quartic interaction, without fine-tuning. The latter situation indicates Fermi surface depletion at the transition point.

Besides, we studied the collective modes. Both the U(2) and the U(1) $\times$
U(1) system has propagating zero sound modes, hidden modes, decaying modes and unstable growing modes. All of these collective modes are potentially detectable through experiments\cite{klein2020hidden}. Notably, at the instability corresponding to an RG fixed, our theory predicts that the Fermi surface oscillation possesses a universal amplitude ratio of $(z^\ast)^{1/\alpha}$, where $z^\ast$ is the ratio $\mu^+/\mu^-$ at the critical point solely depending on quartic interaction strength. This ratio is potentially testable through experiments in the $\alpha<1$ system with $\mu^+\neq \mu^-$.

\section*{Acknowledgements}
We thank Andrey Chubukov, Liujun Zou for insightful discussions.
Research at the Perimeter Institute is supported in part by the
Government of Canada through Industry Canada, and by the Province of
Ontario through the Ministry of Research and Information.

\appendix
\setcounter{tocdepth}{1}
\addtocontents{toc}{\string\setcounter{tocdepth}{2}}
\addtocontents{toc}{\protect\setcounter{tocdepth}{0}}

\section{Two-Flavor Fermi Liquid Theory  \label{app:FL_2_flavor}}

In Fermi liquid theory, the Wigner distribution operator describes the distribution of quasiparticles in phase space. It is defined as
\begin{eqnarray}
W^\pm(\vec{r}_1,\vec{r}_2,t) =\int \frac{d^d \vec{p}_1 d^d \vec{p}_2}{(2\pi)^{2d}}~e^{i(\vec{p}_1 \cdot \vec{r}_1-\vec{p}_2 \cdot \vec{r}_2)} ~ \psi^{\dag}_{\vec{p}_1,\pm}(t) \psi_{\vec{p}_2,\pm}(t) ,
\end{eqnarray} 
where $t$ represents real time. Accordingly, the Wigner density operator is given by
\begin{eqnarray}
    n_{\vec{p}}^\pm(\vec{r},t) &=& \int d^2 \vec{r}' ~e^{-i\vec{p}\cdot \vec{r}'} W^\pm(\vec{r}+\frac{\vec{r}'}{2},\vec{r}-\frac{\vec{r}'}{2},t) \nonumber\\
    &=& \int d^2 \vec{r}' ~e^{-i\vec{p}\cdot \vec{r}'} \left[\int \frac{d^d \vec{p}_1 d^d \vec{p}_2}{(2\pi)^{2d}}~ e^{\frac{i}{2}(\vec{p}_1+\vec{p}_2) \cdot \vec{r}' } e^{i(\vec{p}_1 -\vec{p}_2 )\cdot \vec{r}}  \psi^{\dag}_{\vec{p}_1,\pm}(t) \psi_{\vec{p}_2,\pm}(t)  \right]\nonumber\\
    &=& \int \frac{d^d \vec{q}}{(2\pi)^d}~ e^{ i\vec{q}\cdot  \vec{r}} n_{\vec{p}}^\pm(\vec{q},t) =\int \frac{d^d \vec{q}}{(2\pi)^d}~ e^{ i\vec{q}\cdot  \vec{r}}\psi^{\dag}_{\vec{p}+\frac{\vec{q}}{2},\pm}(t) \psi_{\vec{p}-\frac{\vec{q}}{2},\pm}(t) ,
\end{eqnarray}
with $\vec{p}=\vec{p}_1+\vec{p}_2$ and $\vec{p}_1-\vec{p}_2=\vec{q}$. Transforming to the momentum and real frequency domain yields
\begin{eqnarray}
n_{\vec{p}}^\pm (\vec{q},\omega_R) = \int d^2 r dt ~n_{\vec{p}}^\pm (\vec{r},t) ~e^{-i \vec{q}\cdot \vec{r}- i\omega_R t} = \int \frac{d\Omega_R}{2\pi} \psi^{\dag}_{\vec{p}+\frac{\vec{q}}{2},\pm}(\Omega_R+\omega_R) \psi_{\vec{p}-\frac{\vec{q}}{2},\pm}(\Omega_R) .
\end{eqnarray}
Then, the total energy function is
\begin{eqnarray}
\mathcal{H}^\pm_{\vec{k}}(\omega_R) &=& n^\pm_{\vec{k}}(0,\omega) \varepsilon_{\vec{k}}^\pm \delta(\omega_R)  \\
&+&\int \frac{d^2q}{(2\pi)^2} \frac{ d^{2} \vec{p}  }{(2\pi)^2}~ n^\pm_{\vec{k}}(\vec{q},\omega_R)\left[(\lambda_\pm)_{\vec{k},\vec{p}}(\vec{q})  n^\pm_{\vec{p}}(-\vec{q},\omega_R) 
  + (\lambda_1)_{\vec{k},\vec{p}}(\vec{q})  n^\mp_{\vec{p}}(-\vec{q},\omega_R) \right], \nonumber
  \label{eq:H_k}
\end{eqnarray}
which provides the energy density function presented in Eq.~\eqref{eq:energy_per_particle}.

\subsection{Equilibrium properties \label{app:thermodynamics}}

Using the expression for the energy per particle, $\mathcal{E}_{\vec{k}}^\pm (0,0) \equiv E_{\vec{k}}^\pm$, given in Eq.~\eqref{eq:energy_per_particle}, one can compute 
the density of state at the Fermi surface as 
\begin{eqnarray}
n_F^\pm &=&\frac{1}{V} \sum_{\vec{k}}\delta\left( E_{\vec{k}}^\pm \right)  =\int \frac{d^2 \vec{k}}{(2\pi)^2} \delta \left(\varepsilon^\pm_{\vec{k}} + \lambda_\pm \mathcal{N}^\pm +\lambda_1 \mathcal{N}^\mp \right) \nonumber\\
&=& \int \frac{kdk}{2\pi} \delta \left(c k^{2\alpha}-(\mu')^\pm \right) =  \frac{1}{4\pi \alpha c^{1/\alpha}} \lbrack(\mu')^\pm\rbrack^{1/\alpha-1},
\end{eqnarray}
where the total particle number for each flavor satisfies equation
\begin{eqnarray}
\mathcal{N}^\pm = \int \frac{d^2 \vec{p}}{(2\pi)^2} n_p^\pm (0,0)=\frac{1}{4\pi c^{1/\alpha}} \lbrack(\mu')^\pm\rbrack^{1/\alpha},   
\end{eqnarray}
consistent with the Luttinger theorem.
Provided the chemical potential corrected as $(\mu')^\pm=\mu^\pm -(\lambda_\pm \mathcal{N}^\pm +\lambda_1 \mathcal{N}^\mp)$, we must solve this equation to get $\mathcal{N}^\pm$.
With the vector $\vec{\mathcal{N}} = (\mathcal{N}^+,\mathcal{N}^-)$, the flavor polarization is 
\begin{eqnarray}
\vec{S}= \sum_{\vec{k}}\vec{\sigma}\cdot \vec{\mathcal{N}}= \left( (1-i)\mathcal{N}^-+\mathcal{N}^+ , (1+i)\mathcal{N}^+-\mathcal{N}^- \right)^T.
\end{eqnarray}

\subsection{Linearized Kinetic equation}
\label{app:Kinetic_eqn}

In real time formalism, the action is given by $S=\int dt  \frac{d^2\vec{k}}{(2\pi)^2} \left[\mathcal{H}_{\vec{k}}^+(t) + \mathcal{H}^-_{\vec{k}}(t) \right]$.
The saddle-point equation is 
\begin{eqnarray}
    -i d_t n^\pm_{\vec{k}}(\vec{q},t) &=& \left[\varepsilon^\pm_{\vec{k}+\frac{\vec{q}}{2}}     -\varepsilon^\pm_{\vec{k}-\frac{\vec{q}}{2}}   \right]n^\pm_{ \vec{k} } (\vec{q},t) \\
   &+&2\lambda_\pm\int ~\frac{ d^{2} \vec{p}' d^2\vec{q}' }{(2\pi)^4}~  \Big(  n^\pm_{ \vec{k} -\frac{\vec{q}'}{2}} (\vec{q}-\vec{q}',t) - n^\pm_{ \vec{k} +\frac{\vec{q}' }{2}} (\vec{q}-\vec{q}',t)  \Big)n^\pm_{\vec{p}'}(\vec{q}',t)   \nonumber\\
   &+&2\lambda_1 \int ~\frac{ d^{2} \vec{p}' d^2\vec{q}' }{(2\pi)^4}~  \Big(  n^\pm_{ \vec{k} -\frac{\vec{q}'}{2}} (\vec{q}-\vec{q}',t)  - n^\pm_{ \vec{k} +\frac{\vec{q}' }{2}} (\vec{q}-\vec{q}',t)  \Big) n^\mp_{\vec{p}'}(\vec{q}',t) \nonumber\\
   &\approx& \vec{q} \cdot \nabla_{\vec{k}}  \varepsilon^\pm_{\vec{k}}  n^\pm_{ \vec{k} } (\vec{q},t)-2\int ~\frac{ d^{2} \vec{p}' d^2\vec{q}' }{(2\pi)^4}~   \vec{q}'\cdot \nabla_{\vec{k}} n^\pm_{\vec{k}}(\vec{q}-\vec{q}',t) \left[\lambda_\pm n^\pm_{\vec{p}'}(\vec{q}',t)+\lambda_1 n^\mp_{\vec{p}'}(\vec{q}',t)\right], \nonumber
   \label{eq:EOM_q}
\end{eqnarray}
assuming $\vec{q}$ and $\vec{q}'$ are small and the quartic interactions $\lambda_\pm$ and $\lambda_1$ are constants. The Wigner density can be expanded as 
\begin{eqnarray}
    n^\pm_{ \vec{k} -\frac{\vec{q}'}{2}} (\vec{q}-\vec{q}',t) &\approx& n^\pm_{\vec{k}}(\vec{q}-\vec{q}',t) -\frac{\vec{q}'}{2}\cdot \nabla_{\vec{k}} n^\pm_{\vec{k}}(\vec{q}-\vec{q}',t).
\end{eqnarray}
The equilibrium distribution is $n^0_{\vec{k}}= \Theta(-\varepsilon_{\vec{k}})$. Away from equilibrium, the distribution is defined as $\delta n_{\vec{k}}^\pm (\vec{q},t)$. Expressed in terms of this, in the real frequency domain, we obtain
\begin{eqnarray}
   \left[-\omega_R  +      \vec{q}  \cdot \nabla_{\vec{k}}\varepsilon^\pm_{\vec{k}} \right] \delta n^\pm_{\vec{k}}(\vec{q},\omega_R)
    - 4 \vec{q} \cdot \nabla_{\vec{k}}  (n^0_{\vec{k}})^\pm   \int  \frac{d^2\vec{p} }{(2\pi)^2}   \left[\lambda_\pm    \delta n^\pm_{\vec{p}}(\vec{q},\omega_R)+\lambda_1    \delta n^\mp_{\vec{p}}(\vec{q},\omega_R) \right] =0.
    \label{eq:EOM_q_omega}
\end{eqnarray}

\section{Field-Theoretical Functional RG \label{app:RG}}

In this section, we review the field-theoretical functional RG formalism\cite{ma2024fermi}.
Starting with the action in Eq.~\eqref{eq:action}, 
the counterterms are defined as
\begin{eqnarray}
S_{0,CT}/k_F &=& \int d\omega d\theta dq ~\sum_\sigma \psi^\dag_{\omega,\theta,\vec{q},\sigma} [i \delta_1\omega -(\delta_2 v^\sigma_F q \cos \theta-\delta_3^\sigma \delta\mu^\sigma)]\psi_{\omega,\theta,\vec{q},\sigma},  \nonumber\\
S_{i,CT}/k_F &=& \int d\omega d\Omega d\Omega' d\theta_1 d {\kappa}_1 ~d\theta_2 d {\kappa}_2 ~d^2 \vec{q} ~\sum_{\sigma_{1,2},\sigma'_{1,2}}[A_i F_{i}]_{\theta_1,\theta_2}(\vec{q})~\psi^{\dag}_{\mathcal{K}_1,\sigma_1}\psi^{\dag}_{\mathcal{K}_2,\sigma_2}\psi_{\mathcal{K}'_2 ,\sigma'_2}\psi_{\mathcal{K}'_1,\sigma'_1}.
\end{eqnarray}
Summing the counterterms and the renormalized action yields the bare action,
\begin{eqnarray}
S_{0,B}/k_F &=&\int d\omega_B d\theta dq ~\sum_\sigma \psi^\dag_{B;\omega_B,\theta,\vec{q},\sigma} [i \omega_B -(v_{F}^\sigma q \cos \theta-\delta\mu_B^\sigma)]\psi_{B;\omega_B,\theta,\vec{q},\sigma} , \nonumber\\
S_{i,B}/k_F &=& \int d\omega_B d\Omega_B d\Omega'_B d\theta_1 d {\kappa}_1 ~d\theta_2 d {\kappa}_2 ~d^2 \vec{q} ~\sum_{\sigma_{1,2},\sigma'_{1,2}}[F_{i;B}]_{\theta_1,\theta_2} (\vec{q} )~\psi^{\dag}_{B;\mathcal{K}_1,\sigma_1}\psi^{\dag}_{B;\mathcal{K}_2,\sigma_2}\psi_{B;\mathcal{K}'_2 ,\sigma'_2}\psi_{B;\mathcal{K}'_1,\sigma'_1}, 
\end{eqnarray}
where $i=+,-,1,2$. 
The renormalized and bare variables are related by $X_B = Z_X X$ with $X = \omega,  \mu^\sigma,\psi,\psi^\dag, \lambda_i$ with $Z_\psi=Z_{\psi^\dag}^\ast$. Defining $Z_j=1+\delta_i$ for $j=1,2$, $Z_3^\sigma=1+\delta_3^\sigma$, and $Z_{A_i}=1+A_i$, we obtain from $S_0$: $Z_\omega^2 |Z_\psi|^2 =Z_1$ and $Z_\omega |Z_\psi|^2  Z_{\mu^\sigma}=Z_3^\sigma$. Additionally, from $Z_\omega |Z_\psi|^2 =Z_2$, we derive $Z_\omega =\frac{Z_1}{Z_2}$, $|Z_\psi|^2 = \frac{Z_2^2}{Z_1}$, and $Z_{\mu^\sigma} = \frac{Z_3^\sigma}{Z_2}$. From $S_1$, we obtain $Z_\omega^3 Z_\lambda|Z_\psi|^4=Z_{A_i}$, which gives $Z_\lambda = \frac{ Z_{A_i} }{Z_1 Z_2}$. 

The renormalization conditions at energy scale $E$ are
\begin{eqnarray}
\frac{\partial}{\partial \kappa} \Gamma^{(2)}((k_F+\kappa)\hat{\kappa},\omega=E) \Big|_{\kappa=0} &=& v_F^\pm= 2\alpha  c (\mu^\pm)^{1-1/(2\alpha)} = 2\alpha  \tilde{c} (\tilde{\mu}^\pm)^{1-1/(2\alpha)},  \nonumber\\
-i\frac{\partial}{\partial \omega} \Gamma^{(2)}(k_F\hat{\kappa},\omega=E)  &=& 1 ,\nonumber\\
\left[\Gamma^{(4)}_i\right]_{k_F\hat{\theta},k_F\hat{\theta}'}(\vec{q}, \omega=E )  &=& (\lambda_i)_{k_F\hat{\theta},k_F\hat{\theta}'}(\vec{q}) .
\end{eqnarray}

\subsection{Chemical Potential}

At one loop order, the quadratic counterterm is \begin{eqnarray}
-\delta_3^\pm \delta \mu^\pm = 2\left[ \lambda_{\pm} (0) \Sigma^+ + \lambda_{1}(0) \Sigma^-   \right],
\end{eqnarray}
where $\Sigma^\pm$ are self-energies of the two flavors.
Given that $\varepsilon_{\vec{p}}^\pm=c p^{2\alpha}-\mu^\pm$, $p=[\frac{1}{c} (\varepsilon_{\vec{p}}^\pm+\mu^\pm)]^{\frac{1}{2\alpha}}$ and $dp=\frac{1}{2c\alpha} [\frac{1}{c} (\varepsilon_{\vec{p}}^\pm+\mu^\pm)]^{\frac{1}{2\alpha}-1}d \varepsilon_{\vec{p}}^\pm$, the self-energies are
\begin{eqnarray}
\Sigma^\pm &=&\int \frac{d\omega}{2\pi} \int \frac{d^2 p}{(2\pi)^2}  \frac{e^{i\omega 0^+}}{i\omega-\varepsilon_{\vec{p}}^\pm}  =  \int \frac{d^2 p}{(2\pi)^2} ~n_{\vec{p}}^\pm (0,0)=\int \frac{d^2 p}{(2\pi)^2} \Theta
(-\varepsilon_{\vec{p}}^\pm) 
\nonumber\\
&=&\frac{1}{4\pi c\alpha} \int_{-\mu^\pm}^0  [\frac{1}{c} (\varepsilon^\pm+\mu^\pm)]^{\frac{1}{\alpha}-1}d \varepsilon^\pm =\frac{1}{4\pi c^{1/\alpha}} (\mu^\pm)^{1/\alpha}.
\end{eqnarray}
In terms of the dimensionless variables, 
\begin{eqnarray}
\delta_3^\pm \delta \mu^\pm=-\frac{\omega}{2\pi \tilde{c}^{1/\alpha} }  \left[\tilde{\lambda}_{\pm} (0) (\tilde{\mu}^\pm)^{1/\alpha} + \tilde{\lambda}_{1}(0)  (\tilde{\mu}^\mp)^{1/\alpha}\right].
\end{eqnarray} 
At one loop order, we have $Z_1=Z_2=1$ and then $\omega  \tilde{\mu}^\pm+\delta_3^\pm \delta \mu^\pm = \delta \mu_B^\pm$. 
As a result, we have the $\beta$ functions as
\begin{eqnarray}
\frac{d  \tilde{\mu}^\pm}{d\ln E} &=&-  \tilde{\mu}^\pm -  \frac{d [\delta_3^\pm \delta \tilde{\mu}^\pm  ]_{\omega =E}}{d \ln E} ,
\end{eqnarray}
which are presented in Eq.~\eqref{eq:beta_function_chemical}.

\subsection{Quartic Interaction}

The quartic counterterms are 
\begin{eqnarray}
[A_\pm \lambda_\pm]_{\vec{k}_1,\vec{k}_2}(\vec{q},\omega ) &=& -4 \int \frac{d^2k }{(2\pi)^2} \frac{d\eta}{2\pi}~    \frac{[\lambda_\pm]_{\vec{k}_1,\vec{k}}(\vec{q}) [\lambda_\pm]_{\vec{k},\vec{k}_2}(\vec{q})}{\left[i\eta-\varepsilon_{\vec{k}}^\pm\right]\left[i(\omega+\eta)-\varepsilon_{\vec{k}-\vec{q}}^\pm \right]} \nonumber\\
&-&4 \int \frac{d^2k }{(2\pi)^2} \frac{d\eta}{2\pi}~    \frac{[\lambda_1]_{\vec{k}_1,\vec{k}}(\vec{q}) [\lambda_1]_{\vec{k},\vec{k}_2}(\vec{q})}{\left[i\eta-\varepsilon_{\vec{k}}^\mp\right]\left[i(\omega+\eta)-\varepsilon_{\vec{k}-\vec{q}}^\mp\right]} ,
\nonumber\\
\left[A_1 \lambda_1\right]_{\vec{k}_1,\vec{k}_2}(\vec{q},\omega ) &=& -4 \int \frac{d^2k }{(2\pi)^2} \frac{d\eta}{2\pi}~    \frac{[\lambda_+]_{\vec{k}_1,\vec{k}}(\vec{q}) [\lambda_1]_{\vec{k},\vec{k}_2}(\vec{q})}{\left[i\eta-\varepsilon_{\vec{k}}^+\right]\left[i(\omega+\eta)-\varepsilon_{\vec{k}-\vec{q}}^+ \right]} \nonumber\\
&-&4 \int \frac{d^2k }{(2\pi)^2} \frac{d\eta}{2\pi}~    \frac{[\lambda_-]_{\vec{k}_1,\vec{k}}(\vec{q}) [\lambda_1]_{\vec{k},\vec{k}_2}(\vec{q})}{\left[i\eta-\varepsilon_{\vec{k}}^-\right]\left[i(\omega+\eta)-\varepsilon_{\vec{k}-\vec{q}}^-\right]} ,
\nonumber\\
\left[A_2 \lambda_2\right]_{\vec{k}_1,\vec{k}_2}(\vec{q},\omega) &=& -4 \int \frac{d^2k }{(2\pi)^2} \frac{d\eta}{2\pi}~    \frac{[\lambda_2]_{\vec{k}_1,\vec{k}}(\vec{q}) [\lambda_2]_{\vec{k},\vec{k}_2}(\vec{q})}{\left[i\eta-\varepsilon_{\vec{k}}^+\right]\left[i(\omega+\eta)-\varepsilon_{\vec{k}-\vec{q}}^- \right]} .
\end{eqnarray}
Assume $\lambda_i$ is independent of $\vec{k}_{1,2}$ and depends only on $|\vec{q}|$, which is very small. We get
\begin{eqnarray}
A_\pm \lambda_\pm(\vec{q},\omega ) &=& \frac{1}{v_F^\pm}\mathcal{K}_{\pm}(\vec{q},\omega) [\lambda_\pm(\vec{q})]^2 + \frac{1}{v_F^\mp}\mathcal{K}_{\mp}(\vec{q},\omega) [\lambda_1(\vec{q})]^2, \nonumber\\
A_1 \lambda_1(\vec{q},\omega ) &=& \left[\frac{1}{v_F^+}\mathcal{K}_{+}(\vec{q},\omega) \lambda_+(\vec{q})+\frac{1}{v_F^-} \mathcal{K}_{-}(\vec{q},\omega)\lambda_-(\vec{q})\right] [\lambda_1(\vec{q})],
\nonumber\\
A_2 \lambda_2(\vec{q},\omega) &=& \left[\mathcal{K}_{+-}(\vec{q})+\mathcal{K}_{-+}(\vec{q}) \right] [\lambda_2(\vec{q})]^2 ,
\end{eqnarray}
where
\begin{eqnarray}
\mathcal{K}_\pm(\vec{q},\omega ) &=& 4 v_F^\pm \int \frac{d^2k }{(2\pi)^2} ~\frac{\textrm{sgn}(\varepsilon_{\vec{k}}^\pm)-\textrm{sgn}(\varepsilon_{\vec{k}-\vec{q}}^\pm)}{i\omega+\varepsilon_{\vec{k}}^\pm-\varepsilon_{\vec{k}-\vec{q}}^\pm} =\frac{2}{\pi} \left(\frac{\mu^\pm}{c } \right)^{\frac{1}{2\alpha}}\left[1-\frac{1}{\sqrt{\frac{(v_F^\pm)^2q^2}{\omega^2}+1}}\right]  ,
\nonumber\\
\mathcal{K}_{+-}(\vec{q},\omega)  &=& 4 \int \frac{d^2k }{(2\pi)^2} ~\frac{\textrm{sgn}(\varepsilon_{\vec{k}}^+)-\textrm{sgn}(\varepsilon_{\vec{k}-\vec{q}}^-)}{i\omega+\varepsilon_{\vec{k}}^+-\varepsilon_{\vec{k}-\vec{q}}^-} =\mathcal{K}_{-+}(-\vec{q},\omega) .
\end{eqnarray}
Defining the dimensionless Landau parameters as
\begin{eqnarray}
    F_\pm (\vec{q})&=&\frac{2}{\pi}  (\mu^\pm/c)^{1/(2\alpha)}\frac{1}{v_F^\pm}\lambda_\pm (\vec{q}) =\frac{2}{\pi} \frac{(\mu^{\pm})^{1/\alpha-1}}{2\alpha c^{1/\alpha}}\lambda_\pm (\vec{q}), \nonumber\\
    F_{1,2}(\vec{q})&=&\frac{2}{\pi}  (\mu^+\mu^-/c^2)^{1/(4\alpha)}\frac{1}{\sqrt{v_F^+v_F^-}}\lambda_{1,2}(\vec{q}),
\end{eqnarray}
and with $\mathcal{K}'_\pm (\vec{q},\omega)= \frac{\pi}{2} (\mu^\pm/c)^{-1/(2\alpha)}\mathcal{K}_\pm (\vec{q},\omega)$, the counterterms can be expressed in terms of these Landau parameters as
\begin{eqnarray}
A_+ F_+(\vec{q},\omega) &=& \mathcal{K}'_{+}(\vec{q},\omega) [F_+(\vec{q})]^2 + \mathcal{K}'_{-}(\vec{q},\omega) [F_1(\vec{q})]^2, \nonumber\\
A_- F_-(\vec{q},\omega) &=& \mathcal{K}'_{-}(\vec{q},\omega) [F_-(\vec{q})]^2 + \mathcal{K}'_{+}(\vec{q},\omega) [F_1(\vec{q})]^2, \nonumber\\
A_1 F_1(\vec{q},\omega) &=& \left[\mathcal{K}'_{+}(\vec{q},\omega) F_+(\vec{q}) +\mathcal{K}'_{-}(\vec{q},\omega)F_-(\vec{q})\right] F_1(\vec{q}).
\end{eqnarray}
At one loop order, $Z_1=Z_2=1$ and $F_i (\vec{q})+A_i F_i (\vec{q},\omega)=F_{i;B} (\vec{q})$. The $\beta$ functionals for these Landau parameters are
\begin{eqnarray}
\frac{dF_\pm(\vec{q})}{d\ln E} &=& -\frac{d A_\pm F_\pm (\vec{q},E)}{d \ln E}
\nonumber\\
&=& \left[- \frac{d}{d\ln E} \mathcal{K}'_\pm (\vec{q},E) \right] [F_\pm(\vec{q})]^2 + \left[- \frac{d}{d\ln E} \mathcal{K}'_\mp (\vec{q},E) \right] [F_1(\vec{q})]^2 ,\nonumber\\
\frac{dF_1(\vec{q})}{d\ln E} &=& -\frac{d A_1 F_1 (\vec{q},E)}{d \ln E}\nonumber\\
&=&\left\{ \left[- \frac{d}{d\ln E} \mathcal{K}'_+ (\vec{q},E) \right] F_+(\vec{q}) + \left[- \frac{d}{d\ln E} \mathcal{K}'_- (\vec{q},E) \right] F_-(\vec{q})\right\} F_1(\vec{q}),
\end{eqnarray}
which gives Eq.~\eqref{eq:beta_function_quartic}.

\section{Solving RG Equations \label{app:solving_quartic_RG}}

To solve the differential equations in Eq. \eqref{eq:fp_quartic}, we define $\tilde{U}(\tilde{q})=\tilde{R}_+(\tilde{q})\tilde{R}_-(\tilde{q})$. It satisfies
\begin{eqnarray}
\frac{d \tilde{U}(\tilde{q})}{ d\ln \tilde{q}} &=& \frac{d\tilde{R}_+(\tilde{q})}{d\ln \tilde{q}}\tilde{R}_-(\tilde{q}) + \tilde{R}_+(\tilde{q}) \frac{d\tilde{R}_-(\tilde{q})}{d\ln \tilde{q}} \nonumber\\
&=& - \left[\tilde{\kappa}_+(\tilde{q})\tilde{R}_+(\tilde{q})+ \tilde{\kappa}_-(\tilde{q})\tilde{R}_-(\tilde{q})  \right] \tilde{F}_1(\tilde{q}) \left[1-\tilde{U}(\tilde{q}) \right].
\end{eqnarray}
Consequently, we obtain the relation between $\tilde{U}(\tilde{q})$ and $\tilde{F}_1(\tilde{q})$ as
\begin{eqnarray}
\frac{d\tilde{U} }{d\tilde{F}_1 } =\frac{1-\tilde{U}  }{\tilde{F}_1 } \Rightarrow \tilde{F}_1(\tilde{q}) \left[1- \tilde{U}(\tilde{q})\right] = {F}_1(0) \left[1-  {U}(0)\right]= \textrm{Const}.
\end{eqnarray}
This equality holds for arbitrary $\tilde{q}$ and the constant is determined by the initial condition. Using this relation, we can solve the differential equations for $\tilde{R}_\pm(\vec{q})$,
\begin{eqnarray}
-\frac{d \tilde{R}_+(\tilde{q})}{d\ln \tilde{q}} &=& \tilde{\kappa}_-(\tilde{q})  {F}_1(0) \left[1- {U}(0)\right] ,\nonumber\\
-\frac{d\tilde{R}_-(\tilde{q})}{d\ln\tilde{q}} &=& \tilde{\kappa}_+(\tilde{q})  {F}_1(0) \left[1- {U}(0)\right] ,
\end{eqnarray}
and get
\begin{eqnarray}
\tilde{R}_+(\tilde{q}) &=&  \tilde{R}_+(E\tilde{q}/\Lambda) - {F}_1(0) \left[1-  {U}(0)\right] \mathcal{I}_E^\Lambda (v_F^-\tilde{q}),
\nonumber\\ 
\tilde{R}_-(\tilde{q}) &=&  \tilde{R}_-(E\tilde{q}/\Lambda) - {F}_1(0) \left[1-  {U}(0)\right]
\mathcal{I}_E^\Lambda (v_F^+\tilde{q}).
\end{eqnarray}
For finite $R_\pm(0)$ and $F_1(0)$, $\tilde{R}_\pm(\tilde{q})$ remains finite for all $\tilde{q}$. As $\tilde{q}$ increases, if $ {F}_1(0) [1- {U}(0)]>0$, $\tilde{R}_\pm(\tilde{q})$ monotonically decreases whereas if $ {F}_1(0) [1- {U}(0)]<0$, $\tilde{R}_\pm(\tilde{q})$ is monotonically increasing.

We can further solve for $\tilde{F}_1(\tilde{q})$ as
\begin{eqnarray}
[\tilde{F}_1(\tilde{q}) ]^{-1}- [ {F}_1(E\tilde{q}/\Lambda) ]^{-1} &=&\int_{E\tilde{q}/\Lambda}^{\tilde{q}}d\ln \tilde{q} \Big[\tilde{\kappa}_+ (\tilde{q}) \left( \tilde{R}_+(E\tilde{q}/\Lambda) - {F}_1(0) \left[1-  {U}(0)\right]\mathcal{I}_E^\Lambda (v_F^-\tilde{q})\right) \nonumber\\
&+&\tilde{\kappa}_-(\tilde{q}) \left( \tilde{R}_-(E\tilde{q}/\Lambda) - {F}_1(0) \left[1-  {U}(0)\right]\mathcal{I}_E^\Lambda (v_F^+\tilde{q})\right) \Big].
\end{eqnarray}
With
\begin{eqnarray}
\mathcal{I}^\Lambda_E (v^\pm_F \tilde{q}) &\equiv &\int_{E\tilde{q}/\Lambda}^{\tilde{q}} d\ln \tilde{q}~ \tilde{\kappa}_\pm (\tilde{q}) = \left[\frac{1}{\sqrt{1+\frac{(v_F^\pm)^2E^2\tilde{q}^2}{\Lambda^2}}}-\frac{1}{\sqrt{1+(v_F^\pm)^2\tilde{q}^2}}\right] ,  \\
\mathcal{J}^\Lambda_E (v^\pm_F \tilde{q}) &\equiv&\int_{E\tilde{q}/\Lambda}^{\tilde{q}} d\ln \tilde{q} \tilde{\kappa}_\pm (\tilde{q}) \mathcal{I}^\Lambda_E (v^\mp_F \tilde{q})
=\frac{(v_F^\pm)^2}{(v_F^\pm)^2-(v_F^\mp)^2} \left[\sqrt{\frac{1+(v_F^\mp)^2\tilde{q}^2}{1+(v_F^\pm)^2\tilde{q}^2}}-\sqrt{\frac{1+(v_F^\mp)^2\tilde{q}^2\frac{E^2}{\Lambda^2}}{1+(v_F^\pm)^2\tilde{q}^2 \frac{E^2}{\Lambda^2}}}\right] \nonumber\\
&+& \frac{(v_F^\pm)^2}{  (v_F^\pm)^2-(v^\mp_F)^2\frac{E^2}{\Lambda^2}} \left[\sqrt{1+(v_F^\mp)^2\tilde{q}^2\frac{E^2}{\Lambda^2}}\mathcal{I}_E^\Lambda(v_F^\pm \tilde{q})-\sqrt{1+(v_F^\mp)^2\tilde{q}^2\frac{E^4}{\Lambda^4}}\mathcal{I}_E^\Lambda(v_F^\pm \tilde{q}E/\Lambda)
\right],
\end{eqnarray}
we have
\begin{eqnarray}
[\tilde{F}_1(\tilde{q}) ]^{-1}  &=& [ {F}_1(E\tilde{q}/\Lambda) ]^{-1}+{R}_+(E\tilde{q}/\Lambda)\mathcal{I}_E^\Lambda (v_F^+ \tilde{q})+{R}_-(E\tilde{q}/\Lambda)\mathcal{I}_E^\Lambda (v_F^- \tilde{q}) \nonumber\\
&-& F_1(E\tilde{q}/\Lambda)\left[1-U(E\tilde{q}/\Lambda)\right] \left[\mathcal{J}_E^\Lambda (v_F^+ \tilde{q})+\mathcal{J}_E^\Lambda (v_F^- \tilde{q})\right].
\label{eq:solution_F1}
\end{eqnarray}
Provided $R_\pm(E\tilde{q}/\Lambda)$ independent of $\tilde{q}$,
in the limit of $E \rightarrow 0$ or $\Lambda \rightarrow \infty$, the expression simplifies to
\begin{eqnarray}
[\tilde{F}_1(\tilde{q}) ]^{-1}  &=& [ {F}_1(0) ]^{-1}+\left[ {R}_+(0) - {F}_1(0) [1-  {U}(0)]  \right]\mathcal{I}_0^\Lambda (v_F^+ \tilde{q}) \nonumber\\
&+& \left[  {R}_-(0) - {F}_1(0) [1-  {U}(0)]\right]\mathcal{I}_0^\Lambda(v_F^- \tilde{q}) \nonumber\\
&+&  {F}_1(0) [1-  {U}(0)] \frac{(v_F^+)^2}{(v_F^+)^2-(v_F^-)^2} \left[\frac{\mathcal{I}^\Lambda_0(1/v_F^+\tilde{q})-\mathcal{I}^\Lambda_0(1/v_F^-\tilde{q})}{1-\mathcal{I}^\Lambda_0(1/v_F^-\tilde{q})}
\right] \nonumber\\
&+&   {F}_1(0) [1- {U}(0) ]\frac{(v_F^-)^2}{(v_F^-)^2-(v_F^+)^2} \left[\frac{\mathcal{I}^\Lambda_0(1/v_F^-\tilde{q})-\mathcal{I}^\Lambda_0(1/v_F^+\tilde{q})}{1-\mathcal{I}^\Lambda_0(1/v_F^+\tilde{q})}
\right]  \\
&=& \frac{1}{ {F}_1(0) } \Big\{  \left(1 + F_+(0) \mathcal{I}^\Lambda_0 \left[\frac{1}{v_F^+\tilde{q}}\right] \right)\left(1+ F_-(0) \mathcal{I}^\Lambda_{0}\left[\frac{1}{v_F^-\tilde{q}}\right] \right)\nonumber\\
&-&[F_1(0)]^2 \mathcal{I}^\Lambda_0 \left[\frac{1}{v_F^+\tilde{q}}\right]\mathcal{I}^\Lambda_0 \left[\frac{1}{v_F^-\tilde{q}}\right] \Big\}.\nonumber
\end{eqnarray}
Note that the right hand side of this result is equivalent to Eq.~\eqref{eq:LKE_imbalanced} multiplied by $\frac{1}{ {F}_1(0) }$ after a Wick rotation $E \rightarrow i E$.
When the right hand side of Eq.~\eqref{eq:solution_F1} is zero,
$\tilde{F}_1(\tilde{q})$ diverges, indicating an instability. 

ver
\bibliography{ref.bib}

\begin{thebibliography}{64}%
\makeatletter
\providecommand \@ifxundefined [1]{%
 \@ifx{#1\undefined}
}%
\providecommand \@ifnum [1]{%
 \ifnum #1\expandafter \@firstoftwo
 \else \expandafter \@secondoftwo
 \fi
}%
\providecommand \@ifx [1]{%
 \ifx #1\expandafter \@firstoftwo
 \else \expandafter \@secondoftwo
 \fi
}%
\providecommand \natexlab [1]{#1}%
\providecommand \enquote  [1]{``#1''}%
\providecommand \bibnamefont  [1]{#1}%
\providecommand \bibfnamefont [1]{#1}%
\providecommand \citenamefont [1]{#1}%
\providecommand \href@noop [0]{\@secondoftwo}%
\providecommand \href [0]{\begingroup \@sanitize@url \@href}%
\providecommand \@href[1]{\@@startlink{#1}\@@href}%
\providecommand \@@href[1]{\endgroup#1\@@endlink}%
\providecommand \@sanitize@url [0]{\catcode `\\12\catcode `\$12\catcode
  `\&12\catcode `\#12\catcode `\^12\catcode `\_12\catcode `\%12\relax}%
\providecommand \@@startlink[1]{}%
\providecommand \@@endlink[0]{}%
\providecommand \url  [0]{\begingroup\@sanitize@url \@url }%
\providecommand \@url [1]{\endgroup\@href {#1}{\urlprefix }}%
\providecommand \urlprefix  [0]{URL }%
\providecommand \Eprint [0]{\href }%
\providecommand \doibase [0]{https://doi.org/}%
\providecommand \selectlanguage [0]{\@gobble}%
\providecommand \bibinfo  [0]{\@secondoftwo}%
\providecommand \bibfield  [0]{\@secondoftwo}%
\providecommand \translation [1]{[#1]}%
\providecommand \BibitemOpen [0]{}%
\providecommand \bibitemStop [0]{}%
\providecommand \bibitemNoStop [0]{.\EOS\space}%
\providecommand \EOS [0]{\spacefactor3000\relax}%
\providecommand \BibitemShut  [1]{\csname bibitem#1\endcsname}%
\let\auto@bib@innerbib\@empty
\bibitem [{\citenamefont {Cardy}(1996)}]{cardy1996scaling}%
  \BibitemOpen
  \bibfield  {author} {\bibinfo {author} {\bibfnamefont {J.}~\bibnamefont
  {Cardy}},\ }\href@noop {} {\emph {\bibinfo {title} {Scaling and
  renormalization in statistical physics}}},\ Vol.~\bibinfo {volume} {5}\
  (\bibinfo  {publisher} {Cambridge university press},\ \bibinfo {year}
  {1996})\BibitemShut {NoStop}%
\bibitem [{\citenamefont {Sachdev}(1999)}]{sachdev1999quantum}%
  \BibitemOpen
  \bibfield  {author} {\bibinfo {author} {\bibfnamefont {S.}~\bibnamefont
  {Sachdev}},\ }\bibfield  {title} {\bibinfo {title} {Quantum phase
  transitions},\ }\href@noop {} {\bibfield  {journal} {\bibinfo  {journal}
  {Physics world}\ }\textbf {\bibinfo {volume} {12}},\ \bibinfo {pages} {33}
  (\bibinfo {year} {1999})}\BibitemShut {NoStop}%
\bibitem [{\citenamefont {Fradkin}(2013)}]{fradkin2013field}%
  \BibitemOpen
  \bibfield  {author} {\bibinfo {author} {\bibfnamefont {E.}~\bibnamefont
  {Fradkin}},\ }\href@noop {} {\emph {\bibinfo {title} {Field theories of
  condensed matter physics}}}\ (\bibinfo  {publisher} {Cambridge University
  Press},\ \bibinfo {year} {2013})\BibitemShut {NoStop}%
\bibitem [{\citenamefont {Neto}\ and\ \citenamefont
  {Fradkin}(1994{\natexlab{a}})}]{neto1994bosonization}%
  \BibitemOpen
  \bibfield  {author} {\bibinfo {author} {\bibfnamefont {A.~C.}\ \bibnamefont
  {Neto}}\ and\ \bibinfo {author} {\bibfnamefont {E.}~\bibnamefont {Fradkin}},\
  }\bibfield  {title} {\bibinfo {title} {Bosonization of fermi liquids},\
  }\href@noop {} {\bibfield  {journal} {\bibinfo  {journal} {Physical Review
  B}\ }\textbf {\bibinfo {volume} {49}},\ \bibinfo {pages} {10877} (\bibinfo
  {year} {1994}{\natexlab{a}})}\BibitemShut {NoStop}%
\bibitem [{\citenamefont {Neto}\ and\ \citenamefont
  {Fradkin}(1994{\natexlab{b}})}]{neto1994bosonization1}%
  \BibitemOpen
  \bibfield  {author} {\bibinfo {author} {\bibfnamefont {A.~C.}\ \bibnamefont
  {Neto}}\ and\ \bibinfo {author} {\bibfnamefont {E.}~\bibnamefont {Fradkin}},\
  }\bibfield  {title} {\bibinfo {title} {Bosonization of the low energy
  excitations of fermi liquids},\ }\href@noop {} {\bibfield  {journal}
  {\bibinfo  {journal} {Physical review letters}\ }\textbf {\bibinfo {volume}
  {72}},\ \bibinfo {pages} {1393} (\bibinfo {year}
  {1994}{\natexlab{b}})}\BibitemShut {NoStop}%
\bibitem [{\citenamefont {Dupuis}\ and\ \citenamefont
  {Chitov}(1996)}]{dupuis1996renormalization}%
  \BibitemOpen
  \bibfield  {author} {\bibinfo {author} {\bibfnamefont {N.}~\bibnamefont
  {Dupuis}}\ and\ \bibinfo {author} {\bibfnamefont {G.}~\bibnamefont
  {Chitov}},\ }\bibfield  {title} {\bibinfo {title} {Renormalization-group
  approach to fermi-liquid theory},\ }\href@noop {} {\bibfield  {journal}
  {\bibinfo  {journal} {Physical Review B}\ }\textbf {\bibinfo {volume} {54}},\
  \bibinfo {pages} {3040} (\bibinfo {year} {1996})}\BibitemShut {NoStop}%
\bibitem [{\citenamefont {Dupuis}(1998)}]{dupuis1998fermi}%
  \BibitemOpen
  \bibfield  {author} {\bibinfo {author} {\bibfnamefont {N.}~\bibnamefont
  {Dupuis}},\ }\bibfield  {title} {\bibinfo {title} {Fermi liquid theory: a
  renormalization group approach},\ }\href@noop {} {\bibfield  {journal}
  {\bibinfo  {journal} {The European Physical Journal B-Condensed Matter and
  Complex Systems}\ }\textbf {\bibinfo {volume} {3}},\ \bibinfo {pages} {315}
  (\bibinfo {year} {1998})}\BibitemShut {NoStop}%
\bibitem [{\citenamefont {Dupuis}(2000)}]{dupuis2000unified}%
  \BibitemOpen
  \bibfield  {author} {\bibinfo {author} {\bibfnamefont {N.}~\bibnamefont
  {Dupuis}},\ }\bibfield  {title} {\bibinfo {title} {A unified description of
  static and dynamic properties of fermi liquids},\ }\href@noop {} {\bibfield
  {journal} {\bibinfo  {journal} {International Journal of Modern Physics B}\
  }\textbf {\bibinfo {volume} {14}},\ \bibinfo {pages} {379} (\bibinfo {year}
  {2000})}\BibitemShut {NoStop}%
\bibitem [{\citenamefont {{Haldane}}(2005)}]{Haldane2005luttinger}%
  \BibitemOpen
  \bibfield  {author} {\bibinfo {author} {\bibfnamefont {F.~D.~M.}\
  \bibnamefont {{Haldane}}},\ }\bibfield  {title} {\bibinfo {title}
  {{Luttinger's Theorem and Bosonization of the Fermi Surface}},\ }\href@noop
  {} {\bibfield  {journal} {\bibinfo  {journal} {arXiv e-prints}\ ,\ \bibinfo
  {eid} {cond-mat/0505529}} (\bibinfo {year} {2005})},\ \Eprint
  {https://arxiv.org/abs/cond-mat/0505529} {arXiv:cond-mat/0505529
  [cond-mat.str-el]} \BibitemShut {NoStop}%
\bibitem [{\citenamefont {{Delacretaz}}\ \emph {et~al.}(2022)\citenamefont
  {{Delacretaz}}, \citenamefont {{Du}}, \citenamefont {{Mehta}},\ and\
  \citenamefont {{Thanh Son}}}]{Delacretaz2022nonlinear}%
  \BibitemOpen
  \bibfield  {author} {\bibinfo {author} {\bibfnamefont {L.~V.}\ \bibnamefont
  {{Delacretaz}}}, \bibinfo {author} {\bibfnamefont {Y.-H.}\ \bibnamefont
  {{Du}}}, \bibinfo {author} {\bibfnamefont {U.}~\bibnamefont {{Mehta}}},\ and\
  \bibinfo {author} {\bibfnamefont {D.}~\bibnamefont {{Thanh Son}}},\
  }\bibfield  {title} {\bibinfo {title} {{Nonlinear Bosonization of Fermi
  Surfaces: The Method of Coadjoint Orbits}},\ }\href@noop {} {\bibfield
  {journal} {\bibinfo  {journal} {arXiv e-prints}\ ,\ \bibinfo {eid}
  {arXiv:2203.05004}} (\bibinfo {year} {2022})},\ \Eprint
  {https://arxiv.org/abs/2203.05004} {arXiv:2203.05004 [cond-mat.str-el]}
  \BibitemShut {NoStop}%
\bibitem [{\citenamefont
  {{Khveshchenko}}(2024{\natexlab{a}})}]{Khveshchenko2024pre}%
  \BibitemOpen
  \bibfield  {author} {\bibinfo {author} {\bibfnamefont {D.~V.}\ \bibnamefont
  {{Khveshchenko}}},\ }\bibfield  {title} {\bibinfo {title} {{(Pre-)Modern
  (Non-)Fermi liquids}},\ }\href {https://doi.org/10.48550/arXiv.2409.02316}
  {\bibfield  {journal} {\bibinfo  {journal} {arXiv e-prints}\ ,\ \bibinfo
  {eid} {arXiv:2409.02316}} (\bibinfo {year} {2024}{\natexlab{a}})},\ \Eprint
  {https://arxiv.org/abs/2409.02316} {arXiv:2409.02316 [cond-mat.str-el]}
  \BibitemShut {NoStop}%
\bibitem [{\citenamefont
  {{Khveshchenko}}(2024{\natexlab{b}})}]{Khveshchenko2024strange}%
  \BibitemOpen
  \bibfield  {author} {\bibinfo {author} {\bibfnamefont {D.~V.}\ \bibnamefont
  {{Khveshchenko}}},\ }\bibfield  {title} {\bibinfo {title} {{Strange
  sounds}},\ }\href {https://doi.org/10.1016/j.physleta.2024.130006} {\bibfield
   {journal} {\bibinfo  {journal} {Physics Letters A}\ }\textbf {\bibinfo
  {volume} {527}},\ \bibinfo {eid} {130006} (\bibinfo {year}
  {2024}{\natexlab{b}})},\ \Eprint {https://arxiv.org/abs/2404.01534}
  {arXiv:2404.01534 [cond-mat.str-el]} \BibitemShut {NoStop}%
\bibitem [{\citenamefont {Shankar}(1994)}]{SHANKAR}%
  \BibitemOpen
  \bibfield  {author} {\bibinfo {author} {\bibfnamefont {R.}~\bibnamefont
  {Shankar}},\ }\bibfield  {title} {\bibinfo {title} {Renormalization-group
  approach to interacting fermions},\ }\href
  {https://doi.org/10.1103/RevModPhys.66.129} {\bibfield  {journal} {\bibinfo
  {journal} {Rev. Mod. Phys.}\ }\textbf {\bibinfo {volume} {66}},\ \bibinfo
  {pages} {129} (\bibinfo {year} {1994})}\BibitemShut {NoStop}%
\bibitem [{\citenamefont {Polchinski}(1992)}]{POLCHINSKI1}%
  \BibitemOpen
  \bibfield  {author} {\bibinfo {author} {\bibfnamefont {J.}~\bibnamefont
  {Polchinski}},\ }\bibfield  {title} {\bibinfo {title} {{Effective Field
  Theory and the Fermi Surface}},\ }\href@noop {} {\bibfield  {journal}
  {\bibinfo  {journal} {ArXiv High Energy Physics - Theory e-prints}\ }
  (\bibinfo {year} {1992})},\ \Eprint {https://arxiv.org/abs/hep-th/9210046}
  {hep-th/9210046} \BibitemShut {NoStop}%
\bibitem [{\citenamefont {Borges}\ \emph {et~al.}(2023)\citenamefont {Borges},
  \citenamefont {Borissov}, \citenamefont {Singh}, \citenamefont {Schlief},\
  and\ \citenamefont {Lee}}]{borges2023field}%
  \BibitemOpen
  \bibfield  {author} {\bibinfo {author} {\bibfnamefont {F.}~\bibnamefont
  {Borges}}, \bibinfo {author} {\bibfnamefont {A.}~\bibnamefont {Borissov}},
  \bibinfo {author} {\bibfnamefont {A.}~\bibnamefont {Singh}}, \bibinfo
  {author} {\bibfnamefont {A.}~\bibnamefont {Schlief}},\ and\ \bibinfo {author}
  {\bibfnamefont {S.-S.}\ \bibnamefont {Lee}},\ }\bibfield  {title} {\bibinfo
  {title} {Field-theoretic functional renormalization group formalism for
  non-fermi liquids and its application to the antiferromagnetic quantum
  critical metal in two dimensions},\ }\href@noop {} {\bibfield  {journal}
  {\bibinfo  {journal} {Annals of Physics}\ }\textbf {\bibinfo {volume}
  {450}},\ \bibinfo {pages} {169221} (\bibinfo {year} {2023})}\BibitemShut
  {NoStop}%
\bibitem [{\citenamefont {Ma}\ and\ \citenamefont {Lee}(2024)}]{ma2024fermi}%
  \BibitemOpen
  \bibfield  {author} {\bibinfo {author} {\bibfnamefont {H.}~\bibnamefont
  {Ma}}\ and\ \bibinfo {author} {\bibfnamefont {S.-S.}\ \bibnamefont {Lee}},\
  }\bibfield  {title} {\bibinfo {title} {Fermi liquids beyond the
  forward-scattering limit: The role of nonforward scattering for scale
  invariance and instabilities},\ }\href@noop {} {\bibfield  {journal}
  {\bibinfo  {journal} {Physical Review B}\ }\textbf {\bibinfo {volume}
  {109}},\ \bibinfo {pages} {045143} (\bibinfo {year} {2024})}\BibitemShut
  {NoStop}%
\bibitem [{\citenamefont {Landau}(1959)}]{landau1959theory}%
  \BibitemOpen
  \bibfield  {author} {\bibinfo {author} {\bibfnamefont {L.}~\bibnamefont
  {Landau}},\ }\bibfield  {title} {\bibinfo {title} {On the theory of the fermi
  liquid},\ }\href@noop {} {\bibfield  {journal} {\bibinfo  {journal} {Sov.
  Phys. JETP}\ }\textbf {\bibinfo {volume} {8}},\ \bibinfo {pages} {70}
  (\bibinfo {year} {1959})}\BibitemShut {NoStop}%
\bibitem [{\citenamefont {Abrikosov}\ and\ \citenamefont
  {Khalatnikov}(1958)}]{abrikosov1958theory}%
  \BibitemOpen
  \bibfield  {author} {\bibinfo {author} {\bibfnamefont {A.~A.}\ \bibnamefont
  {Abrikosov}}\ and\ \bibinfo {author} {\bibfnamefont {I.~M.}\ \bibnamefont
  {Khalatnikov}},\ }\bibfield  {title} {\bibinfo {title} {Theory of fermi
  liquid. properties of liquid he $^3$ at low temperatures},\ }\href@noop {}
  {\bibfield  {journal} {\bibinfo  {journal} {Uspekhi Fiz. Nauk}\ }\textbf
  {\bibinfo {volume} {66}} (\bibinfo {year} {1958})}\BibitemShut {NoStop}%
\bibitem [{\citenamefont {Nozi{\`e}res}\ and\ \citenamefont
  {Luttinger}(1962)}]{nozieres1962derivation}%
  \BibitemOpen
  \bibfield  {author} {\bibinfo {author} {\bibfnamefont {P.}~\bibnamefont
  {Nozi{\`e}res}}\ and\ \bibinfo {author} {\bibfnamefont {J.~M.}\ \bibnamefont
  {Luttinger}},\ }\bibfield  {title} {\bibinfo {title} {Derivation of the
  landau theory of fermi liquids. i. formal preliminaries},\ }\href@noop {}
  {\bibfield  {journal} {\bibinfo  {journal} {Physical Review}\ }\textbf
  {\bibinfo {volume} {127}},\ \bibinfo {pages} {1423} (\bibinfo {year}
  {1962})}\BibitemShut {NoStop}%
\bibitem [{\citenamefont {Pines}\ and\ \citenamefont
  {Nozi{\`e}res}(2018)}]{pines2018microscopic}%
  \BibitemOpen
  \bibfield  {author} {\bibinfo {author} {\bibfnamefont {D.}~\bibnamefont
  {Pines}}\ and\ \bibinfo {author} {\bibfnamefont {P.}~\bibnamefont
  {Nozi{\`e}res}},\ }\bibfield  {title} {\bibinfo {title} {Microscopic theories
  of the electron liquid},\ }in\ \href@noop {} {\emph {\bibinfo {booktitle}
  {Theory of quantum liquids}}}\ (\bibinfo  {publisher} {CRC Press},\ \bibinfo
  {year} {2018})\ pp.\ \bibinfo {pages} {270--344}\BibitemShut {NoStop}%
\bibitem [{\citenamefont {Abrikosov}\ \emph {et~al.}(2012)\citenamefont
  {Abrikosov}, \citenamefont {Gorkov},\ and\ \citenamefont
  {Dzyaloshinski}}]{abrikosov2012methods}%
  \BibitemOpen
  \bibfield  {author} {\bibinfo {author} {\bibfnamefont {A.~A.}\ \bibnamefont
  {Abrikosov}}, \bibinfo {author} {\bibfnamefont {L.~P.}\ \bibnamefont
  {Gorkov}},\ and\ \bibinfo {author} {\bibfnamefont {I.~E.}\ \bibnamefont
  {Dzyaloshinski}},\ }\href@noop {} {\emph {\bibinfo {title} {Methods of
  quantum field theory in statistical physics}}}\ (\bibinfo  {publisher}
  {Courier Corporation},\ \bibinfo {year} {2012})\BibitemShut {NoStop}%
\bibitem [{\citenamefont {Chubukov}\ \emph {et~al.}(2006)\citenamefont
  {Chubukov}, \citenamefont {Maslov},\ and\ \citenamefont
  {Millis}}]{chubukov2023nonanalytic}%
  \BibitemOpen
  \bibfield  {author} {\bibinfo {author} {\bibfnamefont {A.~V.}\ \bibnamefont
  {Chubukov}}, \bibinfo {author} {\bibfnamefont {D.~L.}\ \bibnamefont
  {Maslov}},\ and\ \bibinfo {author} {\bibfnamefont {A.~J.}\ \bibnamefont
  {Millis}},\ }\bibfield  {title} {\bibinfo {title} {Nonanalytic corrections to
  the specific heat of a three-dimensional fermi liquid},\ }\href
  {https://doi.org/10.1103/PhysRevB.73.045128} {\bibfield  {journal} {\bibinfo
  {journal} {Phys. Rev. B}\ }\textbf {\bibinfo {volume} {73}},\ \bibinfo
  {pages} {045128} (\bibinfo {year} {2006})}\BibitemShut {NoStop}%
\bibitem [{\citenamefont {Chubukov}\ and\ \citenamefont
  {Maslov}(2004)}]{chubukov2023singular}%
  \BibitemOpen
  \bibfield  {author} {\bibinfo {author} {\bibfnamefont {A.~V.}\ \bibnamefont
  {Chubukov}}\ and\ \bibinfo {author} {\bibfnamefont {D.~L.}\ \bibnamefont
  {Maslov}},\ }\bibfield  {title} {\bibinfo {title} {Singular corrections to
  the fermi-liquid theory},\ }\href
  {https://doi.org/10.1103/PhysRevB.69.121102} {\bibfield  {journal} {\bibinfo
  {journal} {Phys. Rev. B}\ }\textbf {\bibinfo {volume} {69}},\ \bibinfo
  {pages} {121102} (\bibinfo {year} {2004})}\BibitemShut {NoStop}%
\bibitem [{\citenamefont {{Das Sarma}}\ and\ \citenamefont
  {Liao}(2021)}]{DASSARMA2021168495}%
  \BibitemOpen
  \bibfield  {author} {\bibinfo {author} {\bibfnamefont {S.}~\bibnamefont {{Das
  Sarma}}}\ and\ \bibinfo {author} {\bibfnamefont {Y.}~\bibnamefont {Liao}},\
  }\bibfield  {title} {\bibinfo {title} {Know the enemy: 2d fermi liquids},\
  }\href {https://doi.org/https://doi.org/10.1016/j.aop.2021.168495} {\bibfield
   {journal} {\bibinfo  {journal} {Annals of Physics}\ }\textbf {\bibinfo
  {volume} {435}},\ \bibinfo {pages} {168495} (\bibinfo {year} {2021})},\
  \bibinfo {note} {special issue on Philip W. Anderson}\BibitemShut {NoStop}%
\bibitem [{\citenamefont {Chubukov}\ \emph {et~al.}(2018)\citenamefont
  {Chubukov}, \citenamefont {Klein},\ and\ \citenamefont
  {Maslov}}]{chubukov2018fermi}%
  \BibitemOpen
  \bibfield  {author} {\bibinfo {author} {\bibfnamefont {A.~V.}\ \bibnamefont
  {Chubukov}}, \bibinfo {author} {\bibfnamefont {A.}~\bibnamefont {Klein}},\
  and\ \bibinfo {author} {\bibfnamefont {D.~L.}\ \bibnamefont {Maslov}},\
  }\bibfield  {title} {\bibinfo {title} {Fermi-liquid theory and pomeranchuk
  instabilities: fundamentals and new developments},\ }\href@noop {} {\bibfield
   {journal} {\bibinfo  {journal} {Journal of Experimental and Theoretical
  Physics}\ }\textbf {\bibinfo {volume} {127}},\ \bibinfo {pages} {826}
  (\bibinfo {year} {2018})}\BibitemShut {NoStop}%
\bibitem [{\citenamefont {Chubukov}\ \emph {et~al.}(2005)\citenamefont
  {Chubukov}, \citenamefont {Maslov}, \citenamefont {Gangadharaiah},\ and\
  \citenamefont {Glazman}}]{chubukov2005thermodynamics}%
  \BibitemOpen
  \bibfield  {author} {\bibinfo {author} {\bibfnamefont {A.~V.}\ \bibnamefont
  {Chubukov}}, \bibinfo {author} {\bibfnamefont {D.~L.}\ \bibnamefont
  {Maslov}}, \bibinfo {author} {\bibfnamefont {S.}~\bibnamefont
  {Gangadharaiah}},\ and\ \bibinfo {author} {\bibfnamefont {L.~I.}\
  \bibnamefont {Glazman}},\ }\bibfield  {title} {\bibinfo {title}
  {Thermodynamics of a fermi liquid beyond the low-energy limit},\ }\href@noop
  {} {\bibfield  {journal} {\bibinfo  {journal} {Physical review letters}\
  }\textbf {\bibinfo {volume} {95}},\ \bibinfo {pages} {026402} (\bibinfo
  {year} {2005})}\BibitemShut {NoStop}%
\bibitem [{\citenamefont {Delacr{\'e}taz}\ \emph {et~al.}(2025)\citenamefont
  {Delacr{\'e}taz}, \citenamefont {Chowdhury},\ and\ \citenamefont
  {Mehta}}]{delacretaz2025symmetry}%
  \BibitemOpen
  \bibfield  {author} {\bibinfo {author} {\bibfnamefont {L.~V.}\ \bibnamefont
  {Delacr{\'e}taz}}, \bibinfo {author} {\bibfnamefont {S.~D.}\ \bibnamefont
  {Chowdhury}},\ and\ \bibinfo {author} {\bibfnamefont {U.}~\bibnamefont
  {Mehta}},\ }\bibfield  {title} {\bibinfo {title} {Symmetry and causality
  constraints on fermi liquids},\ }\href@noop {} {\bibfield  {journal}
  {\bibinfo  {journal} {arXiv preprint arXiv:2501.02073}\ } (\bibinfo {year}
  {2025})}\BibitemShut {NoStop}%
\bibitem [{\citenamefont {Borges}\ and\ \citenamefont
  {Lee}(2023)}]{borges2023emergence}%
  \BibitemOpen
  \bibfield  {author} {\bibinfo {author} {\bibfnamefont {F.}~\bibnamefont
  {Borges}}\ and\ \bibinfo {author} {\bibfnamefont {S.-S.}\ \bibnamefont
  {Lee}},\ }\bibfield  {title} {\bibinfo {title} {Emergence of curved
  momentum-spacetime and its effect on cyclotron motion in the
  antiferromagnetic quantum critical metal},\ }\href@noop {} {\bibfield
  {journal} {\bibinfo  {journal} {Physical Review B}\ }\textbf {\bibinfo
  {volume} {108}},\ \bibinfo {pages} {245112} (\bibinfo {year}
  {2023})}\BibitemShut {NoStop}%
\bibitem [{\citenamefont {Kukreja}\ \emph {et~al.}(2024)\citenamefont
  {Kukreja}, \citenamefont {Besharat},\ and\ \citenamefont
  {Lee}}]{kukreja2024projective}%
  \BibitemOpen
  \bibfield  {author} {\bibinfo {author} {\bibfnamefont {S.}~\bibnamefont
  {Kukreja}}, \bibinfo {author} {\bibfnamefont {A.}~\bibnamefont {Besharat}},\
  and\ \bibinfo {author} {\bibfnamefont {S.-S.}\ \bibnamefont {Lee}},\
  }\bibfield  {title} {\bibinfo {title} {Projective fixed points for non-fermi
  liquids: A case study of the ising-nematic quantum critical metal},\
  }\href@noop {} {\bibfield  {journal} {\bibinfo  {journal} {Physical Review
  B}\ }\textbf {\bibinfo {volume} {110}},\ \bibinfo {pages} {155142} (\bibinfo
  {year} {2024})}\BibitemShut {NoStop}%
\bibitem [{\citenamefont {Raines}\ and\ \citenamefont
  {Chubukov}(2024)}]{raines2024two}%
  \BibitemOpen
  \bibfield  {author} {\bibinfo {author} {\bibfnamefont {Z.~M.}\ \bibnamefont
  {Raines}}\ and\ \bibinfo {author} {\bibfnamefont {A.~V.}\ \bibnamefont
  {Chubukov}},\ }\bibfield  {title} {\bibinfo {title} {Two-dimensional stoner
  transitions beyond mean field},\ }\href@noop {} {\bibfield  {journal}
  {\bibinfo  {journal} {Physical Review B}\ }\textbf {\bibinfo {volume}
  {110}},\ \bibinfo {pages} {235433} (\bibinfo {year} {2024})}\BibitemShut
  {NoStop}%
\bibitem [{\citenamefont {Raines}\ \emph
  {et~al.}(2024{\natexlab{a}})\citenamefont {Raines}, \citenamefont {Glazman},\
  and\ \citenamefont {Chubukov}}]{raines2024unconventional}%
  \BibitemOpen
  \bibfield  {author} {\bibinfo {author} {\bibfnamefont {Z.~M.}\ \bibnamefont
  {Raines}}, \bibinfo {author} {\bibfnamefont {L.~I.}\ \bibnamefont
  {Glazman}},\ and\ \bibinfo {author} {\bibfnamefont {A.~V.}\ \bibnamefont
  {Chubukov}},\ }\bibfield  {title} {\bibinfo {title} {Unconventional
  discontinuous transitions in a two-dimensional system with spin and valley
  degrees of freedom},\ }\href@noop {} {\bibfield  {journal} {\bibinfo
  {journal} {Physical Review B}\ }\textbf {\bibinfo {volume} {110}},\ \bibinfo
  {pages} {155402} (\bibinfo {year} {2024}{\natexlab{a}})}\BibitemShut
  {NoStop}%
\bibitem [{\citenamefont {Raines}\ \emph
  {et~al.}(2024{\natexlab{b}})\citenamefont {Raines}, \citenamefont {Glazman},\
  and\ \citenamefont {Chubukov}}]{raines2024unconventional1}%
  \BibitemOpen
  \bibfield  {author} {\bibinfo {author} {\bibfnamefont {Z.~M.}\ \bibnamefont
  {Raines}}, \bibinfo {author} {\bibfnamefont {L.~I.}\ \bibnamefont
  {Glazman}},\ and\ \bibinfo {author} {\bibfnamefont {A.~V.}\ \bibnamefont
  {Chubukov}},\ }\bibfield  {title} {\bibinfo {title} {Unconventional
  discontinuous transitions in isospin systems},\ }\href@noop {} {\bibfield
  {journal} {\bibinfo  {journal} {Physical review letters}\ }\textbf {\bibinfo
  {volume} {133}},\ \bibinfo {pages} {146501} (\bibinfo {year}
  {2024}{\natexlab{b}})}\BibitemShut {NoStop}%
\bibitem [{\citenamefont {Wang}\ \emph {et~al.}(2017)\citenamefont {Wang},
  \citenamefont {Nahum}, \citenamefont {Metlitski}, \citenamefont {Xu},\ and\
  \citenamefont {Senthil}}]{wang2022deconfined}%
  \BibitemOpen
  \bibfield  {author} {\bibinfo {author} {\bibfnamefont {C.}~\bibnamefont
  {Wang}}, \bibinfo {author} {\bibfnamefont {A.}~\bibnamefont {Nahum}},
  \bibinfo {author} {\bibfnamefont {M.~A.}\ \bibnamefont {Metlitski}}, \bibinfo
  {author} {\bibfnamefont {C.}~\bibnamefont {Xu}},\ and\ \bibinfo {author}
  {\bibfnamefont {T.}~\bibnamefont {Senthil}},\ }\bibfield  {title} {\bibinfo
  {title} {Deconfined quantum critical points: Symmetries and dualities},\
  }\href {https://doi.org/10.1103/PhysRevX.7.031051} {\bibfield  {journal}
  {\bibinfo  {journal} {Phys. Rev. X}\ }\textbf {\bibinfo {volume} {7}},\
  \bibinfo {pages} {031051} (\bibinfo {year} {2017})}\BibitemShut {NoStop}%
\bibitem [{\citenamefont {Kaplan}\ \emph {et~al.}(2009)\citenamefont {Kaplan},
  \citenamefont {Lee}, \citenamefont {Son},\ and\ \citenamefont
  {Stephanov}}]{kaplan2022conformality}%
  \BibitemOpen
  \bibfield  {author} {\bibinfo {author} {\bibfnamefont {D.~B.}\ \bibnamefont
  {Kaplan}}, \bibinfo {author} {\bibfnamefont {J.-W.}\ \bibnamefont {Lee}},
  \bibinfo {author} {\bibfnamefont {D.~T.}\ \bibnamefont {Son}},\ and\ \bibinfo
  {author} {\bibfnamefont {M.~A.}\ \bibnamefont {Stephanov}},\ }\bibfield
  {title} {\bibinfo {title} {Conformality lost},\ }\href
  {https://doi.org/10.1103/PhysRevD.80.125005} {\bibfield  {journal} {\bibinfo
  {journal} {Phys. Rev. D}\ }\textbf {\bibinfo {volume} {80}},\ \bibinfo
  {pages} {125005} (\bibinfo {year} {2009})}\BibitemShut {NoStop}%
\bibitem [{\citenamefont {Gorbenko}\ \emph
  {et~al.}(2018{\natexlab{a}})\citenamefont {Gorbenko}, \citenamefont
  {Rychkov},\ and\ \citenamefont {Zan}}]{Gorbenko2018a}%
  \BibitemOpen
  \bibfield  {author} {\bibinfo {author} {\bibfnamefont {V.}~\bibnamefont
  {Gorbenko}}, \bibinfo {author} {\bibfnamefont {S.}~\bibnamefont {Rychkov}},\
  and\ \bibinfo {author} {\bibfnamefont {B.}~\bibnamefont {Zan}},\ }\bibfield
  {title} {\bibinfo {title} {Walking, weak first-order transitions, and complex
  cfts},\ }\bibfield  {journal} {\bibinfo  {journal} {Journal of High Energy
  Physics}\ }\textbf {\bibinfo {volume} {2018}},\ \href
  {https://doi.org/10.1007/jhep10(2018)108} {10.1007/jhep10(2018)108} (\bibinfo
  {year} {2018}{\natexlab{a}})\BibitemShut {NoStop}%
\bibitem [{\citenamefont {Gorbenko}\ \emph
  {et~al.}(2018{\natexlab{b}})\citenamefont {Gorbenko}, \citenamefont
  {Rychkov},\ and\ \citenamefont {Zan}}]{Gorbenko2018b}%
  \BibitemOpen
  \bibfield  {author} {\bibinfo {author} {\bibfnamefont {V.}~\bibnamefont
  {Gorbenko}}, \bibinfo {author} {\bibfnamefont {S.}~\bibnamefont {Rychkov}},\
  and\ \bibinfo {author} {\bibfnamefont {B.}~\bibnamefont {Zan}},\ }\bibfield
  {title} {\bibinfo {title} {{Walking, Weak first-order transitions, and
  Complex CFTs II. Two-dimensional Potts model at $Q>4$}},\ }\href
  {https://doi.org/10.21468/SciPostPhys.5.5.050} {\bibfield  {journal}
  {\bibinfo  {journal} {SciPost Phys.}\ }\textbf {\bibinfo {volume} {5}},\
  \bibinfo {pages} {050} (\bibinfo {year} {2018}{\natexlab{b}})}\BibitemShut
  {NoStop}%
\bibitem [{\citenamefont {Ma}\ and\ \citenamefont {He}(2019)}]{Ma2019}%
  \BibitemOpen
  \bibfield  {author} {\bibinfo {author} {\bibfnamefont {H.}~\bibnamefont
  {Ma}}\ and\ \bibinfo {author} {\bibfnamefont {Y.-C.}\ \bibnamefont {He}},\
  }\bibfield  {title} {\bibinfo {title} {Shadow of complex fixed point:
  Approximate conformality of $q>4$ potts model},\ }\href
  {https://doi.org/10.1103/PhysRevB.99.195130} {\bibfield  {journal} {\bibinfo
  {journal} {Phys. Rev. B}\ }\textbf {\bibinfo {volume} {99}},\ \bibinfo
  {pages} {195130} (\bibinfo {year} {2019})}\BibitemShut {NoStop}%
\bibitem [{\citenamefont {Haldar}\ \emph {et~al.}(2023)\citenamefont {Haldar},
  \citenamefont {Tavakol}, \citenamefont {Ma},\ and\ \citenamefont
  {Scaffidi}}]{Haldar2023}%
  \BibitemOpen
  \bibfield  {author} {\bibinfo {author} {\bibfnamefont {A.}~\bibnamefont
  {Haldar}}, \bibinfo {author} {\bibfnamefont {O.}~\bibnamefont {Tavakol}},
  \bibinfo {author} {\bibfnamefont {H.}~\bibnamefont {Ma}},\ and\ \bibinfo
  {author} {\bibfnamefont {T.}~\bibnamefont {Scaffidi}},\ }\bibfield  {title}
  {\bibinfo {title} {Hidden critical points in the two-dimensional model: Exact
  numerical study of a complex conformal field theory},\ }\bibfield  {journal}
  {\bibinfo  {journal} {Physical Review Letters}\ }\textbf {\bibinfo {volume}
  {131}},\ \href {https://doi.org/10.1103/physrevlett.131.131601}
  {10.1103/physrevlett.131.131601} (\bibinfo {year} {2023})\BibitemShut
  {NoStop}%
\bibitem [{\citenamefont {Tang}\ \emph {et~al.}(2024)\citenamefont {Tang},
  \citenamefont {Ma}, \citenamefont {Tang}, \citenamefont {He},\ and\
  \citenamefont {Zhu}}]{tang2024reclaiming}%
  \BibitemOpen
  \bibfield  {author} {\bibinfo {author} {\bibfnamefont {Y.}~\bibnamefont
  {Tang}}, \bibinfo {author} {\bibfnamefont {H.}~\bibnamefont {Ma}}, \bibinfo
  {author} {\bibfnamefont {Q.}~\bibnamefont {Tang}}, \bibinfo {author}
  {\bibfnamefont {Y.-C.}\ \bibnamefont {He}},\ and\ \bibinfo {author}
  {\bibfnamefont {W.}~\bibnamefont {Zhu}},\ }\bibfield  {title} {\bibinfo
  {title} {Reclaiming the lost conformality in a non-hermitian quantum 5-state
  potts model},\ }\href@noop {} {\bibfield  {journal} {\bibinfo  {journal}
  {Physical Review Letters}\ }\textbf {\bibinfo {volume} {133}},\ \bibinfo
  {pages} {076504} (\bibinfo {year} {2024})}\BibitemShut {NoStop}%
\bibitem [{\citenamefont {Gukov}(2017)}]{gukov2017rg}%
  \BibitemOpen
  \bibfield  {author} {\bibinfo {author} {\bibfnamefont {S.}~\bibnamefont
  {Gukov}},\ }\bibfield  {title} {\bibinfo {title} {Rg flows and
  bifurcations},\ }\href@noop {} {\bibfield  {journal} {\bibinfo  {journal}
  {Nuclear Physics B}\ }\textbf {\bibinfo {volume} {919}},\ \bibinfo {pages}
  {583} (\bibinfo {year} {2017})}\BibitemShut {NoStop}%
\bibitem [{\citenamefont {Ma}(2023)}]{ma2023quenched}%
  \BibitemOpen
  \bibfield  {author} {\bibinfo {author} {\bibfnamefont {H.}~\bibnamefont
  {Ma}},\ }\bibfield  {title} {\bibinfo {title} {Quenched random mass disorder
  in the large n theory of vector bosons},\ }\href@noop {} {\bibfield
  {journal} {\bibinfo  {journal} {SciPost Physics}\ }\textbf {\bibinfo {volume}
  {14}},\ \bibinfo {pages} {039} (\bibinfo {year} {2023})}\BibitemShut
  {NoStop}%
\bibitem [{\citenamefont {Pomeranchuk}\ \emph {et~al.}(1958)\citenamefont
  {Pomeranchuk} \emph {et~al.}}]{pomeranchuk1958stability}%
  \BibitemOpen
  \bibfield  {author} {\bibinfo {author} {\bibfnamefont {I.~I.}\ \bibnamefont
  {Pomeranchuk}} \emph {et~al.},\ }\bibfield  {title} {\bibinfo {title} {On the
  stability of a fermi liquid},\ }\href@noop {} {\bibfield  {journal} {\bibinfo
   {journal} {Sov. Phys. JETP}\ }\textbf {\bibinfo {volume} {8}},\ \bibinfo
  {pages} {361} (\bibinfo {year} {1958})}\BibitemShut {NoStop}%
\bibitem [{\citenamefont {Oganesyan}\ \emph {et~al.}(2001)\citenamefont
  {Oganesyan}, \citenamefont {Kivelson},\ and\ \citenamefont
  {Fradkin}}]{oganesyan2001quantum}%
  \BibitemOpen
  \bibfield  {author} {\bibinfo {author} {\bibfnamefont {V.}~\bibnamefont
  {Oganesyan}}, \bibinfo {author} {\bibfnamefont {S.~A.}\ \bibnamefont
  {Kivelson}},\ and\ \bibinfo {author} {\bibfnamefont {E.}~\bibnamefont
  {Fradkin}},\ }\bibfield  {title} {\bibinfo {title} {Quantum theory of a
  nematic fermi fluid},\ }\href@noop {} {\bibfield  {journal} {\bibinfo
  {journal} {Physical Review B}\ }\textbf {\bibinfo {volume} {64}},\ \bibinfo
  {pages} {195109} (\bibinfo {year} {2001})}\BibitemShut {NoStop}%
\bibitem [{\citenamefont {Dell’Anna}\ and\ \citenamefont
  {Metzner}(2006)}]{dell2006fermi}%
  \BibitemOpen
  \bibfield  {author} {\bibinfo {author} {\bibfnamefont {L.}~\bibnamefont
  {Dell’Anna}}\ and\ \bibinfo {author} {\bibfnamefont {W.}~\bibnamefont
  {Metzner}},\ }\bibfield  {title} {\bibinfo {title} {Fermi surface
  fluctuations and single electron excitations near pomeranchuk instability in
  two dimensions},\ }\href@noop {} {\bibfield  {journal} {\bibinfo  {journal}
  {Physical Review B—Condensed Matter and Materials Physics}\ }\textbf
  {\bibinfo {volume} {73}},\ \bibinfo {pages} {045127} (\bibinfo {year}
  {2006})}\BibitemShut {NoStop}%
\bibitem [{\citenamefont {Wu}\ \emph {et~al.}(2007)\citenamefont {Wu},
  \citenamefont {Sun}, \citenamefont {Fradkin},\ and\ \citenamefont
  {Zhang}}]{wu2007fermi}%
  \BibitemOpen
  \bibfield  {author} {\bibinfo {author} {\bibfnamefont {C.}~\bibnamefont
  {Wu}}, \bibinfo {author} {\bibfnamefont {K.}~\bibnamefont {Sun}}, \bibinfo
  {author} {\bibfnamefont {E.}~\bibnamefont {Fradkin}},\ and\ \bibinfo {author}
  {\bibfnamefont {S.-C.}\ \bibnamefont {Zhang}},\ }\bibfield  {title} {\bibinfo
  {title} {Fermi liquid instabilities in the spin channel},\ }\href@noop {}
  {\bibfield  {journal} {\bibinfo  {journal} {Physical Review B—Condensed
  Matter and Materials Physics}\ }\textbf {\bibinfo {volume} {75}},\ \bibinfo
  {pages} {115103} (\bibinfo {year} {2007})}\BibitemShut {NoStop}%
\bibitem [{\citenamefont {Chubukov}\ and\ \citenamefont
  {Maslov}(2009)}]{chubukov2009spin}%
  \BibitemOpen
  \bibfield  {author} {\bibinfo {author} {\bibfnamefont {A.~V.}\ \bibnamefont
  {Chubukov}}\ and\ \bibinfo {author} {\bibfnamefont {D.~L.}\ \bibnamefont
  {Maslov}},\ }\bibfield  {title} {\bibinfo {title} {Spin conservation and
  fermi liquid near a ferromagnetic quantum critical point},\ }\href@noop {}
  {\bibfield  {journal} {\bibinfo  {journal} {Physical review letters}\
  }\textbf {\bibinfo {volume} {103}},\ \bibinfo {pages} {216401} (\bibinfo
  {year} {2009})}\BibitemShut {NoStop}%
\bibitem [{\citenamefont {Maslov}\ and\ \citenamefont
  {Chubukov}(2010)}]{maslov2010fermi}%
  \BibitemOpen
  \bibfield  {author} {\bibinfo {author} {\bibfnamefont {D.~L.}\ \bibnamefont
  {Maslov}}\ and\ \bibinfo {author} {\bibfnamefont {A.~V.}\ \bibnamefont
  {Chubukov}},\ }\bibfield  {title} {\bibinfo {title} {Fermi liquid near
  pomeranchuk quantum criticality},\ }\href@noop {} {\bibfield  {journal}
  {\bibinfo  {journal} {Physical Review B—Condensed Matter and Materials
  Physics}\ }\textbf {\bibinfo {volume} {81}},\ \bibinfo {pages} {045110}
  (\bibinfo {year} {2010})}\BibitemShut {NoStop}%
\bibitem [{\citenamefont {Mross}\ \emph {et~al.}(2010)\citenamefont {Mross},
  \citenamefont {McGreevy}, \citenamefont {Liu},\ and\ \citenamefont
  {Senthil}}]{mross2010controlled}%
  \BibitemOpen
  \bibfield  {author} {\bibinfo {author} {\bibfnamefont {D.~F.}\ \bibnamefont
  {Mross}}, \bibinfo {author} {\bibfnamefont {J.}~\bibnamefont {McGreevy}},
  \bibinfo {author} {\bibfnamefont {H.}~\bibnamefont {Liu}},\ and\ \bibinfo
  {author} {\bibfnamefont {T.}~\bibnamefont {Senthil}},\ }\bibfield  {title}
  {\bibinfo {title} {Controlled expansion for certain non-fermi-liquid
  metals},\ }\href@noop {} {\bibfield  {journal} {\bibinfo  {journal} {Physical
  Review B—Condensed Matter and Materials Physics}\ }\textbf {\bibinfo
  {volume} {82}},\ \bibinfo {pages} {045121} (\bibinfo {year}
  {2010})}\BibitemShut {NoStop}%
\bibitem [{\citenamefont {Hartnoll}\ \emph {et~al.}(2014)\citenamefont
  {Hartnoll}, \citenamefont {Mahajan}, \citenamefont {Punk},\ and\
  \citenamefont {Sachdev}}]{hartnoll2014transport}%
  \BibitemOpen
  \bibfield  {author} {\bibinfo {author} {\bibfnamefont {S.~A.}\ \bibnamefont
  {Hartnoll}}, \bibinfo {author} {\bibfnamefont {R.}~\bibnamefont {Mahajan}},
  \bibinfo {author} {\bibfnamefont {M.}~\bibnamefont {Punk}},\ and\ \bibinfo
  {author} {\bibfnamefont {S.}~\bibnamefont {Sachdev}},\ }\bibfield  {title}
  {\bibinfo {title} {Transport near the ising-nematic quantum critical point of
  metals in two dimensions},\ }\href@noop {} {\bibfield  {journal} {\bibinfo
  {journal} {Physical Review B}\ }\textbf {\bibinfo {volume} {89}},\ \bibinfo
  {pages} {155130} (\bibinfo {year} {2014})}\BibitemShut {NoStop}%
\bibitem [{\citenamefont {Metlitski}\ and\ \citenamefont
  {Sachdev}(2010)}]{metlitski2010quantum}%
  \BibitemOpen
  \bibfield  {author} {\bibinfo {author} {\bibfnamefont {M.~A.}\ \bibnamefont
  {Metlitski}}\ and\ \bibinfo {author} {\bibfnamefont {S.}~\bibnamefont
  {Sachdev}},\ }\bibfield  {title} {\bibinfo {title} {Quantum phase transitions
  of metals in two spatial dimensions. i. ising-nematic order},\ }\href@noop {}
  {\bibfield  {journal} {\bibinfo  {journal} {Physical Review B—Condensed
  Matter and Materials Physics}\ }\textbf {\bibinfo {volume} {82}},\ \bibinfo
  {pages} {075127} (\bibinfo {year} {2010})}\BibitemShut {NoStop}%
\bibitem [{\citenamefont {Wu}\ \emph {et~al.}(2018)\citenamefont {Wu},
  \citenamefont {Klein},\ and\ \citenamefont {Chubukov}}]{wu2018conditions}%
  \BibitemOpen
  \bibfield  {author} {\bibinfo {author} {\bibfnamefont {Y.-M.}\ \bibnamefont
  {Wu}}, \bibinfo {author} {\bibfnamefont {A.}~\bibnamefont {Klein}},\ and\
  \bibinfo {author} {\bibfnamefont {A.~V.}\ \bibnamefont {Chubukov}},\
  }\bibfield  {title} {\bibinfo {title} {Conditions for l= 1 pomeranchuk
  instability in a fermi liquid},\ }\href@noop {} {\bibfield  {journal}
  {\bibinfo  {journal} {Physical Review B}\ }\textbf {\bibinfo {volume} {97}},\
  \bibinfo {pages} {165101} (\bibinfo {year} {2018})}\BibitemShut {NoStop}%
\bibitem [{\citenamefont {Stoner}(1938)}]{stoner1938collective}%
  \BibitemOpen
  \bibfield  {author} {\bibinfo {author} {\bibfnamefont {E.~C.}\ \bibnamefont
  {Stoner}},\ }\bibfield  {title} {\bibinfo {title} {Collective electron
  ferromagnetism},\ }\href@noop {} {\bibfield  {journal} {\bibinfo  {journal}
  {Proceedings of the Royal Society of London. Series A. Mathematical and
  Physical Sciences}\ }\textbf {\bibinfo {volume} {165}},\ \bibinfo {pages}
  {372} (\bibinfo {year} {1938})}\BibitemShut {NoStop}%
\bibitem [{\citenamefont {Stoner}(1939)}]{stoner1939collective}%
  \BibitemOpen
  \bibfield  {author} {\bibinfo {author} {\bibfnamefont {E.~C.}\ \bibnamefont
  {Stoner}},\ }\bibfield  {title} {\bibinfo {title} {Collective electron
  ferromagnetism ii. energy and specific heat},\ }\href@noop {} {\bibfield
  {journal} {\bibinfo  {journal} {Proceedings of the Royal Society of London.
  Series A. Mathematical and Physical Sciences}\ }\textbf {\bibinfo {volume}
  {169}},\ \bibinfo {pages} {339} (\bibinfo {year} {1939})}\BibitemShut
  {NoStop}%
\bibitem [{\citenamefont {Ghazaryan}\ \emph {et~al.}(2021)\citenamefont
  {Ghazaryan}, \citenamefont {Holder}, \citenamefont {Serbyn},\ and\
  \citenamefont {Berg}}]{ghazaryan2021unconventional}%
  \BibitemOpen
  \bibfield  {author} {\bibinfo {author} {\bibfnamefont {A.}~\bibnamefont
  {Ghazaryan}}, \bibinfo {author} {\bibfnamefont {T.}~\bibnamefont {Holder}},
  \bibinfo {author} {\bibfnamefont {M.}~\bibnamefont {Serbyn}},\ and\ \bibinfo
  {author} {\bibfnamefont {E.}~\bibnamefont {Berg}},\ }\bibfield  {title}
  {\bibinfo {title} {Unconventional superconductivity in systems with annular
  fermi surfaces: Application to rhombohedral trilayer graphene},\ }\href@noop
  {} {\bibfield  {journal} {\bibinfo  {journal} {Physical review letters}\
  }\textbf {\bibinfo {volume} {127}},\ \bibinfo {pages} {247001} (\bibinfo
  {year} {2021})}\BibitemShut {NoStop}%
\bibitem [{\citenamefont {Zhou}\ \emph {et~al.}(2021)\citenamefont {Zhou},
  \citenamefont {Xie}, \citenamefont {Ghazaryan}, \citenamefont {Holder},
  \citenamefont {Ehrets}, \citenamefont {Spanton}, \citenamefont {Taniguchi},
  \citenamefont {Watanabe}, \citenamefont {Berg}, \citenamefont {Serbyn} \emph
  {et~al.}}]{zhou2021half}%
  \BibitemOpen
  \bibfield  {author} {\bibinfo {author} {\bibfnamefont {H.}~\bibnamefont
  {Zhou}}, \bibinfo {author} {\bibfnamefont {T.}~\bibnamefont {Xie}}, \bibinfo
  {author} {\bibfnamefont {A.}~\bibnamefont {Ghazaryan}}, \bibinfo {author}
  {\bibfnamefont {T.}~\bibnamefont {Holder}}, \bibinfo {author} {\bibfnamefont
  {J.~R.}\ \bibnamefont {Ehrets}}, \bibinfo {author} {\bibfnamefont {E.~M.}\
  \bibnamefont {Spanton}}, \bibinfo {author} {\bibfnamefont {T.}~\bibnamefont
  {Taniguchi}}, \bibinfo {author} {\bibfnamefont {K.}~\bibnamefont {Watanabe}},
  \bibinfo {author} {\bibfnamefont {E.}~\bibnamefont {Berg}}, \bibinfo {author}
  {\bibfnamefont {M.}~\bibnamefont {Serbyn}}, \emph {et~al.},\ }\bibfield
  {title} {\bibinfo {title} {Half-and quarter-metals in rhombohedral trilayer
  graphene},\ }\href@noop {} {\bibfield  {journal} {\bibinfo  {journal}
  {Nature}\ }\textbf {\bibinfo {volume} {598}},\ \bibinfo {pages} {429}
  (\bibinfo {year} {2021})}\BibitemShut {NoStop}%
\bibitem [{\citenamefont {Zhou}\ \emph {et~al.}(2022)\citenamefont {Zhou},
  \citenamefont {Holleis}, \citenamefont {Saito}, \citenamefont {Cohen},
  \citenamefont {Huynh}, \citenamefont {Patterson}, \citenamefont {Yang},
  \citenamefont {Taniguchi}, \citenamefont {Watanabe},\ and\ \citenamefont
  {Young}}]{zhou2022isospin}%
  \BibitemOpen
  \bibfield  {author} {\bibinfo {author} {\bibfnamefont {H.}~\bibnamefont
  {Zhou}}, \bibinfo {author} {\bibfnamefont {L.}~\bibnamefont {Holleis}},
  \bibinfo {author} {\bibfnamefont {Y.}~\bibnamefont {Saito}}, \bibinfo
  {author} {\bibfnamefont {L.}~\bibnamefont {Cohen}}, \bibinfo {author}
  {\bibfnamefont {W.}~\bibnamefont {Huynh}}, \bibinfo {author} {\bibfnamefont
  {C.~L.}\ \bibnamefont {Patterson}}, \bibinfo {author} {\bibfnamefont
  {F.}~\bibnamefont {Yang}}, \bibinfo {author} {\bibfnamefont {T.}~\bibnamefont
  {Taniguchi}}, \bibinfo {author} {\bibfnamefont {K.}~\bibnamefont
  {Watanabe}},\ and\ \bibinfo {author} {\bibfnamefont {A.~F.}\ \bibnamefont
  {Young}},\ }\bibfield  {title} {\bibinfo {title} {Isospin magnetism and
  spin-polarized superconductivity in bernal bilayer graphene},\ }\href@noop {}
  {\bibfield  {journal} {\bibinfo  {journal} {Science}\ }\textbf {\bibinfo
  {volume} {375}},\ \bibinfo {pages} {774} (\bibinfo {year}
  {2022})}\BibitemShut {NoStop}%
\bibitem [{\citenamefont {Chatterjee}\ \emph {et~al.}(2022)\citenamefont
  {Chatterjee}, \citenamefont {Wang}, \citenamefont {Berg},\ and\ \citenamefont
  {Zaletel}}]{chatterjee2022inter}%
  \BibitemOpen
  \bibfield  {author} {\bibinfo {author} {\bibfnamefont {S.}~\bibnamefont
  {Chatterjee}}, \bibinfo {author} {\bibfnamefont {T.}~\bibnamefont {Wang}},
  \bibinfo {author} {\bibfnamefont {E.}~\bibnamefont {Berg}},\ and\ \bibinfo
  {author} {\bibfnamefont {M.~P.}\ \bibnamefont {Zaletel}},\ }\bibfield
  {title} {\bibinfo {title} {Inter-valley coherent order and isospin
  fluctuation mediated superconductivity in rhombohedral trilayer graphene},\
  }\href@noop {} {\bibfield  {journal} {\bibinfo  {journal} {Nature
  communications}\ }\textbf {\bibinfo {volume} {13}},\ \bibinfo {pages} {6013}
  (\bibinfo {year} {2022})}\BibitemShut {NoStop}%
\bibitem [{\citenamefont {Liu}\ \emph {et~al.}(2022)\citenamefont {Liu},
  \citenamefont {Zhang}, \citenamefont {Watanabe}, \citenamefont {Taniguchi},\
  and\ \citenamefont {Li}}]{liu2022isospin}%
  \BibitemOpen
  \bibfield  {author} {\bibinfo {author} {\bibfnamefont {X.}~\bibnamefont
  {Liu}}, \bibinfo {author} {\bibfnamefont {N.~J.}\ \bibnamefont {Zhang}},
  \bibinfo {author} {\bibfnamefont {K.}~\bibnamefont {Watanabe}}, \bibinfo
  {author} {\bibfnamefont {T.}~\bibnamefont {Taniguchi}},\ and\ \bibinfo
  {author} {\bibfnamefont {J.}~\bibnamefont {Li}},\ }\bibfield  {title}
  {\bibinfo {title} {Isospin order in superconducting magic-angle twisted
  trilayer graphene},\ }\href@noop {} {\bibfield  {journal} {\bibinfo
  {journal} {Nature Physics}\ }\textbf {\bibinfo {volume} {18}},\ \bibinfo
  {pages} {522} (\bibinfo {year} {2022})}\BibitemShut {NoStop}%
\bibitem [{\citenamefont {Dong}\ \emph {et~al.}(2023)\citenamefont {Dong},
  \citenamefont {Davydova}, \citenamefont {Ogunnaike},\ and\ \citenamefont
  {Levitov}}]{dong2023isospin}%
  \BibitemOpen
  \bibfield  {author} {\bibinfo {author} {\bibfnamefont {Z.}~\bibnamefont
  {Dong}}, \bibinfo {author} {\bibfnamefont {M.}~\bibnamefont {Davydova}},
  \bibinfo {author} {\bibfnamefont {O.}~\bibnamefont {Ogunnaike}},\ and\
  \bibinfo {author} {\bibfnamefont {L.}~\bibnamefont {Levitov}},\ }\bibfield
  {title} {\bibinfo {title} {Isospin-and momentum-polarized orders in bilayer
  graphene},\ }\href@noop {} {\bibfield  {journal} {\bibinfo  {journal}
  {Physical Review B}\ }\textbf {\bibinfo {volume} {107}},\ \bibinfo {pages}
  {075108} (\bibinfo {year} {2023})}\BibitemShut {NoStop}%
\bibitem [{\citenamefont {Jungwirth}\ \emph {et~al.}(2024)\citenamefont
  {Jungwirth}, \citenamefont {Fernandes}, \citenamefont {Fradkin},
  \citenamefont {MacDonald}, \citenamefont {Sinova},\ and\ \citenamefont
  {Smejkal}}]{jungwirth2024supefluid}%
  \BibitemOpen
  \bibfield  {author} {\bibinfo {author} {\bibfnamefont {T.}~\bibnamefont
  {Jungwirth}}, \bibinfo {author} {\bibfnamefont {R.}~\bibnamefont
  {Fernandes}}, \bibinfo {author} {\bibfnamefont {E.}~\bibnamefont {Fradkin}},
  \bibinfo {author} {\bibfnamefont {A.}~\bibnamefont {MacDonald}}, \bibinfo
  {author} {\bibfnamefont {J.}~\bibnamefont {Sinova}},\ and\ \bibinfo {author}
  {\bibfnamefont {L.}~\bibnamefont {Smejkal}},\ }\bibfield  {title} {\bibinfo
  {title} {From supefluid 3he to altermagnets},\ }\href@noop {} {\bibfield
  {journal} {\bibinfo  {journal} {arXiv preprint arXiv:2411.00717}\ } (\bibinfo
  {year} {2024})}\BibitemShut {NoStop}%
\bibitem [{\citenamefont {Balents}\ \emph {et~al.}(2020)\citenamefont
  {Balents}, \citenamefont {Dean}, \citenamefont {Efetov},\ and\ \citenamefont
  {Young}}]{balents2020superconductivity}%
  \BibitemOpen
  \bibfield  {author} {\bibinfo {author} {\bibfnamefont {L.}~\bibnamefont
  {Balents}}, \bibinfo {author} {\bibfnamefont {C.~R.}\ \bibnamefont {Dean}},
  \bibinfo {author} {\bibfnamefont {D.~K.}\ \bibnamefont {Efetov}},\ and\
  \bibinfo {author} {\bibfnamefont {A.~F.}\ \bibnamefont {Young}},\ }\bibfield
  {title} {\bibinfo {title} {Superconductivity and strong correlations in
  moir{\'e} flat bands},\ }\href@noop {} {\bibfield  {journal} {\bibinfo
  {journal} {Nature Physics}\ }\textbf {\bibinfo {volume} {16}},\ \bibinfo
  {pages} {725} (\bibinfo {year} {2020})}\BibitemShut {NoStop}%
\bibitem [{\citenamefont {Andrei}\ \emph {et~al.}(2021)\citenamefont {Andrei},
  \citenamefont {Efetov}, \citenamefont {Jarillo-Herrero}, \citenamefont
  {MacDonald}, \citenamefont {Mak}, \citenamefont {Senthil}, \citenamefont
  {Tutuc}, \citenamefont {Yazdani},\ and\ \citenamefont
  {Young}}]{andrei2021marvels}%
  \BibitemOpen
  \bibfield  {author} {\bibinfo {author} {\bibfnamefont {E.~Y.}\ \bibnamefont
  {Andrei}}, \bibinfo {author} {\bibfnamefont {D.~K.}\ \bibnamefont {Efetov}},
  \bibinfo {author} {\bibfnamefont {P.}~\bibnamefont {Jarillo-Herrero}},
  \bibinfo {author} {\bibfnamefont {A.~H.}\ \bibnamefont {MacDonald}}, \bibinfo
  {author} {\bibfnamefont {K.~F.}\ \bibnamefont {Mak}}, \bibinfo {author}
  {\bibfnamefont {T.}~\bibnamefont {Senthil}}, \bibinfo {author} {\bibfnamefont
  {E.}~\bibnamefont {Tutuc}}, \bibinfo {author} {\bibfnamefont
  {A.}~\bibnamefont {Yazdani}},\ and\ \bibinfo {author} {\bibfnamefont {A.~F.}\
  \bibnamefont {Young}},\ }\bibfield  {title} {\bibinfo {title} {The marvels of
  moir{\'e} materials},\ }\href@noop {} {\bibfield  {journal} {\bibinfo
  {journal} {Nature Reviews Materials}\ }\textbf {\bibinfo {volume} {6}},\
  \bibinfo {pages} {201} (\bibinfo {year} {2021})}\BibitemShut {NoStop}%
\bibitem [{\citenamefont {Lucas}\ and\ \citenamefont
  {Das~Sarma}(2018)}]{lucas2018electronic}%
  \BibitemOpen
  \bibfield  {author} {\bibinfo {author} {\bibfnamefont {A.}~\bibnamefont
  {Lucas}}\ and\ \bibinfo {author} {\bibfnamefont {S.}~\bibnamefont
  {Das~Sarma}},\ }\bibfield  {title} {\bibinfo {title} {Electronic sound modes
  and plasmons in hydrodynamic two-dimensional metals},\ }\href@noop {}
  {\bibfield  {journal} {\bibinfo  {journal} {Physical Review B}\ }\textbf
  {\bibinfo {volume} {97}},\ \bibinfo {pages} {115449} (\bibinfo {year}
  {2018})}\BibitemShut {NoStop}%
\bibitem [{\citenamefont {Klein}\ \emph {et~al.}(2020)\citenamefont {Klein},
  \citenamefont {Maslov},\ and\ \citenamefont {Chubukov}}]{klein2020hidden}%
  \BibitemOpen
  \bibfield  {author} {\bibinfo {author} {\bibfnamefont {A.}~\bibnamefont
  {Klein}}, \bibinfo {author} {\bibfnamefont {D.~L.}\ \bibnamefont {Maslov}},\
  and\ \bibinfo {author} {\bibfnamefont {A.~V.}\ \bibnamefont {Chubukov}},\
  }\bibfield  {title} {\bibinfo {title} {Hidden and mirage collective modes in
  two dimensional fermi liquids},\ }\href@noop {} {\bibfield  {journal}
  {\bibinfo  {journal} {npj Quantum Materials}\ }\textbf {\bibinfo {volume}
  {5}},\ \bibinfo {pages} {55} (\bibinfo {year} {2020})}\BibitemShut {NoStop}%
\end{thebibliography}%

\end{document}